\documentclass[english,12pt,a4paper]{article}
\usepackage[T1]{fontenc}
\usepackage[utf8]{inputenc}
\usepackage[english]{babel}
\usepackage[pdftex]{graphicx}
\usepackage[width=\textwidth]{caption}
\usepackage{amsmath,amsopn,amssymb,bm, dsfont,mathrsfs}
\usepackage[top=1.25in, bottom=1.25in, left=1.25in, right=1.25in]{geometry}
\usepackage{array,float, placeins,flafter, longtable,booktabs,pdflscape}
\usepackage{verbatim}
\usepackage{color}
\usepackage{natbib}
\usepackage{lmodern}
\usepackage{tabularx}
\usepackage{xr}
\usepackage{mathtools}
\usepackage[shortlabels]{enumitem}
\usepackage{relsize}
\usepackage{url}

\usepackage{threeparttable}
\usepackage{tablefootnote}
\makeatletter

\newtheorem{thm}{Theorem}

\newtheorem{prop}{Proposition}
\newtheorem{lem}{Lemma}
\newtheorem{hyp}{Assumption}

\newcommand{\Supp}{\text{Supp}}
\newcommand{\norm}[1]{\left\Vert#1\right\Vert}
\newcommand{\abs}[1]{\left\vert#1\right\vert}
\newcommand{\I}{\mathcal{I}}

\newcommand{\ind}[1]{\mathds{1}\left\{#1\right\}}

\newcommand{\eps}{\varepsilon}
\newcommand{\deriv}[2]{\partial #1/\partial #2}

\newcommand{\indep}{\perp \!\!\! \perp}
\newcommand{\convP}{\stackrel{P}{\longrightarrow}}
\newcommand{\convL}{\stackrel{d}{\longrightarrow}}
\newcommand{\sgn}{\text{sgn}}
\newcommand{\st}[1]{\texttt{#1}}

\renewcommand{\section}{\@startsection{section}{2}{0mm}{-1\baselineskip}{1\baselineskip}{\normalfont\large\bfseries}}
\renewcommand{\subsection}{\@startsection{subsection}{2}{0mm}{-1\baselineskip}{0.7\baselineskip}{\normalfont\normalsize\bfseries}}
\renewcommand{\subsubsection}{\@startsection{subsubsection}{3}{0mm}{-0.8\baselineskip}{0.4\baselineskip}{\normalfont\normalsize\itshape}}

\newcommand*{\storecounter}[2]{%
  \edef\@currentlabel{\the\value{#1}}
  \label{#2}
}

\date{First version: January 18, 2022. This version: \today}

\begin{document}

\title{Difference-in-Differences for Continuous Treatments and Instruments with Stayers\thanks{We thank Matias Cattaneo, Max Farrell, Joachim Freyberger, Thomas Le Barbanchon, and Andres Santos for helpful comments. Chaisemartin was funded by the European Union (ERC, REALLYCREDIBLE, GA Number 101043899). Views and opinions expressed are those of the authors and do not reflect those of the European Union or the European Research Council Executive Agency. Neither the European Union nor the granting authority can be held responsible for them. We have used Claude and ChatGPT to improve the wording of our abstract and introduction. We have used Gemini for part A.4.2 of the proof of Theorem 3.}}

\author{Cl\'{e}ment de Chaisemartin\thanks{Sciences Po Paris, clement.dechaisemartin@sciencespo.fr%
} \and Xavier D'Haultf\oe{}uille%
\thanks{CREST-ENSAE, xavier.dhaultfoeuille@ensae.fr.}
\and Doulo Sow\thanks{Princeton University, doulo.sow@princeton.edu.} \and Gonzalo
Vazquez-Bare\thanks{University of California, Santa Barbara,
gvazquez@econ.ucsb.edu.} \and F\'{e}lix Pasquier\thanks{CREST-ENSAE, felix.pasquier@ensae.fr.}}

\maketitle ~\vspace{-1.9cm}

\begin{abstract}
When studying the effects of taxes, tariffs, or prices using panel data, the treatment is often continuously distributed in every period. We develop difference-in-differences (DID) estimators for such settings. We partition units into switchers, whose treatment changes between consecutive periods, and stayers, whose treatment remains constant. Under a parallel-trends assumption, we show that the slopes of switchers' potential outcomes with respect to the treatment are nonparametrically identified by DID comparisons between switchers and stayers sharing the same baseline treatment level. Conditioning on the baseline treatment is key, as it ensures that the underlying parallel-trends assumption accommodates time-varying treatment effects. We then study two weighted averages of these slopes, discuss their respective advantages, and propose for each a doubly robust, semiparametrically efficient, and $\sqrt{n}$-consistent estimator. Finally, we extend our framework to instrumental variables and illustrate it by estimating the effects of gasoline taxes on prices and fuel consumption.
\end{abstract}

\section{Introduction}

To estimate a treatment's effect, researchers often run a two-way fixed
effects (TWFE) regression:
$$Y_{i,t}=\alpha_i+\gamma_t+\beta_{TWFE}D_{i,t}+u_{i,t},$$
where $D_{i,t}$ denotes the treatment status of unit $i$ at time $t$, and
$\alpha_i$ and $\gamma_t$ are unit and period fixed effects. TWFE regressions
are widely used: \cite{de2020difference} find that 26 of the 100
most-cited papers published in the American Economic Review from 2015 to 2019
use this specification, while \cite{chiu2023and} find that 52 papers
published in the top three political science journals from 2017 to 2022 do
likewise. Researchers have long assumed that TWFE regressions rely solely on
a parallel-trends assumption --- namely, that in the absence of treatment,
all units would have experienced the same evolution in outcomes. However,
recent work has shown that this assumption alone is insufficient to guarantee
a well-behaved estimand: $\beta_{TWFE}$ may recover a non-convex weighted
sum of treatment effects across units and periods \cite[see,
e.g.,][]{deChaisemartin2018twowayfev4arxiv,goodman2021difference,borusyak2020revisiting},
and can therefore be negative even when the treatment effect is positive for
every unit-period pair $(i,t)$.

\medskip
In response, several ``heterogeneity-robust difference-in-differences''
estimators --- consistent for convex combinations of $(i,t)$-specific effects
under a parallel-trends assumption --- have been proposed. However, none
apply to treatments continuously distributed at all periods, despite their prevalence in
practice. In economics, taxes \citep[see][]{li2014gasoline} and tariffs
\citep[see][]{fajgelbaum2020return} are often continuously distributed across
all periods; similar examples arise in political science and epidemiology
\citep{peterson2021paper,aklin2021side,mcnamara2022estimating}. With
continuous treatments, TWFE
regressions may estimate highly non-convex combinations of effects ---
with two time periods, exactly 50\% of $(i,t)$-specific effects receive
negative weight \citep{de2021more}. Developing robust estimators for such
treatments is therefore of first-order importance.

\medskip
We assume a two-period panel. From period one to two, some units ---
the \textit{switchers} --- change their treatment, while others ---
the \textit{stayers} --- do not. We propose a parallel-trends assumption
requiring that switchers and stayers with the same period-one treatment level would
have experienced the same average outcome evolution absent any change in
switchers' treatment. This assumption has three key features. First, by
conditioning on the period-one treatment level, it imposes no restriction on
treatment-effect heterogeneity: indeed, assuming parallel trends across units with different period-one treatment levels would essentially rule out time-varying treatment effects.
Second, when a pre-switch
period (period zero) is available, the assumption can be assessed with a
pre-trend test comparing the period-zero-to-one outcome evolutions of
period-one-to-two switchers and stayers. Third, it is compatible with a Roy model of selection into treatment, under some restrictions on the potential outcomes.

\medskip
We consider two target parameters. The first, the Average of Slopes (AS), is the average slope of switchers' period-two potential outcome function between their period-one and period-two treatment levels. The second, the Weighted Average of Slopes (WAS), weights switchers in proportion to the absolute magnitude of their treatment change. The AS and WAS serve complementary purposes. Under shape restrictions on the potential outcome function, the AS can identify or bound the effects of treatment changes other than those observed, while the WAS can be used to conduct cost-benefit analyses of the treatment changes that actually occurred. Under our parallel-trends assumption—which can be assessed with a pre-trend test—both parameters are nonparametrically identified by difference-in-differences estimands comparing the outcome evolutions of switchers and stayers with the same period-one treatment level. This contrasts with other causal parameters studied in the literature, such as the dose-response function and the average marginal effect, whose identification relies on assumptions that cannot be assessed through pre-trend tests.

\medskip
With a continuous treatment, the sample never contains switchers and stayers with exactly the same
period-one treatment.
Then, the estimands identifying the AS and WAS depend on
nonparametric regression functions. We therefore derive doubly-robust moment
conditions and propose corresponding doubly-robust, semiparametrically efficient estimators. The AS
estimator converges at the parametric rate provided switchers cannot
experience arbitrarily small treatment changes; the WAS estimator always
converges at the parametric rate and, under some conditions, achieves a
strictly lower asymptotic variance than the AS estimator. Our estimators are computed by the
\texttt{did\_multiplegt\_stat} Stata and Python packages.\footnote{\label{foot:discrete}Of course, our estimators can also be used when the treatment variable has mass points, and in particular when the treatment is discrete. In that case, it is likely that switchers can be matched to stayers with the same period-one treatment, so estimation is easier. The corresponding estimators are computed by \texttt{did\_multiplegt\_stat} when the \texttt{exact\_match} option is specified.}

\medskip
We extend these results in several directions. First, we consider an
instrumental-variable (IV) setting in which the parallel-trends assumption is
made with respect to an instrument rather than the treatment. For instance,
when estimating the price elasticity of demand, prices may respond to demand
shocks, violating parallel trends; taxes can instead serve as an instrument. We first show that a parallel-trends assumption with
respect to an instrument restricts treatment-effect heterogeneity. With this caveat in mind, we then show that
the reduced-form WAS effect of the instrument on the outcome divided by the first-stage WAS
effect of the instrument on the treatment equals a weighted average of outcome slopes with respect to the
treatment, with more weight given to switchers with larger first-stage
effects. Second, we extend our framework to more than two periods: for each
consecutive pair $(t-1,t)$, we compare outcome evolutions of units switching
and not switching treatment, controlling for their period-$t-1$ treatment,
and then aggregate across $t$. Third, we propose a pre-trends estimator that
compares $t-2$-to-$t-1$ outcome evolutions of $t-1$-to-$t$ switchers and
stayers. Fourth, as our baseline estimators assume
away dynamic effects, we propose a modified version robust to dynamic effects
up to a pre-specified treatment lag. Finally, we propose estimators
accommodating control variables.

\medskip
As an illustration, we apply our estimators to the yearly 1966--2008
US state-level panel of \cite{li2014gasoline}, estimating the effect of
gasoline taxes on consumption and prices. The WAS estimates yield
significantly negative effects on consumption and significantly positive
effects on prices. The AS estimates are close in magnitude but substantially
less precise --- their standard errors are nearly three times larger ---
and are statistically insignificant. Our pre-trend estimates
are small and insignificant, lending credibility to the parallel-trends
assumption.

\medskip
Throughout, we assume that stayers exist in the data. While this typically
holds, this fails in settings where all units change their treatment every
period --- for instance, when $D_{i,t}$ is average temperature in location $i$ and year $t$. In such
cases, one may redefine the treatment to restore the presence of stayers
(e.g., as the number of days exceeding a temperature threshold). However, such redefinitions are not innocuous: they
implicitly assume that, conditional on the redefined treatment, variation in
the original treatment does not affect the outcome. This may be defensible
when external evidence supports the chosen treatment definition (e.g. there is laboratory evidence showing that a crop's growth is only impaired when temperature goes above a threshold), but there may be settings
where no such evidence exists. We encourage researchers to be explicit about, and
to defend, the structural outcome model underlying their treatment definition.

\subsection*{Related Literature.}

Our paper builds on several strands of the panel data literature.
\cite{chamberlain1982multivariate} appears to be the first to have proposed
an estimator of the AS parameter; under the assumption of no counterfactual
time trend, that estimator reduces to a before-after estimator.
\cite{graham2012identification} propose DID estimators of the AS (see their
Equation (21)) when the treatment is continuously distributed at every
period. They rely on a linear effect assumption and require that treatment
effects evolve uniformly across units. Our estimator of the AS imposes no
such restrictions. Beyond the AS, our main contribution to this literature is
to introduce the WAS and to clarify the distinct purposes served by the two
parameters.

\medskip
With respect to the heterogeneity-robust DID literature, we make two
contributions. First, in the non-IV case, we propose estimators applicable
when units receive heterogeneous treatment doses at baseline, complementing a
literature that has focused almost exclusively on the case where all units
are untreated at baseline
\citep{abraham2018,callaway2020difference,liu2024practical,borusyak2020revisiting,callaway2021difference,de2024two}.
One prior exception is \cite{deChaisemartin2018twowayfev4arxiv}, who, in
their web appendix, consider designs with heterogeneous baseline doses of a
non-binary \textit{discrete} treatment, and propose DID estimators comparing
switchers and stayers with exactly the same baseline treatment. This approach
is no longer feasible with a continuously distributed treatment, which
motivates the estimation strategy developed here. A contemporaneous exception
is \cite{de2020difference}, who propose estimators robust to dynamic effects
up to any lag. While they primarily focus on discrete treatments, Section 1.10
of their web appendix sketches an extension to continuous treatments building
on the current paper; however, the estimators therein are parametric and not
doubly-robust, and their asymptotic distributions are not derived. Furthermore, while
allowing for effects of lagged treatments up to any lag is appealing to avoid misspecification, doing so can also lead to much more imprecise estimators: in our empirical application, all the estimators of \cite{de2020difference} are insignificant, with standard errors around 10 times larger than those of the estimators in the current paper. Another approach, proposed by \cite{de2025treatmenteffect} after the first version of this paper, relies on a distributed-lag linear model for potential outcomes, with coefficients that may vary across groups. With respect to the WAS and AS estimators, this approach accommodates effects of lagged taxes, up to a prespecified number of lags. On the other hand it assumes that treatment effects are linear and time-invariant. In our empirical application, both approaches lead to qualitatively similar conclusions regarding the effect of
contemporaneous gasoline taxes on consumption and prices, which is reassuring because each approach has its pros and cons.\footnote{Our results are not directly related to those of
\citet{TVB}, who consider designs where switchers and stayers start from
different treatment levels but focus on discrete rather than continuous
treatments, and who analyze standard TWFE estimators rather than proposing
nonparametric heterogeneity-robust DID estimators.}

\medskip
Second, in the IV case, we extend \cite{de2010note} and
\cite{hudson2017interpreting}, whose results are limited to two-period
designs with a binary instrument and treatment. We also appear to be the
first to note that a parallel-trends assumption with respect to an instrument
restricts treatment-effect heterogeneity.

\paragraph{Organization.} Section \ref{sec:setup} presents the setup,
our main assumptions, and a key identification result. Sections
\ref{sec:AS} and \ref{sec:WAS} cover identification and estimation of the AS
and WAS, respectively. Section \ref{sec:extensions} presents extensions,
some of which are detailed in the web appendix. Section \ref{sec:appli}
presents the empirical application. The appendix collects all proofs.

\section{Setup, assumptions, and building-block identification result}
\label{sec:setup}

\subsection{Setup}

A representative unit is drawn from an infinite super population, and
observed at two time periods. This unit could be an individual or a firm, but
it could also be a geographical unit, like a county or a region.\footnote{In
that case, one may want to weight the estimation by counties' or regions'
populations. Extending the estimators we propose to allow for such weighting
is a mechanical extension.} All expectations below are taken with respect to
the distribution of variables in the super population. We are interested in
the effect of a continuous and scalar treatment variable on that unit's
outcome. Let $D_1$  (resp. $D_2$) denote the unit's treatment at period 1
(resp. 2), let $\mathcal{D}_1$  (resp. $\mathcal{D}_2$) denote the support of $D_1$  (resp. $D_2$), and let $\mathcal{D}$ denote the support of $(D_1,D_2)$. Let $S=1\{D_2\ne
D_1\}$ be an indicator equal to 1 if the unit's treatment changes from period
one to two, i.e. if they are a switcher.

\medskip
For all $(d_1,d_2)\in \mathcal{D}$, let $Y_1(d_1,d_2)$ and $Y_2(d_1,d_2)$
respectively denote the unit's potential outcomes at periods 1 and 2 if $(D_1,D_2)=(d_1,d_2)$, and let $Y_1$ and $Y_2$ denote their observed outcomes.

\medskip
In what follows, all equalities and inequalities involving random variables
are required to hold almost surely. Finally, for any random variable observed
at the two time periods $(X_1,X_2)$, let $\Delta X=X_2-X_1$ denote the change
of $X$ from period $1$ to $2$.

\subsection{Assumptions}

We make the following assumptions.

\begin{hyp}\label{hyp:noanticipation} (No anticipation)
For all $(d_1,d_2)\in \mathcal{D}$, $Y_1(d_1,d_2)=Y_1(d_1).$
\end{hyp}
\begin{hyp}\label{hyp:nodynamic} (No dynamic effects)
For all $(d_1,d_2)\in \mathcal{D}$, $Y_2(d_1,d_2)=Y_2(d_2).$
\end{hyp}
Assumption \ref{hyp:noanticipation} requires that the period-one outcome does not depend on units' period-two treatment, thus ruling out anticipation effects, a commonly-made
assumption in the DID literature. Assumption \ref{hyp:nodynamic} requires that the period-two outcome does not depend on units' period-one treatment, thus ruling out dynamic effects. In Section \ref{sec:extensions}, we propose an extension of our
estimators, which relaxes Assumption \ref{hyp:nodynamic} by allowing for dynamic effects
up to a pre-specified treatment lag. In what follows, we let $\Delta Y(d_1)=Y_2(d_1)-Y_1(d_1)$ represent the unit's counterfactual outcome evolution if their treatment is equal to $d_1$ at periods one and two.

\begin{hyp}\label{hyp:parallel_trends} (Parallel trends)
$\forall$ $d_1\in \mathcal{D}_1$,  $E(\Delta Y(d_1)|D_1=d_1,D_2)=E(\Delta
Y(d_1)|D_1=d_1)$.
\end{hyp}
Assumption \ref{hyp:parallel_trends} is a conditional parallel-trends assumption, requiring that units with different period-two treatments but the same period-one treatment have the same expected outcome evolution if their treatment does not change.

\begin{hyp}\label{hyp:regularity} (Bounded treatment and regularity of potential outcomes)
\begin{enumerate}
\item $\mathcal{D}_1$ and $\mathcal{D}_2$ are bounded subsets of
    $\mathbb{R}$.
\item $\sup_{d_1\in\mathcal{D}_1} E[(\Delta Y(d_1))^2|D_1=d_1]<\infty$.
\item $\forall$ $t\in \{1,2\}$ and $\forall$ $(d,d')\in \mathcal{D}_t^2$,
    there is a random variable $\overline{Y}\ge 0$ such that
    $|Y_t(d)-Y_t(d')|\leq \overline{Y}|d-d'|$, with $\sup_{(d_1,d_2)\in
    \mathcal{D}}E[\overline{Y}^2|D_1=d_1, D_2=d_2]<\infty$.
\end{enumerate}
\end{hyp}
Assumption \ref{hyp:regularity} ensures that all the expectations below are well defined. It requires that the set of values that the period-one and period-two treatments can take be bounded. It also requires that the potential outcome functions be Lipschitz (with a unit-specific Lipschitz constant). This holds if $d\mapsto Y_2(d)$ is differentiable with respect to $d$ and has a bounded derivative. Assumption \ref{hyp:regularity} also holds if $D_t$ is finitely supported (and $E[Y_t(d)^2]<\infty$ for all $d\in\mathcal{D}_t$).

\medskip
Finally, for estimation and inference we assume we observe an iid sample with
the same distribution as $(Y_{1},Y_{2},D_{1},D_{2})$:
\begin{hyp}\label{hyp:iid} (iid sample)
We observe $(Y_{i,1},Y_{i,2},D_{i,1},D_{i,2})_{1\leq i\leq n}$, which are
independent and identically distributed vectors with the same probability
distribution as $(Y_{1},Y_{2},D_{1},D_{2})$.
\end{hyp}
Importantly, Assumption \ref{hyp:iid} allows for the possibility that $Y_{1}$
and $Y_{2}$ (resp. $D_{1}$ and $D_{2}$) are serially correlated, as is
commonly assumed in DID studies \citep[see][]{bertrand2004}.

\subsection{Rationalizing Assumption \ref{hyp:parallel_trends} under a Roy selection model}
\label{sub:Roy}
In this section, we outline an outcome equation and a Roy selection model under which Assumption \ref{hyp:parallel_trends} holds. Let $\gamma_t(d)=E(Y_t(d))$ and $u_t(d)=Y_t(d)-E(Y_t(d))$. Assume that there exist two mean-zero random variables $A(d)$ and $\eps_t$ such that
\begin{equation}\label{eq:additive_separability_shock}
u_t(d)=A(d)+\eps_t:
\end{equation}
$u_t(d)$ is additively separable into a random variable $A(d)$, that can vary across treatment states but is time-invariant, and another random variable $\eps_t$, that can vary over time but does not vary across treatment states. If \eqref{eq:additive_separability_shock} holds,
\begin{equation}
Y_t(d)= A(d)+ \gamma_t(d)+\eps_t.
	\label{eq:pot_outc_model}
\end{equation}
Next, assume the unit chooses its treatment according to the following Roy model:
\begin{equation}
D_t :=\arg\max_{d\in\mathcal{D}_t} Y_t(d) - C_t(d) - C^s_t\ind{d\ne D_{t-1}},	
	\label{eq:modele_d}
\end{equation}
where $C_t(.)$ denotes the cost of treatment $d$, and $C^s_t$ denotes a switching cost, incurred when changing treatment from $t-1$ to $t$.\footnote{Thanks to the switching cost, there can be stayers under \eqref{eq:modele_d}: under support conditions, $P(D_t=D_{t-1})>0$. Also, we can have switchers even if $C_t$ and $C^s_t$ do not change with $t$, thanks to $\gamma_t(d)$ in $Y_t(d)$.} Plugging \eqref{eq:pot_outc_model} into \eqref{eq:modele_d},
\begin{equation}
D_t =\arg\max_{d\in\mathcal{D}_t} A(d)+ \gamma_t(d) - C_t(d) - C^s_t\ind{d\ne D_{t-1}},	
	\label{eq:modele_d_pot_outc_model}
\end{equation}
so $D_t$ is a function of $(A(d),(C_t(d))_d$  and $C_t^s$.
Finally, assume that $\eps_t$ satisfies a parallel-trends assumption conditional on $(A(d),C_{1}(d),C_{1}^s,C_{2}(d),C_{2}^s):$
\begin{equation}\label{eq:strong_exo}
E[\eps_2-\eps_1|A(d),C_{1}(d),C_{1}^s,C_{2}(d),C_{2}^s]=0.
\end{equation}
Then, $E[\eps_2-\eps_1|D_2,D_1]=0$. Combined with \eqref{eq:pot_outc_model}, the previous display implies that Assumption \ref{hyp:parallel_trends} holds.

\medskip
By ensuring that $D_t$ does not depend on idiosyncratic shocks affecting potential outcomes, \eqref{eq:additive_separability_shock} is the key condition ensuring that Assumption \ref{hyp:parallel_trends} can be rationalized under a Roy selection model. \eqref{eq:additive_separability_shock} is not innocuous. For instance, it implies that all units experience the same change in their treatment effect over time:
$$Y_2(d)-Y_2(d')-(Y_1(d)-Y_1(d'))=E(Y_2(d)-Y_2(d'))-E(Y_1(d)-Y_1(d')).$$
If \eqref{eq:additive_separability_shock} fails, it may still be possible to rationalize Assumption \ref{hyp:parallel_trends} under a Roy model, though doing so may require making assumptions that restrict how the idiosyncratic shocks can enter into the selection equation \citep{ghanem2022selection}.

\subsection{Building-block identification result}

Assumption \ref{hyp:parallel_trends} implies the following lemma, our
building-block identification result.
\begin{lem}\label{lem:building_block}
If Assumptions \ref{hyp:noanticipation}-\ref{hyp:parallel_trends} hold, then $\forall$ $(d_1,d_2)\in
\mathcal{D}$ such that $d_1\ne d_2$ and $P(S=1|D_1=d_1)<1$,
\begin{align*}
\text{TE}(d_1,d_2|d_1,d_2):=&E\left(\frac{Y_2(d_2) - Y_2(d_1)}{d_2-d_1}\middle|D_1=d_1,D_2=d_2\right)\\
=&E\left(\frac{\Delta Y - E(\Delta Y | D_1=d_1, S=0)}{d_2-d_1}\middle|D_1=d_1,D_2=d_2\right).
\end{align*}
\end{lem}

\medskip
Intuitively, under Assumption \ref{hyp:parallel_trends} the counterfactual
outcome evolution switchers would have experienced if their treatment had not
changed is identified by the outcome evolution of stayers with the same
period-one treatment. If a unit's treatment changes from two to five, we can
recover its counterfactual outcome evolution if its treatment had not
changed, by using the average outcome evolution of all stayers with a
baseline treatment of two. Then, a DID estimand comparing switchers' and
stayers' outcome evolutions identifies $E\left(Y_2(d_2) -
Y_2(d_1)\middle|D_1=d_1,D_2=d_2\right)$, and we can scale that effect by
$d_2-d_1$ to identify a slope.

\medskip
In a canonical DID design where $\mathcal{D}_1=\{0\}$ and $\mathcal{D}_2\in \{0,1\}$, Lemma \ref{lem:building_block} only applies to $(d_1,d_2)=(0,1)$, $\text{TE}(0,1|0,1)$ reduces to the average treatment on the treated (ATT),
and the estimand reduces to the canonical DID estimand comparing the outcome evolutions of treated and untreated units. Also, if all units are untreated at period one ($\mathcal{D}_1=\{0\}$), we have
$$\text{TE}(0,d_2|0,d_2)=E\left(\frac{Y_2(d_2) - Y_2(0)}{d_2}\middle|D_2=d_2\right),$$ an effect related to the $\Delta_{d0|D=d}$ effect in \cite{fricke2017identification}, or to the $ATT(d|d)$ effect in \cite{callaway2021difference}. The effects we consider are extensions of those effects.

\medskip
In the next sections, our target parameters will be two averages of the slopes $\text{TE}(d_1,d_2|d_1,d_2)$, across both $d_1$ and $d_2$, which can be estimated nonparametrically at the $\sqrt{n}-$parametric rate. We do not propose estimators of $(d_1,d_2)\mapsto
\text{TE}(d_1,d_2|d_1,d_2)$, for two reasons. First, variability in $\text{TE}(d_1,d_2|d_1,d_2)$ across values of $(d_1,d_2)$ conflates a dose-response relationship that may be of interest, and a selection bias, typically not of interest, due to the fact that units with different period one and two treatments may have different treatment effects \citep{callaway2021difference}. Moreover, Lemma
\ref{lem:building_block} shows that estimating $\text{TE}(d_1,d_2|d_1,d_2)$ requires estimating two bivariate nonparametric regressions. Unless one is willing to make parametric functional-form assumptions, under usual smoothness assumptions the resulting estimator converges at most at the $n^{1/3}$ rate \citep{stone1982optimal},
and may therefore be imprecise given that DID applications often have a small number of units (e.g. the 50 US states in our empirical application). Similarly, while nonparametric estimators of $d_1\mapsto E\left(\frac{Y_2(D_2)- Y_2(d_1)}{D_2-d_1}\middle|D_1=d_1,D_2\ne d_1\right)$ or $\delta \mapsto E\left(\frac{Y_2(D_2)- Y_2(D_1)}{D_2-D_1}\middle|D_2-D_1=\delta >0\right)$ may converge, under suitable smoothness assumptions, at a faster rate than $n^{1/3}$, they cannot converge at the $\sqrt{n}$ rate.\footnote{The \st{did\_multiplegt\_stat} package can still compute discretized versions of $d_1\mapsto E\left[(Y_2(D_2)- Y_2(d_1))/(D_2-d_1)\middle|D_1=d_1,D_2\ne d_1\right]$ (resp. $\delta \mapsto E[(Y_2(D_2)- Y_2(D_1))/(D_2-D_1)|D_2-D_1=\delta >0]$), when the option \st{by\_baseline} (resp. \st{by\_fd}) is specified, see the help file for further details.} In a recent paper, \cite{haddad2024difference} consider a parallel-trends assumption similar to Assumption \ref{hyp:parallel_trends}, and propose an estimator of a functional parameter similar to $(d_1,d_2)\mapsto \text{TE}(d_1,d_2|d_1,d_2)$.

\medskip
Rather than focusing on averages of the slopes $\text{TE}(d_1,d_2| d_1,d_2)$, one might instead consider alternative parameters, such as the average dose–slope function $(d,d') \mapsto \text{TE}(d,d') := E\big[(Y_2(d) - Y_2(d'))/(d-d')\big]$, or the average marginal effect $E\left(\dot{Y}_2(D_2)\right)$ where $\dot{Y}_2(D_2)=\frac{\text{d}Y_2}{\text{d} d}(D_2)$. In the next section, we show that the target parameters we study address research questions that differ from those captured by $(d,d') \mapsto \text{TE}(d,d')$ and $E\left(\dot{Y}_2(D_2)\right)$; accordingly, the choice of parameter should be driven primarily by the research question of interest. That said, considerations of empirical credibility may also influence this choice. In particular, an important distinction between these parameters lies in their identification. Averages of $\text{TE}(d_1,d_2| d_1,d_2)$ can be identified under assumptions whose plausibility can be partially evaluated using pre-trend tests. By contrast, $\text{TE}(d,d')$ and $E\left(\dot{Y}_2(D_2)\right)$ cannot be identified under assumptions that lend themselves to such validation, as discussed in Section \ref{appendixsec:alternatives}. Although pre-trend tests have limitations \citep{roth2022pretest}, they remain a key element of empirical practice, helping to bolster the credibility of the identifying assumptions required to recover counterfactual outcomes \citep{imbens2024lalonde}.

\subsection{Alternative parallel-trends assumption}\label{sub:alt_PT}

The DID estimands in Lemma \ref{lem:building_block} compare switchers and
stayers with the same period-one treatment. Instead, one could propose
estimands comparing switchers and stayers, without conditioning on their
period-one treatment. To recover the counterfactual outcome trend of a
switcher going from two to five units of treatment, one could use a stayer
with treatment equal to three at both dates. On top of Assumption
\ref{hyp:parallel_trends}, such estimands rest on two supplementary
conditions:
\begin{enumerate}[(i)]
\item $E(\Delta Y(d)|D_1=d)=E(\Delta Y(d))$.
\item $\forall$ $(d,d')\in \mathcal{D}_1^2$,  $E(\Delta Y(d))=E(\Delta
    Y(d'))$.
\end{enumerate}
(i) requires that all units experience the same evolution of their potential
outcome with treatment $d$, while Assumption \ref{hyp:parallel_trends} only
imposes that requirement for units with the same baseline treatment.
Assumption \ref{hyp:parallel_trends} may be more plausible: units with the
same period-one treatment may be more similar and more likely to be on
parallel trends than units with different period-one treatments. (ii) requires the trend affecting all potential outcomes to be the same. For instance, to rationalize a DID estimand comparing a switcher whose treatment increases from two to five units with a stayer whose treatment remains equal to three, one must have $E(\Delta Y(2))=E(\Delta Y(3))$. Rearranging, (ii) is equivalent to
\begin{equation}\label{eq:constant_effects_overtime}
E\left(Y_2(d)-Y_2(d')\right)=E\left(Y_1(d)-Y_1(d')\right),
\end{equation}
so average treatment effects must be constant over time, a strong restriction on treatment-effect heterogeneity. By contrast, Assumption~\ref{hyp:parallel_trends} imposes no restriction on treatment-effect heterogeneity. Relatedly, note that while (i) is implied by our Roy model in Section \ref{sub:Roy}, (ii) is not.

\section{Estimating the average of switchers' slopes}\label{sec:AS}

\subsection{Target parameter}\label{subsec:ASdef}

In this section, our target parameter is
\begin{align}\label{eq:ampos_definition}
\delta_1& :=E\left(\frac{Y_2(D_2) - Y_2(D_1)}{D_2-D_1} \middle| S=1\right),
\end{align}
the average of the slopes of switchers' potential outcome functions, between
their period-one and their period-two treatments. Hereafter, $\delta_1$ is
referred to as the Average of Slopes (AS). The AS was first considered by \cite{chamberlain1982multivariate}. If all units are untreated at period one ($D_1=0$), then the AS coincides with the ACRT$^{\text{loc}}$ parameter in \cite{callaway2021difference}.

\medskip
The AS is a local effect: it only applies to switchers, and it measures the
effect of changing their treatment from its period-one to its period-two
value, not of other changes of their treatment. Still, the AS can be used to
point or partially identify the effect of other treatment changes under shape
restrictions. First, assume that the potential outcomes are linear: for $t\in
\{1,2\}$, $Y_t(d)=Y_t(0)+B_td$, where $B_t$ is a slope that may vary across units and may change over time.
Then, $\delta_1=E\left(B_2\middle| S=1\right)$: the AS is equal to the
average, across switchers, of the slopes of their potential outcome functions
at period 2. Then, the AS is also equal to $E\left(\dot{Y}_2(D_2)|S=1\right)$, the average marginal effect among switchers, and to $(d,d') \mapsto \text{TE}_S(d,d') := E\big[(Y_2(d) - Y_2(d'))/(d-d')|S=1\big]$, the average dose-slope function among switchers.

\medskip
Of course, this only holds under linearity, which may not be a plausible assumption. Suppose instead that $d\mapsto Y_2(d)$ is convex. Then, for any $\epsilon>0$,
$$\epsilon\delta_1\leq E\left(Y_2(D_2+\epsilon)-Y_2(D_2)\middle|S=1\right).$$
Dividing both sides by $\epsilon$ and letting $\epsilon\to0$, if $d\mapsto Y_2(d)$ is differentiable with a derivative bounded by an integrable function, then
$$\delta_1\leq E\left(\dot{Y}_2(D_2)\middle|S=1\right).$$
Accordingly, under convexity the AS yields a lower bound on the effect of increasing the treatment from $D_2$ to any value greater than $D_2$, and on switchers' average marginal effect at $D_2$. Conversely, under convexity the AS yields an upper bound on the effect of reducing the treatment from $D_1$ to any value lower than $D_1$, and on switchers' average marginal effect at $D_1$. Similarly, under concavity the AS yields an upper bound on the effect of increasing the treatment from $D_2$ to greater values and on $E\left(\dot{Y}_2(D_2)\middle|S=1\right)$, and a lower bound on the effect of reducing the treatment from $D_1$ to lower values and on $E\left(\dot{Y}_2(D_1)\middle|S=1\right)$.\footnote{See \cite{d2021nonparametric} for related bounds derived under convexity and concavity.}

\subsection{Identification}\label{subsec:ASidentification}

The AS is identified under the two following assumptions on the design. Hereafter, let $\kappa$ denote the infimum of the support of $|\Delta D|$ given $S=1$.

\begin{hyp}\label{hyp:support_condition0} (Overlap condition for AS identification)
$P(S=1)>0$ and $P(S=0|D_1)>\zeta$ for some $\zeta>0$.
\end{hyp}

\begin{hyp}\label{hyp:noquasi-stayers} (No quasi-stayers) $\kappa>0$.
\end{hyp}

The estimand identifying the AS compares the outcome evolutions of switchers and stayers with the same period-one treatment. Then, this estimand requires that there be no value of $D_1$ such that only switchers have that value, as assumed in Assumption \ref{hyp:support_condition0}. There are applications where Assumption \ref{hyp:support_condition0} fails, because there are no stayers at all. Assumption \ref{hyp:noquasi-stayers} requires that there are no quasi-stayers: the treatment of all switchers changes by at least $\kappa>0$.
The null that $0$ belongs to the support of $|\Delta D|$ can be tested (under regularity conditions), for instance using the test proposed in Section 3.3 of \cite{de2024two}.

\medskip
Define the functions:
\[\mu_0(d):=E[\Delta Y|S=0,D_1=d]\quad \text{ and } \alpha_0(d):=E\left(\left.\frac{S}{\Delta D}\right\vert D_1=d\right)\frac{1}{E(1-S|D_1=d)}.\]

\begin{thm}\label{thm:main0}
If Assumptions \ref{hyp:noanticipation}-\ref{hyp:regularity},
\ref{hyp:support_condition0}, and \ref{hyp:noquasi-stayers} hold, then for
any functions $d\mapsto\alpha(d)$ and $d\mapsto\mu(d)$
defined on $\mathcal{D}_1$ such that $\alpha$ is bounded and $E[|\mu(D_1)|]<\infty$:
\begin{align*}
\delta_1=& E\left(\frac{1}{E[S]}\left(\frac{S}{\Delta D}-(1-S)\alpha(D_1)\right)\left[\Delta Y-\mu_0(D_1)\right]\right)\\
=& E\left(\frac{1}{E[S]}\left(\frac{S}{\Delta D}-(1-S)\alpha_0(D_1)\right)\left[\Delta Y-\mu(D_1)\right]\right).
\end{align*}
\end{thm}

It readily follows from Lemma \ref{lem:building_block} and the law of
iterated expectations that $\delta_1$  is identified by
\begin{equation}\label{eq:reg_adj_estimand}
E\left[\left(\frac{\Delta Y-E[\Delta Y|S=0,D_1]}{\Delta D}\right)\middle|S=1\right]=E\left[\left(\frac{\Delta Y-\mu_0(D_1)}{\Delta D}\right)\middle|S=1\right],
\end{equation}
where the numerator of the ratio inside the expectation is a DID comparing
the outcome evolutions of switchers and stayers with the same $D_1$. Theorem
\ref{thm:main0} implies that a related estimand,
$$E\left(\frac{1}{E[S]}\left(\frac{S}{\Delta D}-(1-S)\alpha_0(D_1)\right)\left[\Delta Y-\mu_0(D_1)\right]\right),$$
also identifies $\delta_1$, and is doubly-robust in the functions
$\mu_0(D_1)$ and $\alpha_0(D_1)$.\footnote{We are grateful to Andres Santos
for pointing out to us that the AS is identified by a doubly-robust
estimand.} This estimand has a similar expression as the doubly-robust ATT estimand under conditional independence \citep{bang2005doubly}. With respect to the estimand based on a regression adjustment in \eqref{eq:reg_adj_estimand}, the doubly-robust estimand can be estimated without imposing strong  assumptions, in particular on the nonparametric estimators of $\mu_0$ and $\alpha_0$ \citep{chernozhukov2018double}.

\medskip
If there are quasi-stayers, the AS is still identified, but estimation may be
challenging. For any $\eta>0$, let $S_\eta=1\{|\Delta D|>\eta\}$ be an
indicator for switchers whose treatment changes by at least $\eta$ from
period one to two.
\begin{thm}\label{thm:main0_quasistayers}
If Assumptions \ref{hyp:noanticipation}-\ref{hyp:regularity} and
\ref{hyp:support_condition0} hold,
\begin{align*}
\delta_1& = \lim_{\eta\downarrow 0} E\left(\frac{\Delta Y - E(\Delta Y | D_1, S=0)}{\Delta D} \middle| S_\eta=1\right).
\end{align*}
\end{thm}
If there are quasi-stayers whose treatment change is arbitrarily close to $0$
(i.e. $f_{|\Delta D||S=1}(0)>0$), the denominator of $(\Delta Y - E(\Delta Y
| D_1, S=0))/\Delta D$ is close to $0$ for them. On the other hand,
\begin{align*}
\Delta Y - E(\Delta Y | D_1, S=0)&=Y_2(D_2)-Y_2(D_1)+\Delta Y(D_1)- E(\Delta Y(D_1)| D_1, S=0)\\
&\approx\Delta Y(D_1)- E(\Delta Y(D_1)| D_1, S=0),
\end{align*}
so the ratio's numerator may not be close to $0$. Then, under weak
conditions,
$$E\left(\left|\frac{\Delta Y - E(\Delta Y | D_1, S=0)}{\Delta D}\right|\; \middle| S=1\right)=+\infty.$$
Therefore, we need to trim quasi-stayers from the estimand in Theorem
\ref{thm:main0}, and let the trimming go to 0, as in
\cite{graham2012identification} who consider a related estimand with some
quasi-stayers. Accordingly, with quasi-stayers the AS is irregularly
identified by a limiting estimand. Then, we conjecture that the AS cannot be
estimated at the $\sqrt{n}-$rate, as in \cite{graham2012identification}, and
as is often the case with target parameters identified by limiting estimands.
Therefore, we do not consider estimation of the AS with quasi stayers.

\subsection{Estimation and inference}\label{subsec:ASestimation}

Let $g_0(d):=E[S/\Delta D|D_1=d]$ and $p_0(d):=E[1-S|D_1=d]$. It follows from Theorem \ref{thm:main0} that estimating $\delta_1$ requires
estimating the nuisance functions $g_0(d)$, $p_0(d)$, and $\mu_0(d)$ in a first
step, before averaging the variable
$$\left(\frac{S_i}{\Delta D_i}-\frac{\hat g(D_{i,1})}{\hat p(D_{i,1})}(1-S_i)\right)\left(\Delta Y_i-\hat\mu(D_{i,1})\right).$$
We propose to use cross-fitting to perform this two-step
estimation. For any set $A$, let $\# A$ denote its cardinality. Let the sample be split randomly into two sub-samples,
$\mathcal{I}_1$ and $\mathcal{I}_2$, and let $I_k=\# \mathcal{I}_k$ for
$k=1,2$. We let $\hat g^{(2)}$, $\hat p^{(2)}$ and $\hat \mu^{(2)}$ be
nonparametric estimators of $g_0$, $p_0$ and $\mu_0$ based on $\mathcal{I}_2$.
Then the estimator $\hat\delta_{1,\mathsf{DR}}^{(1)}$ is computed in
$\mathcal{I}_1$ as follows:
\[\hat\delta_{1,\mathsf{DR}}^{(1)}=\frac{1}{n_S^{(1)}}\sum_{i\in \mathcal{I}_1}\left(\frac{S_i}{\Delta D_i}-\frac{\hat g^{(2)}(D_{i,1})}{\hat p^{(2)}(D_{i,1})}(1-S_i)\right)\left(\Delta Y_i-\hat\mu^{(2)}(D_{i,1})\right),\]
where $n_S^{(1)}:=\sum_{i\in \mathcal{I}_1} S_i$ and we use the conventions
that $0/0=0$ and $\hat\delta_{1,\mathsf{DR}}^{(1)}=0$ if $n_S^{(1)}=0$. The estimator
$\hat\delta_{1,\mathsf{DR}}^{(2)}$ is computed similarly, permuting the roles
of $\mathcal{I}_1$ and $\mathcal{I}_2$. Finally the cross-fitting
doubly-robust estimator is computed as:
\[\hat{\delta}_{1,\mathsf{DR}}=\frac{n_S^{(1)}}{n_S^{(1)}+n_S^{(2)}}\hat\delta_{1,\mathsf{DR}}^{(1)}+
\frac{n_S^{(2)}}{n_S^{(1)}+n_S^{(2)}}\hat\delta_{1,\mathsf{DR}}^{(2)}.\]
We establish below that under some conditions, $\hat{\delta}_{1,\mathsf{DR}}$ is asymptotically normal, semiparametrically efficient and has asymptotic variance $V_1:=V(\psi_1)$, with
\[\psi_1:=\frac{1}{E[S]}\left\{\left(\frac{S}{\Delta D}-\frac{g_0(D_1)}{p_0(D_1)}(1-S)\right)\left[\Delta Y-\mu_0(D_1)\right]-\delta_1S\right\}.\]
Some algebra shows that $V_1=(E[\Gamma_1^2]-E[S]\delta_1^2)/E[S]^2$, with
\[\Gamma_1:=\left(\frac{S}{\Delta D}-\frac{g_0(D_1)}{p_0(D_1)}(1-S)\right)\left[\Delta Y-\mu_0(D_1)\right].\]
Then, we estimate $V_1$ by $\hat{V}_1= (\sum_{i=1}^n \hat\Gamma_{1i}^2/n - \bar{S} \hat\delta_{1,\mathsf{DR}}^2)/\bar{S}^2$,  where $\bar{S}$ is the sample average of $(S_i)_{i=1,...,n}$,
\begin{align*}
\hat\Gamma_{1i}&=\left(\frac{S_i}{\Delta D_i}-\frac{\hat g^{(3-k(i))}(D_{i,1})}{\hat p^{(3-k(i))}(D_{i,1})}(1-S_i)\right)\left(\Delta Y_i-\hat\mu^{(3-k(i))}(D_{i,1})\right)
\end{align*}
and $k(i)=1\{i\in\mathcal{I}_1\}+2\times1\{i\in\mathcal{I}_2\}$ for all $i\in\{1,...n\}$.

\medskip
To prove the asymptotic results, we first impose the following condition:

\begin{hyp}\label{hyp:subsamples}
$\I_1 \indep (D_{i,2},Y_{i,1},Y_{i,2})_{i=1,...,n}|(D_{i,1})_{i=1,...,n}$ and
$I_1/n\to \pi\in (0,1)$.
\end{hyp}

We also require that the estimators of the nonparametric nuisance functions
satisfy some convergence-rate requirements. For any function $h(\cdot)$, $\hat h^{(1)}(.), \hat h^{(2)}(.)$ estimators of $h(.)$ based on subsample $\mathcal{I}_1$ and $\mathcal{I}_2$ respectively and $k\in\{1,2\}$, define:
\begin{align*}
\norm{\hat{h}-h}_{k,2}& :=\left(\frac{1}{I_k}\sum_{i\in\mathcal{I}_k}(\hat h^{(3-k)}(D_{1,i})-h(D_{1,i}))^2\right)^{1/2},\\
\norm{\hat h-h}_{k,\infty}& :=\sup_{d\in \mathcal{D}_1}\abs{\hat h^{(3-k)}(d)-h(d)}.	
\end{align*}

\begin{hyp}\label{hyp:rate_DR_AS}
For $k=1,2$,
\begin{enumerate}[(a)]
\item\label{hyp:consist_AS} $\norm{\hat g-g_0}_{k,2}=o_P(1)$, $\norm{\hat
    \mu-\mu_0}_{k,2}=o_P(1)$, $\norm{\hat p-p_0}_{k,\infty}=o_P(1)$.
\item\label{hyp:rate_AS} $\sqrt{n}\norm{\hat g-g_0}_{k,2}\times \norm{\hat
    \mu-\mu_0}_{k,2}=o_P(1)$, $\sqrt{n}\norm{\hat
    p-p_0}_{k,2}\times \norm{\hat
    \mu-\mu_0}_{k,2}=o_P(1)$.
\item\label{hyp:for_var_consist} $\max_{i\in \mathcal{I}_k} |\hat g^{3-k}(D_{1,i})|=O_P(1)$.
\end{enumerate}
\end{hyp}

As all nuisance functions are conditional expectations with respect to a
single variable, the convergence-rate requirements in \ref{hyp:rate_AS} are satisfied by
classical nonparametric estimators. For instance, one can use series
estimators with a polynomial order chosen by cross validation within
subsample $\mathcal{I}_k$ to estimate $\hat\mu^{(k)}(d)$, $\hat{g}^{(k)}(d)$,
and $\hat p^{(k)}(d)$. Results in Section 4 of \cite{andrews1991asymptotic}
imply that those estimators satisfy the rate conditions in Assumption
\ref{hyp:rate_DR_AS}, as long as the population functions are twice
differentiable. Note also that the mean-square convergence requirements can
be established under regularity conditions and appropriate choices of tuning
parameters for machine learning estimators such as random forests
\citep{Biau_2012_JMLR,Wager-Athey_2018_JASA,Syrgkanis-Zampetakis_2020} and
neural networks \citep{Chen_White_1999_IEEE, Farrell-etal_2021_ECMA}.\footnote{We rely on
uniform convergence of the propensity score to ensure that the inverse
propensity score is consistent. While uniform convergence may be too strong
for machine learning estimators, this condition may be relaxed using
automatic debiased machine learning, as proposed in
\cite{Cherno-etal_2024_wp}.} In Python, the \texttt{did\_multiplegt\_stat} package has an option to compute a debiased machine learning estimator $\hat{\delta}_{1,\mathsf{DR}}$, where all nuisance functions are estimated with machine learning. Finally, note that if the distribution of $D_1$ has mass points, some nonparametric estimators (e.g., local polynomial estimators) have a lower variance at such points than at other points. This suggests that mass points in the distribution of $D_1$ make Conditions (a)-(b) less restrictive.

\medskip
We impose Assumption \ref{hyp:rate_DR_AS}\ref{hyp:for_var_consist} to obtain consistency of $\hat{V}_1$. This condition can be fulfilled as follows. Suppose initial estimators $\tilde{g}^{(k)}$ ($k=1,2$) satisfy Assumption \ref{hyp:rate_DR_AS}\ref{hyp:consist_AS}-\ref{hyp:rate_AS}. Then, let $\hat{g}^{(k)}(d):=\max(\min(\tilde{g}^{(k)}, M^{(k)}_n), m^{(k)}_n)$, where
$$M^{(k)}_n:=2\max\left(0,\max_{i\in\mathcal{I}_k:S_i\ne 0} 1/\Delta D_i\right),\quad m^{(k)}_n:=2\min\left(0,\min_{i\in\mathcal{I}_k:S_i\ne 0} 1/\Delta D_i\right).$$
Under Assumption \ref{hyp:noquasi-stayers}, $m^{(k)}_n=O_P(1)$ and $M^{(k)}_n=O_P(1)$, ensuring that $\hat{g}^{(k)}$ satisfies Assumption \ref{hyp:rate_DR_AS}\ref{hyp:for_var_consist}. Also, one can show that with probability approaching one,
$$[\min_{d\in\mathcal{D}_1} g_0(d),\max_{d\in\mathcal{D}_1} g_0(d)]\subset [m^{(k)}_n,M^{(k)}_n],$$
and that under this event, $|\hat{g}^{(k)}(d)-g_0(d)|\le |\tilde{g}^{(k)}(d)-g_0(d)|$ for all $d\in\mathcal{D}_1$. Hence, $\hat{g}^{(k)}$ also satisfies Assumption \ref{hyp:rate_DR_AS}\ref{hyp:consist_AS}-\ref{hyp:rate_AS}.

\begin{thm}\label{thm:ASasnormality} If Assumptions \ref{hyp:noanticipation}-\ref{hyp:subsamples} and \ref{hyp:rate_DR_AS}\ref{hyp:consist_AS}-\ref{hyp:rate_AS} hold, $\sqrt{n}(\hat\delta_{1,\mathsf{DR}}-\delta_1)=\frac{1}{\sqrt{n}}\sum_i\psi_{1i}+o_P(1)\convL \mathcal{N}(0,V_1)$ and $\psi_1$ is the efficient influence function. If Assumption \ref{hyp:rate_DR_AS}\ref{hyp:for_var_consist} also holds, $\hat{V}_1\convP V_1$.
\end{thm}

\section{Estimating a weighted average of switchers' slopes}\label{sec:WAS}

\subsection{Target parameter}\label{subsec:WASdef}

In this section, our target parameter is
\begin{align*}
\delta_2 :=&E\left(\frac{|D_2-D_1|}{E(|D_2-D_1||S=1)} \times \frac{Y_2(D_2)-Y_2(D_1)}{D_2-D_1}\middle|S=1\right),
\end{align*}
a weighted average of the slopes of switchers' potential
outcome functions from their period-one to their period-two treatments, where
slopes receive a weight proportional to switchers' absolute treatment change
from period one to two. We refer to $\delta_{2}$ as the Weighted
Average of Slopes (WAS). $\delta_2=\delta_1$ if and only if switchers' slopes are uncorrelated
with $|D_2-D_1|$, conditional on $S=1$.

\medskip
The AS and WAS serve different purposes. Unlike the AS, the WAS may not be used to identify or bound the effect of other treatment changes than the actual changes that took place, even under shape
restrictions on the potential outcome function. Instead, the WAS may be used to conduct a cost-benefit analysis of the treatment changes that took place from period
one to two. To simplify the discussion, let us assume in the remainder of
this paragraph that $D_2\ge D_1$.\footnote{\label{foot:cost_ben} Otherwise, one can conduct the cost-benefit analysis separately among the subpopulations of ``switchers up'' and ``switchers down'' defined below.} Assume also that the outcome is a measure
of output, such as agricultural yields or wages, expressed in monetary units.
Finally, assume that the treatment is costly, with a cost linear in dose,
uniform across units, and known to the analyst: the cost of giving $d$ units
of treatment to a unit at period $t$ is $c_t\times d$ for some known
$(c_t)_{t\in \{1,2\}}$. Then, $D_2$ is beneficial relative to $D_1$ if and
only if $E(Y_2(D_2)-c_2D_2)>E(Y_2(D_1)-c_2D_1)$, namely if and only if
$\delta_2>c_2:$ comparing $\delta_2$ to $c_2$ is sufficient to evaluate if
changing the treatment from $D_1$ to $D_2$ was beneficial. The average marginal effect among switchers $E(Y_2'(D_2)|S=1)$ cannot be used to conduct this cost-benefit analysis. Instead, it can be used to assess if infinitesimal treatment changes around $D_2$ are beneficial or not. Similarly, the average dose-slope function among switchers can be used for different cost-benefit analyses, assessing if universal treatment changes from $d$ to $d'$ are beneficial or not.

\subsection{Identification}\label{subsec:WASidentification}

Let $S_+=1\{D_2-D_1>0\}$ and $S_-=1\{D_2-D_1<0\}$ respectively be indicators
for ``switchers up'' and ``switchers down'' and define the function:
\[\gamma_0(d):=\frac{E(S_+|D_1=d)-E(S_-|D_1=d)}{E(1-S|D_1=d)}.\]
\begin{thm}\label{thm:main}
If Assumptions \ref{hyp:noanticipation}-\ref{hyp:regularity}, and
\ref{hyp:support_condition0} hold, then for any functions  $d\mapsto\gamma(d)$ and $d\mapsto\mu(d)$ defined on $\mathcal{D}_1$ and such that $\gamma$ is bounded and $E[|\mu(D_1)|]<\infty$:
\begin{align*}
\delta_2=& \frac{E\left[\left(S_+-S_--(1-S)\gamma(D_1)\right)(\Delta Y-\mu_0(D_1))\right]}{E[\abs{\Delta D}]}\\
=& \frac{E\left[\left(S_+-S_--(1-S)\gamma_0(D_1)\right)(\Delta Y-\mu(D_1))\right]}{E[\abs{\Delta D}]}.
\end{align*}
\end{thm}
Theorem \ref{thm:main} shows that $\delta_2$ is identified by an estimand that is doubly-robust in
$\mu_0(D_1)$ and $\gamma_0(D_1)$, because it combines regression adjustment with propensity-score reweighting. If $D_2\geq D_1$, the numerator of the doubly-robust estimand is very similar to the doubly-robust ATT estimand under conditional independence \citep{bang2005doubly}, with $D_1$ playing the role of the covariates.

\medskip
Note that $\delta_2$ can be identified under weaker assumptions than Assumption \ref{hyp:parallel_trends}. To simplify the discussion, let us assume that $D_2\geq D_1$: there are no switchers down. Then, let us assume that
\begin{equation}\label{eq:parallel_trends_average}
E(E(\Delta Y(D_1)|D_1,S=1)|S=1)=E(E(\Delta Y(D_1)|D_1,S=0)|S=1).
\end{equation}
Assumption \ref{hyp:parallel_trends} implies that $E(\Delta Y(D_1)|D_1,S=1)$ and $E(\Delta Y(D_1)|D_1,S=0)$ are equal for all values of $D_1$, while \eqref{eq:parallel_trends_average} only requires that the two random variables have the same expectation conditional on $S=1$. If one replaces Assumption \ref{hyp:parallel_trends} by \eqref{eq:parallel_trends_average},
one can still show that
$$\delta_2=\frac{E\left[\left(S_+-S_-\right)(\Delta Y-\mu_0(D_1))\right]}{E[\abs{\Delta D}]}.$$
Similarly, if one replaces Assumption \ref{hyp:parallel_trends} by
\begin{equation}\label{eq:parallel_trends_average2}
E(\Delta Y(D_1)|S=1)=E\left[\frac{E(S|D_1)}{E(1-S|D_1)}\frac{E(1-S)}{E(S)}E(\Delta Y(D_1)|S=0)\right],
\end{equation}
then one can still show that
$$\delta_2=\frac{E\left[\left(S_+-S_--(1-S)\gamma_0(D_1)\right)\Delta Y\right]}{E[\abs{\Delta D}]}.$$
However, \eqref{eq:parallel_trends_average} and \eqref{eq:parallel_trends_average2} are not sufficient to ensure that
$\delta_2$ is identified by an estimand that is doubly-robust in  $\mu_0(D_1)$ and $\gamma_0(D_1)$.

\subsection{Estimation and inference}\label{subsec:WASestimation}

Let $h_0(d):=E[S_+-S_-|D_1=d]$. As before, split the sample into
subsamples $\mathcal{I}_1$ and $\mathcal{I}_2$, and let
\[\hat\delta_{2,\mathsf{DR}}^{(1)}:=\frac{1}{\Delta_S^{(1)}}\sum_{i\in \mathcal{I}_1}\left(S_{i+}-S_{i-}-\frac{\hat h^{(2)}(D_{i,1})}{\hat p^{(2)}(D_{i,1})}(1-S_i)\right) (\Delta Y_i -\hat\mu^{(2)}(D_{i,1})),\]
where $\Delta_S^{(1)}:=\sum_{i\in \mathcal{I}_1} \abs{\Delta D_i}$ and we use
the convention that $\hat\delta_{2,\mathsf{DR}}^{(1)}=0$ if
$\Delta_S^{(1)}=0$. The estimator $\hat\delta_{2,\mathsf{DR}}^{(2)}$ is
constructed analogously, permuting the roles of $\mathcal{I}_1$ and
$\mathcal{I}_2$. Next, we let
\[\hat\delta_{2,\mathsf{DR}}:=\frac{\Delta_S^{(1)}}{\Delta_S^{(1)}+\Delta_S^{(2)}}\hat\delta_{2,\mathsf{DR}}^{(1)}+\frac{\Delta_S^{(2)}}{\Delta_S^{(1)}+\Delta_S^{(2)}}\hat\delta_{2,\mathsf{DR}}^{(2)}.\]
We establish below that under assumptions, $\hat{\delta}_{2,\mathsf{DR}}$ is asymptotically normal, with asymptotic variance $V_2:=V(\psi_2)$, with
\begin{align*}
\psi_{2}&:=\frac{1}{E(\abs{\Delta D})}\left\{\left(S_+-S_--E(S_+-S_-|D_1)\frac{(1-S)}{E(1-S|D_1)}\right)\right. \\
&\hspace{2.6cm}\times
(\Delta Y -E(\Delta Y | D_{1}, S=0))-\delta_{2}\abs{\Delta D}\bigg\}.
\end{align*}
One can show that
$V_2=\left\{E[\Gamma_2^2]-2\delta_2E[\Gamma_2\abs{\Delta D}]+\delta_2^2E[\abs{\Delta D}^2]\right\}/E(\abs{\Delta D})^2$, where

\[\Gamma_2=\left(S_{+}-S_{-} -\frac{h_0(D_1)}{p_0(D_1)}(1-S)\right)
(\Delta Y -\mu_0(D_1)).\]
Then, we estimate $V_2$ by
\begin{align*}
\hat V_2&=\frac{1}{\overline{\abs{\Delta D}}^2}\left\{\frac{1}{n}\sum_{i=1}^n \hat\Gamma_{2i}^2-2\hat\delta_{2,\mathsf{DR}}\frac{1}{n}\sum_{i=1}^n\abs{\Delta D_i}\hat\Gamma_{2i}+\hat\delta_{2,\mathsf{DR}}^2\frac{1}{n}\sum_{i=1}^n\abs{\Delta D_i}^2\right\},
\end{align*}
where
\begin{align*}
\hat\Gamma_{2i}&=\left(S_{+i}-S_{-i}-\frac{\hat h^{(3-k(i))}(D_{i,1})}{\hat p^{(3-k(i))}(D_{i,1})}(1-S_i)\right)\left(\Delta Y_i-\hat\mu^{(3-k(i))}(D_{i,1})\right).
\end{align*}

To prove the asymptotic results, we impose the following condition:

\begin{hyp}\label{hyp:rate_DR_WAS}
For $k=1,2$,
\begin{enumerate}[(a)]
\item $\norm{\hat h-h_0}_{k,2}=o_P(1)$, $\norm{\hat
    \mu-\mu_0}_{k,2}=o_P(1)$, $\norm{\hat p-p_0}_{k,\infty}=o_P(1)$.
\item $\sqrt{n}\norm{\hat h-h_0}_{k,2}\times \norm{\hat
    \mu-\mu_0}_{k,2}=o_P(1)$, $\sqrt{n}\norm{\hat
    p-p_0}_{k,2}\times \norm{\hat
    \mu-\mu_0}_{k,2}=o_P(1)$.
\item\label{hyp:for_var_cons_WAS} $\max_{i\in \mathcal{I}_k} |\hat h^{3-k}(D_i)|=O_P(1)$.
\end{enumerate}
\end{hyp}
As above, Condition \ref{hyp:for_var_cons_WAS} is only needed to obtain consistency of $\hat V_2$. Since $h_0(d)\in[-1,1]$ by definition, this condition can always be achieved by constraining any initial estimator to belong to $[-1,1]$.

\begin{thm}\label{thm:WASasnormality_nonmonotonictreatment}
If Assumptions \ref{hyp:noanticipation}-\ref{hyp:support_condition0},
\ref{hyp:subsamples}, and \ref{hyp:rate_DR_WAS}(a)-(b) hold, $\sqrt{n}(\hat\delta_{2,\mathsf{DR}}-\delta_2)=\frac{1}{\sqrt{n}}\sum_i\psi_{2i}+o_P(1)\convL \mathcal{N}(0,V_2)$ and $\psi_2$ is the efficient influence function. Moreover, if Assumption \ref{hyp:rate_DR_WAS}\ref{hyp:for_var_cons_WAS} holds as well, $\hat V_2\convP V_2$.
\end{thm}
The proof of this theorem is very similar to that of Theorem \ref{thm:ASasnormality} and is given in the web appendix. Finally, we show that under some assumptions, the asymptotic variance of the WAS estimator is lower, and potentially much lower, than that of the AS estimator. Let us define $\xi:=(Y_2(D_2)-Y_2(D_1))/(D_2-D_1)$ when $S=1$.
\begin{prop}\label{prop:variance_comparison}
Suppose that Assumptions \ref{hyp:noanticipation}-\ref{hyp:parallel_trends} and \ref{hyp:support_condition0}-\ref{hyp:noquasi-stayers} hold. Then:
\begin{enumerate}
\item If one further supposes that Assumption \ref{hyp:regularity} holds, $\xi=\delta_1(=\delta_2)$ when $S=1$, $V(\Delta Y(D_1) $ $|D_1,D_2)=\sigma^2$ for some real number $\sigma^2>0$, $D_2\ge D_1$, and $\Delta D \indep D_1$, then
\begin{align*}
V_1=&\sigma^2\left[\frac{E(1/(\Delta D)^2|S=1)}{P(S=1)}+\frac{\left(E(1/\Delta D|S=1)\right)^2}{P(S=0)}\right]\\
\ge & \sigma^2\frac{1}{\left(E(\Delta D|S=1)\right)^2}\left[\frac{1}{P(S=1)}+\frac{1}{P(S=0)}\right]=V_2,
\end{align*}
with equality if and only if $V(\Delta D|S=1)=0$.
\item If one further supposes that (i) $V(\Delta Y(D_1)|D_1,D_2)\in[\underline{\sigma}^2,\overline{\sigma}^2]\subset (0,\infty)$, (ii) the density of $|\Delta D|$ given $S=1$ exists and there exists $(c, \underline{f})\in (0,\infty)^2$ such that $f_{|\Delta D|\, |S=1}(x)\ge \underline{f}$ on $[\kappa,\kappa+c]$, and (iii) $cov(\Delta Y(D_1),\xi|S=1,D_1,D_2)=0$ and $V(|\Delta D|(\xi-\delta_2)|S=1)\le M$, then
$$\frac{V_1}{V_2} \ge \frac{\underline{f}^3c^5}{8\kappa(\kappa+c)} \times \frac{\underline{\sigma}^2}{\overline{\sigma}^2} \times \frac{1}{1+1/\zeta+M}.$$
Therefore, $V_1\ge V_2$ if $\kappa$ is small enough and $V_1/V_2\to \infty$ if $\kappa\to 0$, keeping  the other constants fixed.
\end{enumerate}
\end{prop}
Point 1 of Proposition \ref{prop:variance_comparison} states that, under admittedly strong assumptions which imply $\delta_1 = \delta_2$ so that both estimators target the same parameter, the asymptotic variance of $\widehat{\delta}_{1,\mathsf{DR}}$ exceeds that of $\widehat{\delta}_{2,\mathsf{DR}}$. Point 2 provides a (non-sharp) lower bound on the ratio of these asymptotic variances $V_1/V_2$. Point 2 is derived under a regularity condition on the distribution of $\Delta D$ (Condition (ii)), and other conditions (Conditions (i) and (iii)) that are much weaker than the corresponding conditions in Point 1. In particular, one may have $\delta_1\ne \delta_2$ under these conditions. This result implies that $V_1/V_2$ goes to infinity when $\kappa\to 0$.
Thus, there may be settings in which $\delta_{1}$ is the parameter of interest given the research question, yet $\widehat{\delta}_{2,\mathsf{DR}}$ has a lower mean squared error for estimating that parameter. Of course, when $\delta_{1}\ne \delta_{2}$, this cannot occur asymptotically, as $\widehat{\delta}_{1,\mathsf{DR}}$ is consistent for $\delta_{1}$. However, DID estimators are often computed using relatively small samples (e.g., the 50 US states). While Point 2 does not prove that we cannot estimate $\delta_1$ at the root-$n$ rate when $\kappa=0$, it does show  (since $V_1$ and $V_2$ are the semiparametric efficiency bounds for $\delta_1$ and $\delta_2$) that it becomes much more difficult to estimate $\delta_1$ than $\delta_2$ when $\kappa$ is small.

\section{Extensions}
\label{sec:extensions}

\subsection{Instrumental-variable estimators}
\label{subsec:IV}

In this section, we consider instances where Assumption \ref{hyp:parallel_trends} is implausible, but one has an instrument that satisfies a similar parallel-trends condition. For example, we may seek to identify the price elasticity of gasoline, but price changes may not satisfy a parallel trend condition because of simultaneity. On the other hand, parallel trends with respect to gasoline taxes may hold, in which case the latter may be used as an instrument.

\subsubsection{Notation and assumptions}

Let $Z_t$ denote the instrumental variable at period $t\in\{1,2\}$. Throughout this section, we rule out anticipatory and dynamic effects of the instrument on the treatment. Then, let
$D_t(z)$ denote the unit's potential treatment at $t$ if $Z_t=z$. Let also $SC=1\{D_2(Z_2)\ne D_2(Z_1)\}$ be an indicator equal to 1 for switchers-compliers, namely units whose instrument changes from period one to two and whose treatment is affected by that change. Finally, let $\mathcal{Z}_t$ (resp. $\mathcal{Z}$) denote the support of $Z_t$ (resp. of $(Z_1,Z_2)$).

\medskip
We replace Assumption \ref{hyp:parallel_trends} by the following assumption.\footnote{Our notation, where potential outcomes do not depend on $z$, implicitly imposes the usual exclusion restriction.}
\begin{hyp}\label{hyp:parallel_trends_IV} (Reduced-form and first-stage parallel trends)
For all $z\in \mathcal{Z}_1$,
\begin{enumerate}
\item $E(Y_2(D_2(z))-Y_1(D_1(z))|Z_1=z,Z_2,D_1)=E(Y_2(D_2(z))-Y_1(D_1(z))|Z_1=z,D_1)$.
\item $E(D_2(z)-D_1(z)|Z_1=z,Z_2,D_1)=E(D_2(z)-D_1(z)|Z_1=z,D_1)$.
\end{enumerate}
\end{hyp}
Point 1 of Assumption \ref{hyp:parallel_trends_IV} requires that $Y_2(D_2(z))-Y_1(D_1(z))$, units' outcome evolutions in the counterfactual where their instrument does not change from period one to two, be mean independent of $Z_2$, conditional on $Z_1=z$ and $D_1$. This condition imposes some restrictions on treatment effect heterogeneity, and the goal of conditioning on $D_1$ is to alleviate those restrictions. To see this, assume that
\begin{align}
&E(Y_2(D_1(z))-Y_1(D_1(z))|Z_1=z,Z_2,D_1)=E(Y_2(D_1(z))-Y_1(D_1(z))|Z_1=z,D_1).\label{eq:SC_PT_IV1}
\end{align}
\eqref{eq:SC_PT_IV1} requires that $Y_2(D_1(z))-Y_1(D_1(z))$, units' outcome evolutions in the counterfactual where their instrument and their treatment does not change from period one to two, be mean independent of $Z_2$, conditional on $Z_1=z$ and $D_1$. Thanks to the conditioning on $D_1$, \eqref{eq:SC_PT_IV1} is a standard parallel-trends assumption that does not impose any restriction on treatment effect heterogeneity, like Assumption \ref{hyp:parallel_trends}. If $D_1$ was not conditioned upon, \eqref{eq:SC_PT_IV1} would require parallel trends among units with different baseline treatments, which implicitly assumes homogeneous treatment effects over time, as discussed in Section \ref{sub:alt_PT}. Under \eqref{eq:SC_PT_IV1}, Point 1 of Assumption \ref{hyp:parallel_trends_IV} holds if and only if
\begin{align}
&E(Y_2(D_2(z))-Y_2(D_1(z))|Z_1=z,Z_2,D_1)\nonumber\\
=&E(Y_2(D_2(z))-Y_2(D_1(z))|Z_1=z,D_1).\label{eq:SC_PT_IV2}
\end{align}
\eqref{eq:SC_PT_IV2} is a restriction on treatment effect heterogeneity across units. It requires that switching the treatment from $D_1(Z_1)$ to $D_2(Z_1)$, the natural change happening over time without a change in the instrument, has an effect on the outcome that is mean independent of $Z_2$ conditional on $Z_1$ and $D_1$. Thus, it is only if \eqref{eq:SC_PT_IV1} fails that Point 1 of Assumption \ref{hyp:parallel_trends_IV} does not restrict treatment-effect heterogeneity. However, it may not be plausible to assume that Assumption \ref{hyp:parallel_trends_IV} holds but \eqref{eq:SC_PT_IV1} fails.

\medskip
\cite{de2010note} and \cite{hudson2017interpreting} also consider IV-DID estimands, in classical designs with two periods and a binary instrument that turns on for some units at period two. Both papers introduce a ``reduced-form'' parallel-trends assumption similar to Point 1 of Assumption \ref{hyp:parallel_trends_IV}, but without noting that it imposes restrictions on effects' heterogeneity, even in the simple designs considered by those papers.
Turning to Point 2 of Assumption \ref{hyp:parallel_trends_IV}, this condition requires that units' treatment evolutions under $Z_1$ be mean independent of $Z_2$, conditional on $Z_1$ and $D_1$. Because $D_1$ is conditioned upon, this parallel trends condition is equivalent to a sequential exogeneity assumption \citep[see][]{robins1986new,bojinov2020panel}.

\medskip
We also make the following
assumptions.
\begin{hyp}\label{hyp:monotonicity} (Monotonicity and strictly positive first-stage)
i) For all $(z,z')\in \mathcal{Z}_2^2$, $z\ge z' \Rightarrow D_2(z)\ge D_2(z')$, and ii) $E(|D_2(Z_2)-D_2(Z_1)|)>0$.
\end{hyp}
Point i) of Assumption \ref{hyp:monotonicity} is a monotonicity assumption similar to that in \cite{Imbens94}. It requires that increasing the period-two instrument weakly increases the period-two treatment. This condition is plausible when the instrument is taxes and the treatment is prices, as in our application. Point ii) requires that the instrument has a strictly positive first stage.
\begin{hyp}\label{hyp:regularity_IV} (Bounded instrument and regularity of potential outcomes and treatments)
\begin{enumerate}
\item $\mathcal{Z}_1$ and $\mathcal{Z}_2$ are bounded subsets of $\mathbb{R}$.
\item $\sup_{z_1\in\mathcal{Z}_1} E[(Y_2(D_2(z_1))-Y_1(D_1(z_1)))^2|Z_1=z_1]<\infty$, $\sup_{z_1\in\mathcal{Z}_1} E[(\Delta D(z_1))^2|Z_1=z_1]<\infty$.
\item For all $t\in \{1,2\}$ and for all $(z,z')\in \mathcal{Z}_t^2$, there is a random variable $\overline{Y}\ge 0$ such that $|Y_t(D_t(z))-Y_t(D_t(z'))|\leq \overline{Y}|z-z'|$, with $\sup_{(z_1,z_2)\in \mathcal{Z}}E[\overline{Y}^2|Z_1=z_1, Z_2=z_2]<\infty$.
    \item For all $t\in \{1,2\}$ and for all $(z,z')\in \mathcal{Z}_t^2$, there is a random variable $\overline{D}\ge 0$ such that $|D_t(z)-D_t(z')|\leq \overline{D}|z-z'|$, with $\sup_{(z_1,z_2)\in \mathcal{Z}}E[\overline{D}^2|Z_1=z_1, Z_2=z_2]<\infty$.
\end{enumerate}
\end{hyp}
\begin{hyp}\label{hyp:iid_IV} (iid sample)
We observe $(Y_{i,1},Y_{i,2},D_{i,1},D_{i,2},Z_{i,1},Z_{i,2})_{1\leq i\leq n}$, that are independent and identically distributed with the same probability distribution as $(Y_{1},Y_{2},D_{1},D_{2},Z_1,Z_2)$.
\end{hyp}
Assumptions \ref{hyp:regularity_IV} and \ref{hyp:iid_IV} are adaptations of Assumptions \ref{hyp:regularity} and \ref{hyp:iid} to the IV setting.

\subsubsection{Target parameter}

In this section, our target parameter is
\begin{align*}
\delta_{\mathsf{IV}} :=&E\left(\frac{|D_2(Z_2)-D_2(Z_1)|}{E(|D_2(Z_2)-D_2(Z_1)||SC=1)}\times \frac{Y_2(D_2(Z_2))-Y_2(D_2(Z_1))}{D_2(Z_2)-D_2(Z_1)}\middle|SC=1\right).
\end{align*}
$\delta_{\mathsf{IV}}$ is a weighted average of the slopes of compliers-switchers' period-two potential outcome functions, from their period-two treatment under their period-one instrument, to their period-two treatment under their period-two instrument. Slopes receive a weight proportional to the absolute value of compliers-switchers' treatment response to the instrument change. Under Assumption \ref{hyp:monotonicity}, $\delta_{\mathsf{IV}}$ is equal to the reduced-form WAS effect of the instrument on the outcome
$$E\left(\frac{|Z_2-Z_1|}{E(|Z_2-Z_1||Z_2\ne Z_1)}\times \frac{Y_2(D_2(Z_2))-Y_2(D_2(Z_1))}{Z_2-Z_1}\middle|Z_2\ne Z_1\right),$$
divided by the first-stage WAS effect of the instrument on the treatment
$$E\left(\frac{|Z_2-Z_1|}{E(|Z_2-Z_1||Z_2\ne Z_1)}\times \frac{D_2(Z_2)-D_2(Z_1)}{Z_2-Z_1}\middle|Z_2\ne Z_1\right).$$
With a binary instrument, such that $Z_1=0$ and $Z_2 \in \{0,1\}$, our IV-WAS effect coincides with that identified in Corollary 2 of \cite{angrist2000}, in a cross-sectional IV model.

\medskip
We could also consider a reduced-form AS divided by a first-stage AS. The resulting target is a weighted average of
the slopes $(Y_2(D_2(Z_2))-Y_2(D_2(Z_1)))/(D_2(Z_2)-D_2(Z_1))$, with weights proportional to  $(D_2(Z_2)-D_2(Z_1))/
(Z_2-Z_1)$. It seems more natural to us to weight compliers-switchers' slopes by the absolute value of their first-stage than by the slope of their first-stage.\footnote{If the first-stage effect is homogenous and linear, the weights in the IV-AS effect reduce to one, and one recovers a standard AS effect. However, linearity and homogeneity of the first-stage effect are strong assumptions.}

\subsubsection{Identification}

Let $S^I=1\{Z_2-Z_1\ne 0\}$, $S^I_{+}=1\{Z_2-Z_1>0\}$,
$S^I_{-}=1\{Z_2-Z_1<0\}$ and
\begin{align*}
&\mu^U_0(z,d):=E[\Delta U|S^I=0,Z_1=z,D_1=d],\quad U\in\{D,Y\},\\
&\gamma^I_0(z,d):=\frac{E(S^I_+|Z_1=z,D_1=d)-E(S^I_-|Z_1=z,D_1=d)}{E(1-S^I|Z_1=z,D_1=d)}.
\end{align*}
The following condition is similar to Assumption \ref{hyp:support_condition0}.

\begin{hyp}\label{hyp:support_condition_IV} (Overlap conditions for IV-WAS identification)
$P(S^I=1)>0$ and $P(S^I=0|Z_1,D_1)>\zeta$ for some $\zeta>0$.
\end{hyp}

\begin{thm}\label{thm:main_IV}
If Assumptions \ref{hyp:noanticipation}, \ref{hyp:nodynamic}, \ref{hyp:parallel_trends_IV}-\ref{hyp:regularity_IV} and \ref{hyp:support_condition_IV} hold, then
\begin{align*}
\delta_{\mathsf{IV}}=& \frac{E\left[\left(S^I_+-S^I_--(1-S^I)\gamma^I_0(Z_1,D_1)\right)(\Delta Y-\mu^{Y}_0(Z_1,D_1))\right]}{E\left[\left(S^I_+-S^I_--(1-S^I)\gamma^I_0(Z_1,D_1)\right)(\Delta D-\mu^{D}_0(Z_1,D_1))\right]}.
\end{align*}
\end{thm}
The proof of this theorem is given in the web appendix. The estimand identifying $\delta_{\mathsf{IV}}$ is equal to the estimand identifying the reduced-form WAS effect of the instrument on the outcome controlling for $D_1$, divided by the estimand identifying the first-stage WAS effect of the instrument on the treatment controlling for $D_1$.
One can show that the estimand identifying $\delta_{\mathsf{IV}}$ is doubly robust in $\gamma^I_0(Z_1,D_1)$ and $(\mu^{Y}_0(Z_1,D_1),\mu^{D}_0(Z_1,D_1))$.

\subsubsection{Estimation and inference}

Estimating $\delta_{\mathsf{IV}}$ requires estimating the nuisance functions $\mu^{Y}_0(z,d)$, $\mu^{D}_0(z,d)$, $E(S^I_+|Z_1=z,D_1=d)$, $E(S^I_-|Z_1=z,D_1=d)$, and $E(1-S^I|Z_1=z,D_1=d)$, before averaging the variables $(S^I_{i+}-S^I_{i-}-\widehat{\gamma}^I(S^I_i,Z_{i,1},D_{i,1}))(\Delta Y_i-
	\widehat{\mu}^{Y}_0(Z_{i,1},D_{i,1}))$ and $(S^I_{i+}-S^I_{i-}-\widehat{\gamma}^I(S^I_i,Z_{i,1},D_{i,1}))(\Delta D_i-\widehat{\mu}^{D}_0 (Z_{i,1},D_{i,1}))$, and taking the ratio between these two averages. We propose to use cross-fitting to perform this two-step estimation, and we let $\widehat{\delta}_{\mathsf{IV},\mathsf{DR}}$ denote the resulting estimator. As the definition of $\widehat{\delta}_{\mathsf{IV},\mathsf{DR}}$ is similar to that of $\widehat{\delta}_{2,\mathsf{DR}}$, the full definition is omitted to preserve space. Since all the nuisance functions in Theorem \ref{thm:main_IV} are conditional expectations with respect to two variables, one can still use series estimators with a polynomial order chosen by cross validation within each subsample to estimate them. Results in Section 4 of \cite{andrews1991asymptotic} imply that those estimators converge fast enough to have that $\widehat{\delta}_{\mathsf{IV},\mathsf{DR}}$ is $\sqrt{n}-$consistent and asymptotically normal, as long as the population functions are twice differentiable. Finally, for $U\in\{D,Y\}$, let
\begin{align*}
\delta_{U}&=E\left[\sgn(\Delta Z)\left(\Delta U - E(\Delta U |S^I=0,Z_1,D_1)\right)\right],\\
\psi_{U}&=\frac{1}{E(\abs{\Delta Z})}\left\{\left(S^I_+-S^I_--E(S^I_+-S^I_-|Z_1,D_1)\frac{(1-S^I)}{E(1-S^I|Z_1,D_1)}\right)\right. \\
&\hspace{2.3cm}\times (\Delta U -E(\Delta U | Z_{1},D_1, S^I=0))-\delta_{U}\abs{\Delta Z}\bigg\}.
\end{align*}
Then, let $\psi_{\mathsf{IV}}=(\psi_{Y}-\delta_{\mathsf{IV}}\psi_{D})/\delta_{D}$. Under technical conditions similar to those in Section \ref{sec:WAS}, one can show that $\psi_{\mathsf{IV}}$ is the influence function of $\widehat{\delta}_{\mathsf{IV},\mathsf{DR}}$.

\subsection{Alternative target parameters}\label{appendixsec:alternatives}

One may be interested in other parameters than the AS and WAS we consider in the paper, like the average dose-slope
function
$$(d,d')\mapsto \text{TE}(d,d'):=E\left(\frac{Y_2(d) - Y_2(d')}{d-d'}\right),$$
or the average marginal effect $E\left(\dot{Y}_2(D_2)\right)$. What is the appeal
of considering the AS and the WAS, namely averages of $\text{TE}(d_1,d_2|d_1,d_2)$, rather than those
other parameters? Conditional on $D_1=d_1,D_2=d_2$, $Y_2(d_2)$ is observed,
so estimating $\text{TE}(d_1,d_2|d_1,d_2)$ only requires estimating
$Y_2(d_1)$, switchers' counterfactual outcomes at period $2$ if their
treatment had not changed. By definition, $Y_1(d_1)$ is observed. If the data
contains a third period $0$ and the treatment of some units does not change
from period $0$ to $1$, then $Y_0(d_1)$ is also observed for some switchers
and stayers. Then, as explained below, one can run a pre-trend test of Assumption
\ref{hyp:parallel_trends}, by comparing the outcome evolutions of switchers
and stayers from period $0$ to $1$.
This shows that averages of $\text{TE}(d_1,d_2|d_1,d_2)$ can be identified under a parallel-trends assumption whose plausibility can be assessed using a pre-trend test. In contrast, identifying $\text{TE}(d,d')$ requires imputing, for most units, \textit{two} unobserved counterfactual outcomes. This cannot be achieved under an assumption whose plausibility can be similarly assessed using pre-trends, as we observe only \textit{one} potential outcome at each date. Likewise, identifying $E[(Y_2(D_2) - Y_2(d'))/(D_2-d')]$ requires imputing $Y_2(d')$. However, the plausibility of the assumption needed for this imputation cannot be similarly assessed using pre-trends, because $Y_1(d')$ is not observed for all units. Since $\dot{Y}_2(D_2)=\lim_{d'\rightarrow D_2}(Y_2(D_2) - Y_2(d'))/(D_2-d')$, the same limitation applies to $E\left(\dot{Y}_2(D_2)\right)$.

\subsection{Other extensions}

We return to the case where the treatment satisfies a parallel-trends condition.

\subsubsection{Estimators with more than two time periods}\label{subsec:Tgeq3}

This section sketches the extension of our estimators to datasets with three
periods or more. For further details, see Section \ref{appendixsec:Tgeq3} of
the web appendix. With $T>2$ time periods, we propose a simple extension of our estimators. Let $\widehat{\delta}_{1,t,\mathsf{DR}}$ and $\widehat{\delta}_{2,t,\mathsf{DR}}$ be defined exactly as $\hat{\delta}_{1,\mathsf{DR}}$ and $\hat{\delta}_{2,\mathsf{DR}}$ but on a pair of consecutive time periods $(t-1,t)$. We consider a weighted average across $t$ of $\widehat{\delta}_{1,t,\mathsf{DR}}$, with weights proportional to the proportion of switchers from $t-1$ to $t$, yielding an AS estimator with more than two periods, denoted $\widehat{\delta}^{T\geq 3}_{1}$. We also consider a weighted average across $t$ of
$\widehat{\delta}_{2,t,\mathsf{DR}}$, with weights proportional to the
average absolute treatment change from $t-1$ to $t$, yielding a WAS
estimator with more than two periods, denoted $\widehat{\delta}^{T\geq
3}_{2}$.

\medskip
Those estimators are consistent under the following parallel-trends
assumption: $t-1$-to-$t$ switchers and stayers with the same $t-1$ treatment
would have experienced the same average $t-1$-to-$t$ outcome evolutions if
switchers had not switched. Because it is conditional on the lagged treatment, this assumption cannot be ``chained'' across pairs of periods: it only requires that some groups be on parallel trends over consecutive periods, not over the entire duration of the panel.
We propose pre-trend estimators to assess the plausibility of that assumption.
The placebo AS and WAS estimators are analogous to $\widehat{\delta}^{T\geq
3}_{1}$ and $\widehat{\delta}^{T\geq 3}_{2}$, but they compare the
$t-2$-to-$t-1$ outcome evolutions of $t-1$-to-$t$ switchers and stayers,
restricting the sample to $t-2$-to-$t-1$ stayers to avoid that those
comparisons be contaminated by the treatment's effect. If one finds that from
$t-2$-to-$t-1$, $t-1$-to-$t$ switchers and stayers are on parallel trends,
this lends credibility to the parallel-trends assumption underlying
$\widehat{\delta}^{T\geq 3}_{1}$ and $\widehat{\delta}^{T\geq 3}_{2}$.

\medskip
If there are dynamic treatment effects, $\widehat{\delta}^{T\geq 3}_{1}$ and
$\widehat{\delta}^{T\geq 3}_{2}$ may be biased. Those estimators compare the
outcome evolutions of $t-1$-to-$t$ switchers and stayers. But there may be,
say, $t-1$-to-$t$ stayers whose treatment changed from $t-2$ to $t-1$. If
lagged treatments affect units' current outcome, that change could still
affect the $t-1$-to-$t$ outcome evolution of those stayers, thus leading them
to violate the parallel-trends assumption underlying $\widehat{\delta}^{T\geq
3}_{1}$ and $\widehat{\delta}^{T\geq 3}_{2}$. To mitigate this concern, we
propose the following robustness check. One can recompute
$\widehat{\delta}^{T\geq 3}_{1}$ and $\widehat{\delta}^{T\geq 3}_{2}$,
restricting, for each pair of consecutive periods $(t-1,t)$, the estimation
sample to $t-2$-to-$t-1$ stayers, as in the placebo analysis. We show that
the resulting estimator is robust to dynamic effects up to one treatment lag.
Similarly, if one wants to allow for effects of the first and second
treatment lags on the outcome, one just needs to restrict the estimation
sample to $t-3$-to-$t-1$ stayers. However, the more robustness to dynamic
effects one would like to have, the smaller the estimation sample becomes.

\subsubsection{Identification and estimation with covariates}\label{subsec:covariates}

This section sketches how our estimators can be generalized to incorporate covariates, see Section \ref{appendixsec:covariates} of the
web appendix for more details. Consider for
simplicity the case with two periods, $T=2$, where $X_t$ is a $c$-dimensional
vector of covariates. We show in Section \ref{appendixsec:covariates} of the
web appendix that the AS and WAS can be identified and estimated under an
alternative parallel trends assumption that holds conditional on the
first-period covariates $X_1$ (Assumption \ref{hyp:parallel_trendsX}).
Because our estimation and inference methods are fully nonparametric, they
can be readily applied to the case with covariates after redefining the
nuisance functions to depend on the covariates, for instance,
$\mu_0(x,d)=E(\Delta Y|X_1=x,D_1=d,S=0)$ and so on.

\section{Application} 
\label{sec:appli}

\subsection{Data and design}

We use the yearly 1966-to-2008 panel dataset of \cite{li2014gasoline},
covering 48 US states (Alaska and Hawaii are excluded). This is a long panel, but recall that our estimators
assume parallel trends only across pairs of consecutive years. For each state$\times$year cell $(i,t)$, the data
contains $Z_{i,t}$, the total (state plus federal) gasoline tax in cents per
gallon, $D_{i,t}$, the log tax-inclusive price of gasoline, and $Y_{i,t}$,
the log gasoline consumption per adult. We seek to estimate the effect of
taxes on gasoline consumption and prices, and the
price-elasticity of gasoline consumption.\footnote{Instead, \cite{li2014gasoline} jointly estimate the effect of gasoline taxes
and tax-exclusive prices on consumption, using a TWFE regression with two
treatments. Between each pair of periods, the tax-exclusive price
changes in all states, so this treatment does not have stayers and its effect
cannot be estimated using our estimators.}

\medskip
Let $\mathcal{S}$ be the set of switching $(i,t)$ cells such that $Z_{i,t}\ne
Z_{i,t-1}$ but $Z_{i',t}=Z_{i',t-1}$ for some $i'$. The second condition
drops seven pairs of consecutive periods between
which the federal gasoline tax changed, thus implying that taxes changed in all states. $\mathcal{S}$ includes 384 cells, namely 19\% of the 2,016 state$\times$year cells for which $Z_{i,t}-Z_{i,t-1}$  can
be computed. Table \ref{table:balancing} in the web appendix shows that cells in $\mathcal{S}$ are
similar to other cells on a number of observable characteristics. Thus, there is no strong indication that the effects below apply to a selected subgroup.

\medskip
The top panel of Figure \ref{fig:support} below shows the
distribution of $Z_{i,1987}$ for 1987-to-1988 stayers, while the bottom panel
shows the distribution for 1987-to-1988 switchers. The figure shows that
there are many values of $Z_{i,1987}$ such that only one or two states have
that value, so $Z_{i,1987}$ is close to being continuously distributed.
Moreover, all switchers $i$ are such that
$$\min_{i':Z_{i',1988}=Z_{i',1987}}Z_{i',1987}\leq Z_{i,1987}\leq \max_{i':Z_{i',1988}=Z_{i',1987}}Z_{i',1987}.$$
Thus, Assumption \ref{hyp:support_condition0} seems to hold for this pair of
years. $(1987,1988)$ is not atypical: there are many other years where
$Z_{i,t}$ is close to being continuously distributed, and almost 95\%
of cells in $\mathcal{S}$ are such that
$\min_{i':Z_{i',t}=Z_{i',t-1}}Z_{i',t-1}\leq Z_{i,t-1}\leq
\max_{i':Z_{i',t}=Z_{i',t-1}}Z_{i',t-1}$. Dropping the few cells that do not
satisfy this condition barely changes the results presented below.

\begin{figure}[H]
    \centering
    \includegraphics[scale=0.38]{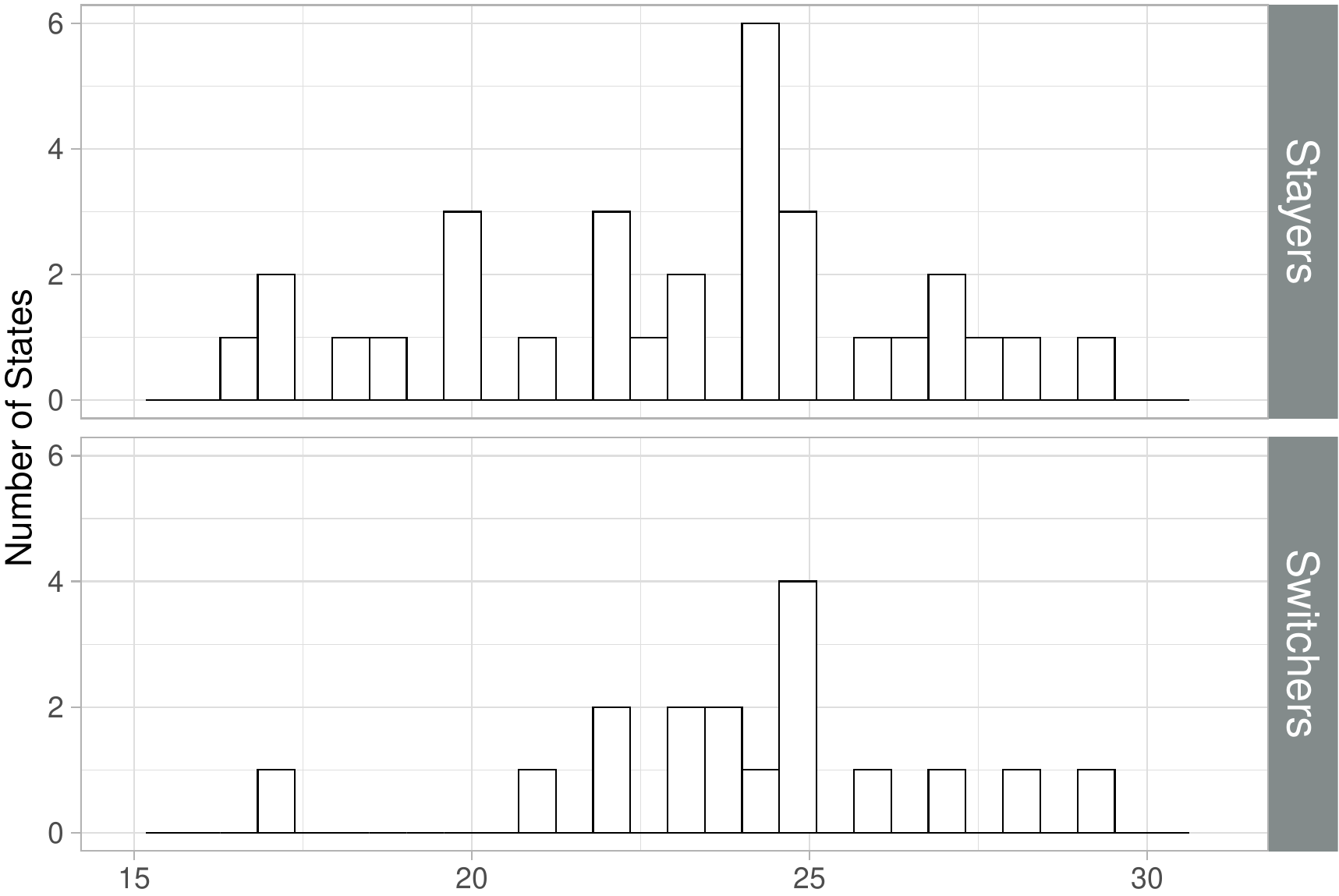}
    \caption{Gasoline tax in 1987 among 1987-to-1988 switchers and stayers}
    \label{fig:support}
    \normalsize
\end{figure}

Turning to the distribution of the tax changes $Z_{i,t}-Z_{i,t-1}$, 90\% of the cells in $\mathcal{S}$ experience an increase in their taxes, and 10\% experience a decrease. The average value of
$|Z_{i,t}-Z_{i,t-1}|$ is equal to 1.61 cents, while prior to the tax change,
switchers' average gasoline price is equal to 112 cents: our estimators
leverage small changes in taxes relative to gasoline prices. Finally,
$\min_{(i,t)\in \mathcal{S}}|Z_{i,t}-Z_{i,t-1}| = 0.05$ : some switchers
experience a very small change in their taxes.\footnote{\label{foot:notest_A7} We do not implement the test of whether $0$ belongs to the support of $|Z_{t}-Z_{t-1}||S_t=1$ proposed in Section 3.3 of \cite{de2024two}. The test therein assumes iid observations, while here $|Z_{i,t}-Z_{i,t-1}|$ is likely to be correlated across $t$.}

\subsection{Reduced-form and first-stage estimates}

Table \ref{table:RFandFS} below shows the AS and WAS estimates
of the reduced-form (Panel A) and first-stage (Panel B) effects of taxes on
gasoline quantities and prices. We control for lagged
prices $D_{t-1}$, to ensure that the resulting IV is robust to
heterogeneous effects over time, as explained in Section \ref{subsec:IV}. Estimates are computed with cross-fitting, using 10 splits. We use a polynomial of order 1 in
$(Z_{t-1},D_{t-1})$ to estimate $E(\Delta Y_t|Z_{t-1},D_{t-1}, S_t=0)$,
$E(\Delta D_t|Z_{t-1},D_{t-1},S_t=0)$, $E[S/\Delta D|Z_{t-1},D_{t-1}]$,
$P(S_{+,t}=1|Z_{t-1},D_{t-1})$, $P(S_{-,t}=1|Z_{t-1},D_{t-1})$, and
$P(S_{t}=0|Z_{t-1},D_{t-1})$. 10-folds cross-validation selects a polynomial of order one
for all conditional expectations, except for $E(\Delta D_t|Z_{t-1},D_{t-1},S_t=0)$ where it selects a polynomial of order two. Thus, polynomials of order one are in line with those selected by cross validation.\footnote{In a previous version of this paper \citep{de2022difference}, we had shown results very similar to those below with polynomials of order two and without cross fitting. With polynomials of order two and cross fitting, estimates become noisy: due to the small sample size, $\widehat{P}(S_{t}=0|Z_{t-1},D_{t-1})$ is close to zero for some $(i,t)$.} Standard errors
clustered at the state level, computed following \eqref{eq:ASIFT>=3} and
\eqref{eq:WASIFT>=3} in the web appendix, are shown between parentheses. Estimations use 1632 (48$\times$ 35) first-difference observations: 7 periods
are excluded as they do not have stayers.

\medskip
In Panel A, the AS estimate indicates that increasing gasoline tax
by 1 cent decreases quantities consumed by 0.43 percent on average for the
switchers. That effect is insignificant. The WAS estimate indicates that increasing gasoline tax
by 1 cent decreases quantities consumed by 0.36 percent, an effect that is significant. In Panel B, the AS
estimate of the first-stage effect is positive and insignificant. The WAS estimate is significant, and indicates that if gasoline tax increases by 1 cent on average, prices increase by 0.58 percent. In both panels, the AS estimates are close to, and
not significantly different from, the WAS estimates, but they are insignificant because they are
markedly less precise: in line with the results of Proposition \ref{prop:variance_comparison}, their standard errors are almost three times larger. In what follows we focus on the WAS estimates.

\begin{table}[H]
\begin{center}
			\begin{threeparttable}
				\caption{Effects of gasoline tax on quantities and prices}
				\label{table:RFandFS}
				\begin{tabular}{lcc}
					\toprule   %
					{
		\begin{tabular}{lc}
\multicolumn{2}{c}{Panel A: Reduced-form effect of taxes on quantities}\\\midrule
AS                  &     -0.0043 (0.0027) \\
WAS                 &     -0.0036  (0.0010)\\
			\midrule
\multicolumn{2}{c}{Panel B: First-stage effect of taxes on prices}\\\midrule
AS                  &      0.0038 (0.0024)\\
WAS                 &      0.0058 (0.0009) \\ \midrule
		Observations        &       1,632 \\
	\end{tabular}
}
\\
					\bottomrule
				\end{tabular}
\end{threeparttable}
\end{center}
\footnotesize
Notes: Panel A (resp. B) shows the AS and WAS estimates of the reduced-form (resp. first-stage) effect of taxes on gasoline quantities (resp. prices). All estimates control for the lag of prices. All conditional expectations are estimated using a polynomial of order 1 in $(Z_{t-1},D_{t-1})$, with  cross-fitting, using ten splits. Standard errors clustered at the state level are shown next to the estimates, between parentheses.
\end{table}

\subsection{Assessing estimates' plausibility}

The estimates in Table \ref{table:RFandFS} rely on the reduced-form and first-stage parallel trends assumptions in Assumption \ref{hyp:parallel_trends_IV}. In this application, these assumptions require that, among states with the same gasoline taxes and prices in year $t-1$, counterfactual changes in gasoline demand and prices in the absence of tax changes be mean independent of gasoline taxes in year $t$. These assumptions may be plausible if states adjust their gasoline taxes primarily for fiscal reasons, as their budgetary needs may be largely unrelated to shocks to gasoline demand and prices. However, they could for instance be violated if policymakers lower gasoline taxes when their state experiences a recession and raise them when it experiences a boom.

\medskip
To assess the plausibility of Assumption \ref{hyp:parallel_trends_IV}, Table \ref{table:placeboRFandFS} reports placebo (or pre-trend) WAS estimates of the reduced-form and first-stage effects. The placebos are analogous to the actual estimates, but they replace $\Delta Y_t$ by $\Delta Y_{t-1}$, and they restrict the sample to states whose taxes did not change between $t-2$ and $t-1$. This reduces the sample size by around a third. The placebos
are small and insignificant, both for quantities and prices. This shows that before switchers change
their gasoline taxes, switchers' and stayers' consumption and
prices do not follow detectably different evolutions. In Table \ref{table:RFandFS_placebosample} of the web appendix we re-estimate the WAS in the placebo subsample. In that subsample, the WAS remains valid if the first lag of taxes affects current gasoline
prices and quantities, as discussed in Section \ref{subsec:Tgeq3}. The estimates therein are close to those in
Table \ref{table:RFandFS}.

\begin{table}[H]
\begin{center}
			\begin{threeparttable}
				\caption{Placebo WAS effects of gasoline tax on quantities and prices}
				\label{table:placeboRFandFS}
				\begin{tabular}{lcc}
					\toprule   %
					{
		\begin{tabular}{lc}
Placebo reduced-form effect of taxes on quantities          &      0.0011 (0.0015) \\
Placebo first-stage effect of taxes on prices          &      0.0011 (0.0015)\\
		 \midrule[1pt]
		
		Observations        &       1,059\\
		
	\end{tabular}
}
\\
					\bottomrule
				\end{tabular}
			\end{threeparttable}
\end{center}
\footnotesize
Notes: The table shows placebo WAS estimates of the reduced-form and first-stage effects of taxes on quantities and prices. The estimates and their standard errors are computed as the actual estimates, replacing $\Delta Y_t$ by $\Delta Y_{t-1}$, and restricting the sample to states whose taxes did not change between $t-2$ and $t-1$.
\end{table}

\subsection{Comparison with other estimates}

We re-estimate the effects of taxes on consumption and prices using two alternative approaches. The first, proposed by \cite{de2020difference}, accommodates continuous treatments, effects of lagged treatments up to any observed lag, and time-varying treatment effects. Results are reported in Table \ref{table:RFandFS_unrestricted_dyn} of the web appendix. The estimates have the same signs as those in Table \ref{table:RFandFS}, but are much less precise. The reason is that allowing for effects up to any lag substantially reduces the estimation sample. Rather than comparing states whose taxes change between $t-1$ and $t$ to states whose taxes do not change, the estimators compare states experiencing their first tax change at $t$ to states whose taxes have not yet changed. This leaves only 46 $(i,t)$ cells experiencing a first tax change, to which the estimated effects apply, and 169 control cells, for a total of 215 observations. Consequently, neither estimate is statistically significant: the standard error of the estimated effect on consumption is 13 times that of the WAS estimate in Table \ref{table:RFandFS},\footnote{The target parameters in \cite{de2020difference} generalize the WAS to settings with unrestricted dynamic effects.} while the standard error for prices is more than nine times as large. Thus, in this application, allowing for effects up to any lag and time-varying effects comes at a substantial cost in terms of both precision and the population of observations to which the estimates apply.

\medskip
The second approach, proposed by \cite{de2025treatmenteffect} after the first version of this paper, relies on a distributed-lag linear model for potential outcomes, with coefficients that may vary across groups. Unlike the WAS and AS estimators, it accommodates effects of lagged taxes, up to a prespecified number of lags. However, it assumes that treatment effects are linear, time-invariant, and additively separable across current and lagged taxes. The time-invariance assumption may be particularly restrictive in our application, given the long time span of the data. Results for models including up to two lags are reported in Table 4 of the web appendix.\footnote{In models with 5 lags, effects on lags 3 to 5 are small and statistically insignificant, and including lags 3 to 5 barely changes the other estimated effects, hence our choice of a model with two lags.} The contemporaneous first-stage effect is similar to our WAS first-stage estimate, while the contemporaneous reduced-form effect is smaller than its WAS counterpart, although still statistically significant. In both equations, the coefficients on the first and second lags are smaller in magnitude than the contemporaneous coefficients, but remain statistically significant. Overall, the distributed-lag estimates support the qualitative conclusions of our baseline analysis regarding the contemporaneous effects of taxes on consumption and prices: although lagged taxes also appear to matter, accounting for their effects does not overturn our main findings. These findings are reassuring given the different assumptions underlying the two approaches: the WAS and AS estimators rule out lagged effects, whereas the distributed-lag estimators impose linearity and time invariance\footnote{The two estimators also weight observations differently: the WAOSS weights each switching state-year by the absolute value of its tax change, whereas the \st{DistLagHet} estimator gives equal weight to each state.}. As shown in the previous paragraph, relaxing both sets of restrictions simultaneously leads to uninformative conclusions.

\subsection{IV estimate}

Table \ref{table:IV} shows an IV-WAS estimate of the price elasticity of gasoline consumption. As the instrument's first stage is not very strong (t-stat=6.42, F-stat=41.27), to construct a confidence interval we invert the Anderson-Rubin test \citep{anderson1949estimation}, clustered at the state level.\footnote{When the \st{ar\_ci} option is specified, the \st{did\_multiplegt\_stat} Stata package computes this confidence interval.} The IV-WAS estimate is negative but not significant. In addition to Assumption \ref{hyp:parallel_trends_IV}, the IV-WAS estimator also relies on the monotonicity condition in Assumption \ref{hyp:monotonicity}. In this application, this condition seems plausible, as an increase in gasoline taxes is unlikely to lead to a decrease in gasoline prices.

\begin{table}[H]
\begin{center}
\caption{Price elasticity of gasoline consumption}
				\label{table:IV}
				\begin{tabular}{lcc}
					\toprule   %
					{
		\begin{tabular}{lc} 

		IV-WAS                  &     -0.6556   \\
		       
		                 & [-2.917,  0.852] \\
	\midrule
		Observations        &       1,632 \\
	\end{tabular}
}
\\
					\bottomrule
				\end{tabular}
\end{center}
\footnotesize
Notes: The table shows an IV-WAS estimate of the price-elasticity of gasoline consumption, controlling for the lag of prices. All conditional expectations are estimated using a polynomial of order 1 in $(Z_{t-1},D_{t-1})$, with cross-fitting, using ten splits. An Anderson-Rubin test-based confidence interval, clustered at the state level, is shown below the estimate.
\end{table}



\section{Conclusion}

We propose new doubly-robust, semiparametrically efficient difference-in-difference (DID) estimators for average slopes of switchers' potential outcomes when treatments are continuously distributed in every time period. We assume that between pairs of consecutive periods, the treatment of some units, the switchers, changes, while the treatment of other units, the stayers, does not change. We propose a parallel-trends assumption on the outcome evolution of switchers and stayers with the same baseline treatment. Under that assumption, two target parameters can be estimated. Our first target is the average slope of switchers' period-two potential outcome function, from their period-one to their period-two treatment, referred to as the AS. Our second target is a weighted average of switchers' slopes, where switchers receive a weight proportional to the absolute value of their treatment change, referred to as the WAS. The AS and WAS serve different purposes. When it comes to estimation, the WAS has two advantages. First, it can be estimated at the parametric rate even if units can experience an arbitrarily small treatment change. Second, under some conditions, the asymptotic variance of the WAS estimator is strictly lower than that of the AS. In our application, we use US-state-level panel data to estimate the effect of gasoline taxes on gasoline consumption. The standard error of the WAS is almost three times smaller than that of the AS, and the two  estimates are close.

\medskip
We also consider the instrumental-variable case, as there are instances where units experiencing/not experiencing a treatment change are unlikely to be on parallel trends, but one has at hand an instrument such that units experiencing/not experiencing an instrument change are more likely to be on parallel trends. Then, we propose widely applicable IV-DID estimators.

\medskip
Throughout, we assume that there are some stayers, namely units whose treatment or instrument does not change between consecutive periods. \cite{de2023nostayers} discuss the extension of the results in this paper to applications without stayers.

\newpage
\bibliography{biblio}

\newpage
\appendix
\section{Proofs}

Hereafter, $\Supp(A)$ denotes the support of the random variable $A$. The proofs of Theorems \ref{thm:WASasnormality_nonmonotonictreatment} and \ref{thm:main_IV} are given in the main appendix.

\subsection{Lemma \ref{lem:building_block}}
\begin{align*}
	&E\left(Y_2(d_2) - Y_2(d_1)\middle|D_1=d_1,D_2=d_2\right)\\
	=&E\left(\Delta Y \middle|D_1=d_1,D_2=d_2\right)-E\left(\Delta Y(d_1)\middle|D_1=d_1,D_2=d_2\right)\\
	=&E\left(\Delta Y \middle|D_1=d_1,D_2=d_2\right)-E\left(\Delta Y(d_1)\middle|D_1=d_1,D_2=d_1\right)\\
	=&E\left(\Delta Y \middle|D_1=d_1,D_2=d_2\right)-E\left(\Delta Y\middle|D_1=d_1,S=0\right)\\
	=&E\left(\Delta Y -E\left(\Delta Y\middle|D_1=d_1,S=0\right)\middle|D_1=d_1,D_2=d_2\right),
\end{align*}
where the second equality follows from Assumption \ref{hyp:parallel_trends}.
This proves the result $_\Box$

\subsection{Theorem \ref{thm:main0}}

For any function $d\mapsto\mu(d)$ and for any function  $d\mapsto\alpha(d)$, let
$$\psi_1(\mu,\alpha)=\frac{1}{E[S]}\left(\frac{S}{\Delta D}-(1-S)\alpha(D_1)\right)\left[\Delta Y-\mu(D_1)\right]-\delta_1.$$
To prove the result, we have to show that $E[\psi_1(\mu_0,\alpha)]=E[\psi_1(\mu,\alpha_0)]=0$. First, note that these expectations are well-defined since $S/\Delta D\le 1/\kappa$ by Assumption \ref{hyp:noquasi-stayers}, $\alpha_0$ is bounded by Assumption \ref{hyp:support_condition0}, $\alpha$ is bounded, $E[|\Delta Y|]<\infty$ and $E[|\mu_0(D_1)|]<\infty$ by Assumptions \ref{hyp:parallel_trends} and \ref{hyp:regularity} and  $E[|\mu(D_1)|]<\infty$.

\medskip
Now, we have
\begin{align*}
E[\psi_1(\mu_0,\alpha)]&=E\left[\left(\frac{\Delta Y-\mu_0(D_1)}{\Delta D}\right)\frac{S}{E[S]}\right]-\delta_1-\frac{1}{E[S]}E\left[(1-S)\alpha(D_1)\left(\Delta Y-\mu_0(D_1)\right)\right]\\
&=E\left[\left(\frac{\Delta Y-\mu_0(D_1)}{\Delta D}\right)\middle|S=1\right]-\delta_1\\
&-\frac{1}{E[S]}E\left[(1-S)\alpha(D_1)\left(E(\Delta Y|S,D_1)-\mu_0(D_1)\right)\right]\\
&=0.
\end{align*}
The last equality follows from the two following facts. First, $\delta_1=E\left[\left(\Delta Y-\mu_0(D_1)\right)/\Delta D\middle|S=1\right]$, by the definitions of $\delta_1$ and $\mu_0(d)$, Lemma \ref{lem:building_block}, and the law of iterated expectations. Second, $(1-S)\left(E(\Delta Y|S,D_1)-\mu_0(D_1)\right)=0$ by the definition of $\mu_0(d)$. Then,
\begin{align*}
E[\psi_1(\mu,\alpha_0)]&=E\left[\left(\frac{\Delta Y-\mu(D_1)}{\Delta D}\right)\frac{S}{E[S]}\right]-\delta_1-\frac{1}{E[S]}E\left[(1-S)\alpha_0(D_1)\left(\Delta Y-\mu(D_1)\right)\right]\\
&=E\left[\left(\frac{\Delta Y-\mu_0(D_1)}{\Delta D}\right)\middle|S=1\right]-\delta_1+\frac{1}{E[S]}E\left[\left(\mu_0(D_1)-\mu(D_1)\right)\frac{S}{\Delta D}\right]\\
&-\frac{1}{E[S]}E\left[(1-S)\alpha_0(D_1)\left(\Delta Y-\mu(D_1)\right)\right]\\
&=\frac{1}{E[S]}E\left[\left(\mu_0(D_1)-\mu(D_1)\right)E\left(\left.\frac{S}{\Delta D}\right\vert D_1\right)\right]\\
&-\frac{1}{E[S]}E\left[\frac{1-S}{E(1-S|D_1)}\left(\Delta Y-\mu(D_1)\right)E\left(\left.\frac{S}{\Delta D}\right\vert D_1\right)\right]\\
&=\frac{1}{E[S]}E\left[\left(\mu_0(D_1)-\mu(D_1)\right)E\left(\left.\frac{S}{\Delta D}\right\vert D_1\right)\right]\\
&-\frac{1}{E[S]}E\left[\left(\mu_0(D_1)-\mu(D_1)\right)E\left(\left.\frac{S}{\Delta D}\right\vert D_1\right)\right]\\
&=0.
\end{align*}
The second quality follows after adding and subtracting $\mu_0(D_1)$ in the first expectation. The third equality follows from the fact that $\delta_1=E\left[\left(\Delta Y-\mu_0(D_1)\right)/\Delta D\middle|S=1\right]$, from the law of iterated expectations, and from the definition of $\alpha_0(d)$. The fourth equality follows from the law of iterated expectations $_\Box$

\subsection{Theorem \ref{thm:main0_quasistayers}}

First, observe that for any $\eta\ge \eta'>0$, $\{S_\eta=1\}\subseteq \{S_{\eta'}=1\}$ and $\{S=1\}=\cup_{\eta>0} \{S_\eta=1\}$. Then, by continuity of probability measures,
\begin{equation}\label{eq:cont_proba}
  \lim_{\eta\downarrow 0}P(S_\eta=1)=P(S=1)>0,
\end{equation}
where the inequality follows by Assumption \ref{hyp:support_condition0}. Thus, there exists $\underline{\eta}>0$ such that for all $\eta\in(0,\underline{\eta})$, $P(S_\eta=1)>0$. Hereafter, we assume that $\eta\in(0,\underline{\eta})$.

\medskip
Next, we prove that $\Supp(D_1|S=1)\subseteq \Supp(D_1|S=0)$. Suppose that $P(D_1\in \text{Supp}(D_1|S = 0)|S=1)<1$. Then, letting $A:=\text{Supp}(D_1|S = 1)\cap \text{Supp}(D_1|S = 0)^c$, we have $P(D_1 \in A|S=1)>0$ and therefore $P(D_1 \in A)>0$. Next,
\begin{align*}
P(S=0|D_1\in A)= & \frac{P(D_1 \in A, S=0)}{P(D_1\in A)}\le \frac{P(D_1\in\text{Supp}(D_1|S = 0)^c,S=0)}{P(D_1\in A)}= 0.	
\end{align*}
On the other hand, by Assumption \ref{hyp:support_condition0},
$$P(S=0|D_1\in A)=E(P(S=0|D_1)|D_1\in A)>  \zeta,$$
so we reach a contradiction. Therefore, $P(D_1\in \text{Supp}(D_1|S = 0)|S=1)=1$. Finally, $\Supp(D_1|S = 1)\subseteq\Supp(D_1|S = 0)$ follows from the definition of the support of a random variable.

\medskip
Now, since $\Supp(D_1|S_\eta=1)\subseteq \Supp(D_1|S=1)$, we obtain that for all $(d_1,d_2)\in \Supp(D_1,D_2|S_\eta=1)$, $d_1\in \Supp(D_1|S=0)$, so $E(Y_2(d_1) - Y_1 (d_1)|D_1=d_1, S=0)=E(Y_2(d_1) - Y_1 (d_1)|D_1=d_1, D_2=d_1)$ is well-defined. Moreover, for almost all such $(d_1,d_2)$,
\begin{align}
E(Y_2(d_1) - Y_1 (d_1)|D_1=d_1, D_2=d_2) = & E(Y_2(d_1) - Y_1 (d_1)|D_1=d_1, D_2=d_1) \notag \\
= & E(\Delta Y | D_1=d_1, S=0),
\label{eq:nonparam_eq1}
\end{align}
where the first equality follows from Assumption \ref{hyp:parallel_trends}. Now, by Point 3 of Assumption \ref{hyp:regularity}, $[Y_2(D_2) - Y_2(D_1)]/\Delta D$ admits an expectation. Moreover,
\begin{align}
& E\left(\frac{Y_2(D_2) - Y_2(D_1)}{\Delta D} \middle| S_\eta=1\right) \notag\\
=& E\left(\frac{E(Y_2(D_2) - Y_1(D_1) | D_1, D_2) - E(Y_2(D_1) - Y_1(D_1) | D_1, D_2)}{\Delta D} \middle| S_\eta=1\right)  \notag \\
=&  E\left(\frac{E(\Delta Y | D_1, D_2) - E(\Delta Y | D_1, S=0)}{\Delta D} \middle| S_\eta=1\right) \notag \\
=&  E\left(\frac{\Delta Y - E(\Delta Y | D_1, S=0)}{\Delta D} \middle| S_\eta=1 \right),\label{eq:expr_given_c}
\end{align}
where the first equality follows from the law of iterated expectations, the second follows from \eqref{eq:nonparam_eq1}, and the third again by the law of iterated expectations. Next,
$$\delta_1 = \Pr(S_\eta=1|S=1) E\left[\frac{Y_2(D_2)-Y_2(D_1)}{\Delta D}\middle|S_\eta=1 \right]  + E\left[(1-S_\eta) \frac{Y_2(D_2)-Y_2(D_1)}{\Delta D}\middle|S=1\right].$$
Moreover,
\begin{align*}
\left|E\left[(1-S_\eta) \frac{Y_2(D_2)-Y_2(D_1)}{\Delta D}\middle|S=1\right]\right| & \le E\left[(1-S_\eta) \left|\frac{Y_2(D_2)-Y_2(D_1)}{\Delta D}\right|\; \middle|S=1\right] \\
& \le E\left[(1-S_\eta)\overline{Y}|S=1\right],
\end{align*}
where the second inequality follows by Assumption \ref{hyp:regularity}. Now, using $S_\eta\le S$ and \eqref{eq:cont_proba} again,
$$\lim_{\eta\downarrow 0} P(S_\eta=1|S=1)=\lim_{\eta\downarrow 0} \frac{P(S_\eta=1)}{P(S=1)}=1.$$
As a result,conditional on $S=1$, $\lim_{\eta\downarrow 0} (1-S_\eta)\overline{Y}=0$ a.s.. Moreover, $(1-S_\eta)\overline{Y}\le \overline{Y}$ with $E[\overline{Y}|S=1]<\infty$. Then, by the dominated convergence theorem,
$$\lim_{\eta\downarrow 0} E\left[(1-S_\eta) \frac{Y_2(D_2)-Y_2(D_1)}{\Delta D}\middle|S=1\right]=0.$$
We finally obtain
\begin{equation}
\delta_1 = \lim_{\eta\downarrow 0} E\left[\frac{Y_2(D_2)-Y_2(D_1)}{\Delta D}\middle|S_\eta=1 \right].
\label{eq:lim_c}
\end{equation}
The result follows by combining \eqref{eq:expr_given_c} and \eqref{eq:lim_c} $_\Box$

\subsection{Theorem \ref{thm:ASasnormality}}

For $t\in \{1,2\}$ and $k\in\{1,2\}$, let $\mathbf{D}^{(k)}_t=(D_{i,t})_{i\in \mathcal{I}_k}$, $\mathbf{D}^{(k)}=(D_{i,1},D_{i,2})_{i\in \mathcal{I}_k}$ and $\bm{U}^{(k)}=(D_{i,1}, D_{i,2},$ $Y_{i,1},Y_{i,2})_{i\in \mathcal{I}_k}$. Also, to reduce notation, we use hereafter $C$ to denote a generic constant whose value may be different in each appearance. Below, we use repeatedly the fact that if, for  sequences of variables $(A_n, B_n)_{n\ge 1}$, $E(A_n|B_n)=0$ and $V(A_n|B_n)=o_P(1)$, then $A_n = o_P(1)$. To see this, note first that by Chebyshev's inequality, $P(|A_n|>M|B_n) = o_P(1)$ for all $M>0$. This implies that the bounded random variable $E(\ind{|A_n|>M}|B_n)$ converges in distribution to 0. Then, by, e.g., Theorem 2.20 in \cite{vanderVaart2000},
$$P(|A_n|>M) = E[E(\ind{|A_n|>M}|B_n)] \to 0.$$

\subsubsection{Asymptotic normality} 
\label{ssub:asymptotic_normality}

The proof follows a similar strategy as that of Theorem 3.2 in \citet{wager2024_book}. We have
\begin{align*}
& \sqrt{I_1}(\hat\delta_{1,\mathsf{DR}}^{(1)}-\delta_1)\\
=&\frac{1}{E[S]+o_P(1)}\cdot\frac{1}{\sqrt{I_1}}\sum_{i\in \mathcal{I}_1} \left\{\left(\frac{S_i}{\Delta D_i}-\frac{\hat g^{(2)}(D_{i,1})}{\hat p^{(2)}(D_{i,1})}(1-S_i)\right)\left(\Delta Y_i-\hat\mu^{(2)}(D_{i,1})\right)-\delta_1 S_i\right\}\\
=&\frac{1}{E[S]+o_P(1)}\cdot\frac{1}{\sqrt{I_1}}\sum_{i\in \mathcal{I}_1} V_i,
\end{align*}
where
\[V_i:=\left(\frac{S_i}{\Delta D_i}-\frac{\hat g^{(2)}(D_{i,1})}{\hat p^{(2)}(D_{i,1})}(1-S_i)\right)\left(\Delta Y_i-\hat\mu^{(2)}(D_{i,1})\right)-\delta_1 S_i.\]
Moreover,
\begin{align*}
V_i&=\left(\frac{S_i}{\Delta D_i}-\frac{g_0(D_{i,1})}{p_0(D_{i,1})}(1-S_i)\right)\left(\Delta Y_i-\mu_0(D_{i,1})\right)-\delta_1 S_i\\
&+\left(\frac{g_0(D_{i,1})}{p_0(D_{i,1})}-\frac{\hat g^{(2)}(D_{i,1})}{\hat p^{(2)}(D_{i,1})}\right)(1-S_i)\left(\Delta Y_i-\mu_0(D_{i,1})\right)\\
&+\left(\frac{g_0(D_{i,1})}{p_0(D_{i,1})}-\frac{\hat g^{(2)}(D_{i,1})}{\hat p^{(2)}(D_{i,1})}\right)(1-S_i)\left(\mu_0(D_{i,1})-\hat\mu^{(2)}(D_{i,1})\right)\\
&+\left(\frac{S_i}{\Delta D_i}-\frac{g_0(D_{i,1})}{p_0(D_{i,1})}(1-S_i)\right)\left(\mu_0(D_{i,1})-\hat\mu^{(2)}(D_{i,1})\right)\\
&=:E[S]\psi_{1i}+T_{1i}+T_{2i}+T_{3i}
\end{align*}
and thus, letting $T_j=\sum_{i\in \mathcal{I}_1} T_{ji}/\sqrt{I_1}$ for $j=1,2,3$, we obtain
\[\sqrt{I_1}(\hat\delta_{1,\mathsf{DR}}^{(1)}-\delta_1)=\frac{1}{E[S]+o_P(1)}\left(\frac{1}{\sqrt{I_1}}\sum_{i\in \mathcal{I}_1}E[S]\psi_{1i}+T_1+T_2+T_3\right).\]

First, consider the term $T_1$. We have
\begin{align*}
E\left[T_1|\I_1,\mathbf{D}^{(1)},\bm{U}^{(2)}\right]&=\frac{1}{\sqrt{I_1}}\sum_{i\in \mathcal{I}_1}\left(\frac{g_0(D_{i,1})}{p_0(D_{i,1})}-\frac{\hat g^{(2)}(D_{i,1})}{\hat p^{(2)}(D_{i,1})}\right)(1-S_i)\left(E[\Delta Y_i|D_{i,1},D_{i,2}]-\mu_0(D_{i,1})\right)\\
&=\frac{1}{\sqrt{I_1}}\sum_{i\in \mathcal{I}_1}\left(\frac{g_0(D_{i,1})}{p_0(D_{i,1})}-\frac{\hat g^{(2)}(D_{i,1})}{\hat p^{(2)}(D_{i,1})}\right)(1-S_i)\left(E[\Delta Y_i|D_{i,1},S_i=0]-\mu_0(D_{i,1})\right)\\
&=0,
\end{align*}
where we recall that $\I_1$ is the set of indices of units in the first subsample (e.g., $\{1, 3, 4...\}$), not their corresponding random variables, and the first equality uses Assumption \ref{hyp:subsamples}. Also,
\begin{align*}
V\left[T_1|\I_1,\mathbf{D}^{(1)},\bm{U}^{(2)}\right]&=\frac{1}{I_1}\sum_{i\in \mathcal{I}_1}\left(\frac{\hat g^{(2)}(D_{i,1})}{\hat p^{(2)}(D_{i,1})}-\frac{g_0(D_{i,1})}{p_0(D_{i,1})}\right)^2(1-S_i)V(\Delta Y|D_{i,1},S_i=0)\\
&\le C \cdot \frac{1}{I_1}\sum_{i\in \mathcal{I}_1}\left(\frac{\hat g^{(2)}(D_{i,1})}{\hat p^{(2)}(D_{i,1})}-\frac{g_0(D_{i,1})}{p_0(D_{i,1})}\right)^2.
\end{align*}
Now, noting that
\begin{align*}
\frac{\hat g}{\hat p}-\frac{g_0}{p_0}=\frac{\hat g-g_0}{\hat p}-\frac{g_0}{p_0}\left(\frac{\hat p-p_0}{\hat p}\right),
\end{align*}
and using the inequality $(x+y)^2 \le 2 (x^2 + y^2)$, we have
\begin{align*}
\left(\frac{\hat g}{\hat p}-\frac{g_0}{p_0}\right)^2\le \frac{2}{\hat p^2}\left[(\hat g-g_0)^2 + \frac{g^2}{p^2}\left(\hat p-p_0\right)^2\right],
\end{align*}
and therefore
\begin{align*}
& \frac{1}{I_1}\sum_{i\in \mathcal{I}_1}\left(\frac{\hat g^{(2)}(D_{i,1})}{\hat p^{(2)}(D_{i,1})}-\frac{g_0(D_{i,1})}{p_0(D_{i,1})}\right)^2 \\
\le&  \frac{2}{I_1}\sum_{i\in \mathcal{I}_1}\left(\frac{(\hat g^{(2)}(D_{i,1})-g_0(D_{i,1}))^2}{\hat p^{(2)}(D_{i,1})^2}+\frac{g_0(D_{i,1})^2}{p_0(D_{i,1})^2}\frac{\left(\hat p^{(2)}(D_{i,1})-p_0(D_{i,1})\right)^2}{\hat p^{(2)}(D_{i,1})^2}\right)\\
\le&  \frac{C}{\inf_d{\hat p^{(2)}(d)^2}}\cdot \left(\frac{1}{I_1}\sum_{i\in \mathcal{I}_1}(\hat g^{(2)}(D_{i,1})-g_0(D_{i,1}))^2+\sup_{d}\abs{\hat p^{(2)}(d)-p_0(d)}^2\right)\\
=&o_P(1).
\end{align*}
The last line follows from two points. First, by Assumption \ref{hyp:rate_DR_AS}-(a), the numerator is an $o_P(1)$. Second,  uniform consistency of $\hat p^{(2)}$ (still by Assumption \ref{hyp:rate_DR_AS}-(a)) and Assumption \ref{hyp:support_condition0} imply that $\inf_d{\hat p^{(2)}(d)^2}$ is bounded from below with probability approaching 1.

\medskip
Turning to $T_2$, we have, by Cauchy-Schwarz inequality,
{\small\begin{align*}
&\abs{T_2}\\
&\le \frac{1}{\sqrt I_1}\sum_{i\in \mathcal{I}_1}\abs{\frac{\hat g^{(2)}(D_{i,1})}{\hat p^{(2)}(D_{i,1})}-\frac{g_0(D_{i,1})}{p_0(D_{i,1})}}\abs{\hat\mu^{(2)}(D_{i,1})-\mu_0(D_{i,1})}\\
&\le \sqrt{\frac{I_1}{n}}\cdot\sqrt{n}\left(\frac{1}{I_1}\sum_{i\in \mathcal{I}_1}\left(\frac{\hat g^{(2)}(D_{i,1})}{\hat p^{(2)}(D_{i,1})}-\frac{g_0(D_{i,1})}{p_0(D_{i,1})}\right)^2\right)^{1/2}\left(\frac{1}{I_1}\sum_{i\in \mathcal{I}_1}\left(\hat\mu^{(2)}(D_{i,1})-\mu_0(D_{i,1})\right)^2\right)^{1/2}\\
&\le \sqrt{\frac{I_1}{n}}\cdot\sqrt{n}\cdot\left(\frac{C}{\inf_d{\hat p^{(2)}(d)^2}}\cdot \left(\frac{1}{I_1}\sum_{i\in \mathcal{I}_1}(\hat g^{(2)}(D_{i,1})-g_0(D_{i,1}))^2+\frac{1}{I_1}\sum_{i\in \mathcal{I}_1}(\hat p^{(2)}(D_{i,1})-p_0(D_{i,1}))^2\right)\right)^{1/2}\\
&\times \left(\frac{1}{I_1}\sum_{i\in \mathcal{I}_1}(\hat\mu^{(2)}(D_{i,1})-\mu_0(D_{i,1}))^2\right)^{1/2}\\
&\le \sqrt{\frac{I_1}{n}}\frac{C}{\inf_d{\hat p^{(2)}(d)^2}}\cdot\sqrt{n}\cdot\left(\cdot \left(\frac{1}{I_1}\sum_{i\in \mathcal{I}_1}(\hat g^{(2)}(D_{i,1})-g_0(D_{i,1}))^2\right)^{1/2}+\left(\frac{1}{I_1}\sum_{i\in \mathcal{I}_1}(\hat p^{(2)}(D_{i,1})-p_0(D_{i,1}))^2\right)^{1/2}\right)\\
&\times \left(\frac{1}{I_1}\sum_{i\in \mathcal{I}_1}(\hat\mu^{(2)}(D_{i,1})-\mu_0(D_{i,1}))^2\right)^{1/2}\\
& = o_P(1),
\end{align*}}
where the last inequality uses again the fact that for two positive real numbers $x$ and $y$, $x^2+y^2\leq 2(x+y)^2$, and the last equality follows from Assumption \ref{hyp:rate_DR_AS}-(b) and again, the fact that $\inf_d{\hat p^{(2)}(d)^2}$ is bounded from below with probability approaching 1.

\medskip
Finally, for $T_3$, note that
\begin{align*}
E\left[T_3|\I_1,\mathbf{D}^{(1)}_1,\bm{U}^{(2)}\right]&=\frac{1}{\sqrt{I_1}}\sum_{i\in \mathcal{I}_1}E\left[\left(\frac{S_i}{\Delta D_i}-\frac{g_0(D_{i,1})}{p_0(D_{i,1})}(1-S_i)\right)\big| D_{i,1}\right]\left(\mu_0(D_{i,1})-\hat\mu^{(2)}(D_{i,1})\right)\\
&=0,
\end{align*}
where the first equality uses Assumption \ref{hyp:subsamples}. Also,
\begin{align*}
V\left[T_3|\I_1,\mathbf{D}^{(1)}_1,\bm{U}^{(2)}\right]&=\frac{1}{I_1}\sum_{i\in \mathcal{I}_1}\left(\mu_0(D_{i,1})-\hat\mu^{(2)}(D_{i,1})\right)^2V\left[\frac{S_i}{\Delta D_i}-\frac{g_0(D_{i,1})}{p_0(D_{i,1})}(1-S_i)\big| D_{i,1}\right]\\
&\le C\cdot \frac{1}{I_1}\sum_{i\in \mathcal{I}_1}\left(\mu_0(D_{i,1})-\hat\mu^{(2)}(D_{i,1})\right)^2=o_P(1).
\end{align*}

Collecting all the results, we obtain
\begin{align*}
\sqrt{I_1}(\hat\delta_{1,\mathsf{DR}}^{(1)}-\delta_1)&=\frac{1}{\sqrt I_1}\sum_{i\in \mathcal{I}_1} \psi_{1i}+o_P(1).
\end{align*}
By analogous calculations, we can show the same result for $\hat\delta_{1,\mathsf{DR}}^{(2)}$. Also, by the law of large numbers, $n_S^{(k)}/I_k\convP  E(S_i)$. Therefore,
$$\frac{n}{I_k} \frac{n_S^{(k)}}{n_S^{(1)}+n_S^{(2)}} \convP 1.$$
As a result,
\begin{align*}
\sqrt{n}(\hat\delta_{1,\mathsf{DR}}-\delta_1)&=\sqrt{n}\left(\frac{I_1}{n} \frac{n}{I_1}\frac{n_S^{(1)}}{n_S^{(1)}+n_S^{(2)}} (\hat\delta_{1,\mathsf{DR}}^{(1)}-\delta_1)+\frac{I_2}{n} \frac{n}{I_2}\frac{n_S^{(2)}}{n_S^{(1)}+n_S^{(2)}} (\hat\delta_{1,\mathsf{DR}}^{(2)}-\delta_1)\right)\\
&=\sqrt{n}\left[\frac{1}{n}\sum_{i\in \mathcal{I}_1}\psi_{1i}+\frac{1}{n}\sum_{i\in \mathcal{I}_2}\psi_{1i}+o_P\left(\frac{1}{\sqrt{I_1}}\right)+o_P\left(\frac{1}{\sqrt{I_2}}\right)\right]\\
&=\frac{1}{\sqrt{n}}\sum_{i=1}^n\psi_{1i}+o_P(1)\convL \mathcal{N}(0,V_1),
\end{align*}
where the convergence in distribution follows from the CLT, using the fact that $E\left[\psi_{1i}^2\right]<\infty$ under Assumptions \ref{hyp:parallel_trends}, \ref{hyp:regularity}, \ref{hyp:support_condition0} and \ref{hyp:noquasi-stayers}.


\subsubsection{Efficiency of $\psi_1$}

To show that the influence function $\psi_1$ is efficient,
first note that our identification assumptions do not impose any restrictions on the observed data, and our regularity conditions are inequalities that do not restrict the set of parametric submodels to be considered. Thus, the tangent space $\mathcal{T}$ is the set of mean-zero, square-integrable functions, implying that $\psi_1\in\mathcal{T}$. Next, we show that $\delta_1$ is pathwise differentiable. Let us consider a parametric submodel indexed by $\theta$, with the true DGP being at $\theta=\theta_0$ (we omit the indexation by $\theta_0$ hereafter). We denote its score function by $q(Y_1,Y_2, D_1,D_2)$. By Theorem \ref{thm:main0} used with $\alpha(D_1)=0$, for all $\theta$,
\begin{align*}
0=E_\theta\left[\left(\frac{\Delta Y-\mu_\theta(D_1)}{\Delta D}\right)S-\delta_1 S \right]=E_\theta\left[\frac{\Delta Y}{\Delta D}S-\frac{S}{\Delta D}\mu_\theta(D_1)-\delta_1 S \right].
\end{align*}
As a result, by the chain rule,
\begin{align*}
0&=\frac{\partial}{\partial \theta}E_\theta\left[\frac{S}{\Delta D}\Delta Y-\frac{S}{\Delta D}\mu_\theta(D_1)-\delta_1(\theta) S \right]\big\vert_{\theta_0}\\
&=-E\left[\frac{S}{\Delta D}\cdot\frac{\partial}{\partial\theta}\mu_\theta(D_1)\big\vert_{\theta_0}\right]-E[S]\frac{\partial}{\partial\theta}\delta_1(\theta)\big\vert_{\theta_0} \\
& +E\left[\left(\frac{S}{\Delta D}\Delta Y-\frac{S}{\Delta D}\mu(D_1)-\delta_1 S\right)q(Y_1,Y_2, D_1,D_2) \right].
\end{align*}
Then, solving for the derivative of $\delta_1$, we obtain
\begin{align*}
E[S]\frac{\partial}{\partial\theta}\delta_1(\theta)\big\vert_{\theta_0}&=E\left[\left(\frac{S}{\Delta D}(\Delta Y-\mu(D_1))-\delta_1 S\right)q(Y_1,Y_2, D_1,D_2)\right] \\
& -E\left[\frac{S}{\Delta D}\cdot\frac{\partial}{\partial\theta}\mu(D_1)\big\vert_{\theta_0}\right]
\end{align*}
and
\[E\left[\frac{S}{\Delta D}\cdot\frac{\partial}{\partial\theta}\mu(D_1)\right]=E\left[E\left[\left.\frac{S}{\Delta D}\right\vert D_1\right]\frac{\partial}{\partial\theta}\mu(D_1)\right].\]
Next, note that for any function $a(D_1)$, $E_\theta\left[a(D_1)(1-S)(\Delta Y-\mu_\theta(D_1))\right]=0$. By differentiating with respect to $\theta$ on both sides, we get
\begin{align*}
0&=\frac{\partial}{\partial \theta} E_\theta\left[a(D_1)(1-S)(\Delta Y-\mu_\theta(D_1))\right]\big\vert_{\theta_0}\\
&=-E\left[a(D_1)(1-S)\cdot\frac{\partial}{\partial\theta}\mu_\theta(D_1)\big\vert_{\theta_0}\right]+E\left[a(D_1)(1-S)(\Delta Y-\mu(D_1))q(Y_1,Y_2, D_1,D_2)\right],
\end{align*}

from which it follows that
\begin{align*}
E\left[a(D_1)(1-S)(\Delta Y-\mu(D_1))q(Y_1,Y_2, D_1,D_2)\right]&=E\left[a(D_1)(1-S)\frac{\partial}{\partial \theta}\mu(D_1)\big\vert_{\theta_0}\right]\\
&=E\left[a(D_1)p_0(D_1)\frac{\partial}{\partial \theta}\mu(D_1)\big\vert_{\theta_0}\right].
\end{align*}
Thus, setting $a(D_1)=E[S/\Delta D|D_1]/p_0(D_1)$, we get
\begin{align*}
E\left[E\left[\left.\frac{S}{\Delta D}\right\vert D_1\right]\frac{1-S}{p_0(D_1)}(\Delta Y-\mu(D_1))q(Y_1,Y_2, D_1,D_2)\right]&=E\left[E\left[\left.\frac{S}{\Delta D}\right\vert D_1\right]\frac{\partial}{\partial \theta}\mu(D_1)\big\vert_{\theta_0}\right].
\end{align*}
Plugging back into the derivative of the parameter finally yields
\begin{align*}
\frac{\partial}{\partial\theta}\delta_1(\theta)\big\vert_{\theta_0}&=E\left[\frac{1}{E[S]}\left(\left(\frac{S}{\Delta D}-E\left[\left.\frac{S}{\Delta D}\right\vert D_1\right]\frac{1-S}{p_0(D_1)}\right)(\Delta Y-\mu(D_1))-\delta_1 S\right)\right.\\
&\times q(Y_1,Y_2, D_1,D_2)\bigg]\\
&=E\left[\psi_1q(Y_1,Y_2,D_1,D_2) \right].
\end{align*}
Hence, $\delta_1$ is pathwise differentiable, and the previous display also implies \cite[see, e.g.][p.363]{vanderVaart2000} that $\psi_1$ is the efficient influence function.

\subsubsection{Consistency of $\hat V_1$} 
\label{ssub:consistency_of_hat_v_1}

By previous results and the continuous mapping theorem, $\hat\delta_{1,\mathsf{DR}}^2/\bar{S}\convP \delta_1^2/E[S]$
and thus
$$\hat{V}_1=\frac{1}{E[S]^2}\left\{\frac{1}{n}\sum_i \hat\Gamma_{1i}^2-\delta_1^2E[S]\right\}+o_P(1).$$
Let us consider the term $\sum_i\hat\Gamma_{1i}^2/n$. Note that:
\begin{align*}
\frac{1}{n}\sum_i\hat\Gamma_{1i}^2&=\frac{1}{n}\sum_i\Gamma_{1i}^2+\frac{1}{n}\sum_i(\hat\Gamma_{1i}^2-\Gamma_{1i}^2)\\
&=\frac{1}{n}\sum_i\Gamma_{1i}^2+\frac{1}{n}\sum_i(\hat\Gamma_{1i}-\Gamma_{1i})^2+\frac{2}{n}\sum_i\Gamma_{1i}(\hat\Gamma_{1i}-\Gamma_{1i}).
\end{align*}
By Cauchy-Schwarz,
\begin{align*}
\abs{\frac{1}{n}\sum_i\Gamma_{1i}(\hat\Gamma_{1i}-\Gamma_{1i})}& \le \left(\frac{1}{n}\sum_i \Gamma_{1i}^2\right)^{1/2}\left(\frac{1}{n}\sum_i(\hat\Gamma_{1i}-\Gamma_{1i})^2\right)^{1/2} \\
& = \left(\frac{1}{n}\sum_i(\hat\Gamma_{1i}-\Gamma_{1i})^2\right)^{1/2} \times O_P(1).
\end{align*}
Thus, it suffices to show that $\frac{1}{n}\sum_i(\hat\Gamma_{1i}-\Gamma_{1i})^2=o_P(1)$. Some algebra shows
{\small \begin{align*}
\hat\Gamma_{1i}-\Gamma_{1i}&=-\frac{S_i}{\Delta D_i}(\hat\mu^{(3-k(i))}(D_{i,1})-\mu(D_{i,1}))-((1-S_i)\Delta Y_i-\mu(D_{i,1}))\left(\frac{\hat g^{(3-k(i))}(D_{i,1})}{\hat p^{(3-k(i))}(D_{i,1})}-\frac{g(D_{i,1})}{p(D_{i,1})}\right)\\
&+\left(\frac{\hat g^{(3-k(i))}(D_{i,1})}{\hat p^{(3-k(i))}(D_{i,1})}-\frac{g(D_{i,1})}{p(D_{i,1})}\right)(\hat\mu^{(3-k(i))}(D_{i,1})-\mu(D_{i,1}))+\frac{g(D_{i,1})}{p(D_{i,1})}(\hat\mu^{(3-k(i))}(D_{i,1})-\mu(D_{i,1})).
\end{align*}}
Using $(\sum_{m=1}^Ma_m)^2\le M \sum_m a_m^2$, this implies
\begin{align*}
\frac{1}{n}\sum_i(\hat\Gamma_{1i}-\Gamma_{1i})^2&\le\frac{4}{n}\sum_i\left(\frac{S_i}{\Delta D_i}\right)^2(\hat\mu^{(k(i))}(D_{i,1})-\mu(D_{i,1}))^2\\
&+\frac{4}{n}\sum_i((1-S_i)\Delta Y_i-\mu(D_{i,1}))^2\left(\frac{\hat g^{(k(i))}(D_{i,1})}{\hat p^{(k(i))}(D_{i,1})}-\frac{g(D_{i,1})}{p(D_{i,1})}\right)^2\\
&+\frac{4}{n}\sum_i\left(\frac{g(D_{i,1})}{p(D_{i,1})}\right)^2(\hat\mu^{(k(i))}(D_{i,1})-\mu(D_{i,1}))^2\\
&+\frac{4}{n}\sum_i\left(\frac{\hat g^{(k(i))}(D_{i,1})}{\hat p^{(k(i))}(D_{i,1})}-\frac{g(D_{i,1})}{p(D_{i,1})}\right)^2(\hat\mu^{(k(i))}(D_{i,1})-\mu(D_{i,1}))^2\\
&=4(T_1+T_2+T_3+T_4).
\end{align*}
Now,
\begin{align*}
T_1&=\frac{n_S^{(1)}}{n}\cdot\frac{1}{n_S^{(1)}}\sum_{i\in\mathcal{I}_1}\left(\frac{S_i}{\Delta D_i}\right)^2(\hat\mu^{(2)}(D_{i,1})-\mu(D_{i,1}))^2\\
&+\frac{n_S^{(2)}}{n}\cdot\frac{1}{n_S^{(2)}}\sum_{i\in\mathcal{I}_1}\left(\frac{S_i}{\Delta D_i}\right)^2(\hat\mu^{(1)}(D_{i,1})-\mu(D_{i,1}))^2\\
&=\frac{n_S^{(1)}}{n}T_1^{(1)}+\frac{n_S^{(2)}}{n}T_1^{(2)}
\end{align*}
and for $k=1,2$,
\begin{align*}
E[T_1^{(k)}|\mathcal{I}_{3-k},\mathbf{D}_1^{(k)}]&=\frac{1}{n_S^{(k)}}\sum_{i\in\mathcal{I}_k}E\left[\left.\left(\frac{S_i}{\Delta D_i}\right)^2\right\vert D_{i,1}\right](\hat\mu^{(3-k)}(D_{i,1})-\mu(D_{i,1}))^2\\
&\le C\cdot \frac{1}{n_S^{(k)}}\sum_{i\in\mathcal{I}_1}(\hat\mu^{(3-k)}(D_{i,1})-\mu(D_{i,1}))^2\convP 0
\end{align*}
by Markov's inequality given bounded moments and consistency of $\hat\mu^{(3-k)}$. Thus $T_1=o_P(1)$. By the same logic, let
\[T_2^{(k)}=\frac{1}{n_S^{(k)}}\sum_{i\in\mathcal{I}_k}((1-S_i)\Delta Y_i-\mu(D_{i,1}))^2\left(\frac{\hat g^{(3-k)}(D_{i,1})}{\hat p^{(3-k)}(D_{i,1})}-\frac{g(D_{i,1})}{p(D_{i,1})}\right)^2\]
so we have that
\begin{align*}
E[T_2^{(k)}|\mathcal{I}_{3-k},\mathbf{D}_1^{(k)}]&=\frac{1}{n_S^{(k)}}\sum_{i\in\mathcal{I}_k}E[((1-S_i)\Delta Y_i-\mu(D_{i,1}))^2|D_{i,1}]\left(\frac{\hat g^{(3-k)}(D_{i,1})}{\hat p^{(3-k)}(D_{i,1})}-\frac{g(D_{i,1})}{p(D_{i,1})}\right)^2\\
&\le C\cdot \frac{1}{n_S^{(k)}}\sum_{i\in\mathcal{I}_k}\left(\frac{\hat g^{(3-k)}(D_{i,1})}{\hat p^{(3-k)}(D_{i,1})}-\frac{g(D_{i,1})}{p(D_{i,1})}\right)^2\convP 0
\end{align*}
as established in the first part above. Thus, $T_2=o_P(1)$. Similarly, for the next term, let
\[T_3^{(k)}=\frac{1}{n_S^{(k)}}\sum_{i\in\mathcal{I}_k}\left(\frac{g(D_{i,1})}{p(D_{i,1})}\right)^2(\hat\mu^{(3-k)}(D_{i,1})-\mu(D_{i,1}))^2\]
and
\begin{align*}
E[T_3^{(k)}|\mathcal{I}_{3-k},\mathbf{D}_1^{(k)}]&=\frac{1}{n_S^{(k)}}\sum_{i\in\mathcal{I}_k}E\left[\left.\left(\frac{g(D_{i,1})}{p(D_{i,1})}\right)^2\right\vert D_{i,1}\right](\hat\mu^{(3-k)}(D_{i,1})-\mu(D_{i,1}))^2\\
&\le C\cdot \frac{1}{n_S^{(k)}}\sum_{i\in\mathcal{I}_k}(\hat\mu^{(3-k)}(D_{i,1})-\mu(D_{i,1}))^2\convP 0
\end{align*}
so $T_3=o_P(1)$. Finally, for the fourth term, let
\[T_4^{(k)}=\frac{1}{n_S^{(k)}}\sum_{i\in\mathcal{I}_k}\left(\frac{\hat g^{(3-k)}(D_{i,1})}{\hat p^{(3-k)}(D_{i,1})}-\frac{g(D_{i,1})}{p(D_{i,1})}\right)^2(\hat\mu^{(3-k)}(D_{i,1})-\mu(D_{i,1}))^2.\]
Now, using $(x-y)^2\le 2(x^2+y^2)$ and boundedness of $g(.)/p(.)$, which holds by Assumptions \ref{hyp:support_condition0} and \ref{hyp:noquasi-stayers}, we have
$$\left(\frac{\hat g^{(3-k)}(D_{i,1})}{\hat p^{(3-k)}(D_{i,1})}-\frac{g(D_{i,1})}{p(D_{i,1})}\right)^2 \le 2\frac{\hat g^{(3-k)}(D_{i,1}){}^2}{\inf_{d\in\mathcal{D}_1} \hat p^{(3-k)}(D_{i,1}){}^2} + C$$
Hence, since $\max_{d\in\mathcal{D}_1} |\hat g^{(3-k)}(D_{i,1})|=O_P(1)$, $\|\hat p - p_0\|_{k,\infty} =o_P(1)$ and $\inf_{d\in\mathcal{D}_1} p_0(d)>0$, we obtain
$$\max_{i\in\mathcal{I}_k}\left(\frac{\hat g^{(3-k)}(D_{i,1})}{\hat p^{(3-k)}(D_{i,1})}-\frac{g(D_{i,1})}{p(D_{i,1})}\right)^2 =O_P(1).$$
As a result,
\[T_4^{(k)}= O_P(1)\times \frac{1}{n_S^{(k)}}\sum_{i\in\mathcal{I}_k}(\hat\mu^{(3-k)}(D_{i,1})-\mu(D_{i,1}))^2 \convP 0.\]
Under these conditions, $\hat{V}_1\convP (E[\Gamma_{1i}^2]/E[S]-\delta_1^2)/E[S]=E[\psi_1^2]$, as required.


\subsection{Theorem \ref{thm:main}}

For any function $d\mapsto\mu(d)$ and for any function  $d\mapsto\gamma(d)$, let
$$\psi_2(\mu,\gamma):=\frac{\left(S_+-S_--(1-S)\gamma(D_1)\right)(\Delta Y-\mu(D_1))}{E[\abs{\Delta D}]}-\delta_2.$$
To prove the result, we have to show that $E[\psi_2(\mu_0,\gamma)]=E[\psi_2(\mu,\gamma_0)]=0$.

\medskip
We have that:
\begin{align*}
E[\psi_2(\mu_0,\gamma)]
&=\frac{E\left[\left(S_+-S_-\right)(\Delta Y-\mu_0(D_1))\right]}{E[\abs{\Delta D}]}-\delta_2-\frac{E\left[(1-S)\gamma(D_1)(\Delta Y-\mu_0(D_1))\right]}{E[\abs{\Delta D}]}\\
&=\frac{E\left[\left(S_+-S_-\right)(\Delta Y-\mu_0(D_1))\right]}{E[\abs{\Delta D}]}-\delta_2-\frac{E\left[(1-S)\gamma(D_1)(E(\Delta Y|S,D_1)-\mu_0(D_1))\right]}{E[\abs{\Delta D}]}\\
&=0.
\end{align*}
The last equality follows from the two following facts. First, $\delta_2=E\big[\left(S_+-S_-\right)(\Delta Y- $ $\mu_0(D_1))\big]/E[\abs{\Delta D}]$, by the definitions of $\delta_2$ and $\mu_0(d)$, Lemma \ref{lem:building_block}, and the law of iterated expectations. Second, $(1-S)\left(E(\Delta Y|S,D_1)-\mu_0(D_1)\right)=0$ by the definition of $\mu_0(d)$.

\medskip
Then,
\begin{align*}
E[\psi_2(\mu,\gamma_0)]&=\frac{E\left[\left(S_+-S_-\right)(\Delta Y-\mu(D_1))\right]}{E[\abs{\Delta D}]}-\delta_2-\frac{E\left[(1-S)\gamma_0(D_1)(\Delta Y-\mu(D_1))\right]}{E[\abs{\Delta D}]}\\
&=\frac{E\left[\left(S_+-S_-\right)(\Delta Y-\mu_0(D_1))\right]}{E[\abs{\Delta D}]}-\delta_2+\frac{E\left[\left(S_+-S_-\right)(\mu_0(D_1)-\mu(D_1))\right]}{E[\abs{\Delta D}]}\\
&-\frac{E\left[(1-S)\gamma_0(D_1)(\Delta Y-\mu(D_1))\right]}{E[\abs{\Delta D}]}\\
&=\frac{E\left[\left(E(S_+|D_1=d)-E(S_-|D_1=d)\right)(\mu_0(D_1)-\mu(D_1))\right]}{E[\abs{\Delta D}]}\\
&-\frac{E\left[\frac{E(S_+|D_1=d)-E(S_-|D_1=d)}{E(1-S|D_1=d)}(1-S)(\Delta Y-\mu(D_1))\right]}{E[\abs{\Delta D}]}\\
&=\frac{E\left[\left(E(S_+|D_1=d)-E(S_-|D_1=d)\right)(\mu_0(D_1)-\mu(D_1))\right]}{E[\abs{\Delta D}]}\\
&-\frac{E\left[\left(E(S_+|D_1=d)-E(S_-|D_1=d)\right)(\mu_0(D_1)-\mu(D_1))\right]}{E[\abs{\Delta D}]}\\
&=0.
\end{align*}
The second quality follows after adding and subtracting $\mu_0(D_1)$ in the first expectation. The third equality follows from the fact that $\delta_2=E\left[\left(S_+-S_-\right)(\Delta Y-\mu_0(D_1))\right]$ $/E[\abs{\Delta D}]$, from the law of iterated expectations, and from the definition of $\gamma_0(d)$. The fourth equality follows from the law of iterated expectations $_\Box$

\subsection{Proposition \ref{prop:variance_comparison}}

\subsubsection{Part 1} 
\label{ssub:part_1}

If $D_2\ge D_1$ and $\Delta D \indep D_1$,
\begin{align*}
\psi_1&=\frac{1}{E(S)}\left\{\left(\frac{S}{\Delta D}-E\left(\frac{S}{\Delta D}\right)\frac{(1-S)}{E[1-S]}\right)[\Delta Y-E(\Delta Y | D_1, S=0)]-\delta_1S\right\},\\
\psi_{2}&=\frac{1}{E(\Delta D)}\left\{\left(S-E(S)\frac{(1-S)}{1-E(S)}\right)\times (\Delta Y -E(\Delta Y | D_{1}, S=0))-\delta_{2}\Delta D\right\}.
\end{align*}
Since $Y_2(D_2)-Y_2(D_1)= \Delta D\delta$, we have $\delta_{1}=\delta_{2}=\delta$, and $\Delta Y=\Delta Y(D_1)+\Delta D \delta$, so after some algebra (noting in particular that $(1-S)\Delta D=0$) the previous display simplifies to
\begin{align*}
\psi_1&=\frac{1}{\Delta D}\left(\frac{S}{E(S)}-\frac{(1-S)}{E[1-S]}\frac{\Delta D}{E(S)}E\left(\frac{S}{\Delta D}\right)\right)\times (\Delta Y(D_1) -E(\Delta Y(D_1) | D_{1}, S=0)).\\
\psi_{2}&=\frac{1}{E(\Delta D)}\left(S-(1-S)\frac{E(S)}{1-E(S)}\right)\times (\Delta Y(D_1) -E(\Delta Y(D_1) | D_{1}, S=0)).
\end{align*}
Then, using Assumption \ref{hyp:parallel_trends}, $E(\psi_1|D_1,D_2)=E(\psi_2|D_1,D_2)=0.$ Moreover, using the law of total variance, the fact that $V(\Delta Y(D_1)|D_1,D_2)=\sigma^2$, and some algebra,
\begin{align*}
V_1=&E(V(\psi_1|D_1,D_2))=\sigma^2E\left(\left[\frac{\frac{S}{\Delta D}-\frac{1-S}{1-E(S)}E\left(\frac{S}{\Delta D}\right)}{E(S)}\right]^2\right)\\
=&\sigma^2\left[\frac{E(1/(\Delta D)^2|S=1)}{P(S=1)}+\frac{\left(E(1/\Delta D|S=1)\right)^2}{P(S=0)}\right],
\end{align*}
and
\begin{align*}
V_2=&E(V(\psi_2|D_1,D_2))=\sigma^2E\left(\left[\frac{S-(1-S)\frac{E(S)}{1-E(S)}}{E(\Delta D)}\right]^2\right)\\
=&\sigma^2\frac{1}{\left(E(\Delta D|S=1)\right)^2}\left[\frac{1}{P(S=1)}+\frac{1}{P(S=0)}\right].
\end{align*}
The inequality follows from the convexity of $x\mapsto x^2$, the convexity of $x\mapsto 1/x$ on $\mathbb{R}^+\setminus\{0\}$ and $\Delta D|S=1\in \mathbb{R}^+\setminus\{0\}$, Jensen's inequality, and $x\mapsto x^2$ increasing on $\mathbb{R}^+$, which together imply that
\begin{align*}
&E(1/(\Delta D)^2|S=1)\ge \left(E(1/\Delta D|S=1)\right)^2\ge \frac{1}{\left(E(\Delta D|S=1)\right)^2}.
\end{align*}
Finally, Jensen's inequality is strict for strictly convex functions, unless the random variable is actually constant. The last claim of the proposition follows.


\subsubsection{Part 2} 
\label{ssub:part_2}

Let $\eta:=\Delta Y(D_1) - \mu_0(D_1)$ and $\xi_k:=\xi-\delta_k$. Remark that
\begin{align*}
	\psi_1E[S] & = \left(\frac{S}{\Delta D} - \frac{g_0(D_1)}{p_0(D_1)}(1-S) \right)(\eta + \Delta D\xi_1),\\
	\psi_2E[S] & = \frac{1}{E[|\Delta D|\, |S=1]}\left(S_+-S_- - \frac{h_0(D_1)}{p_0(D_1)}(1-S)\right)(\eta+\Delta D \xi_2),
\end{align*}
Thus, using $S(1-S)=0$, $(S_+-S_-)^2=S$ and $(a+b)^2\le 2(a^2+b^2)$, we obtain
\begin{align*}
	\psi_1^2 E[S]^2& = \eta^2\left\{\frac{1}{\Delta D^2}S + \left(\frac{g_0(D_1)}{p_0(D_1)}\right)^2(1-S) \right\} + S\xi_1^2 + 2S\eta \xi_1/\Delta D,\\
	\psi_2^2 E[S]^2 & \le \frac{2\eta^2}{E[|\Delta D|\, |S=1]^2}\left\{S + \left(\frac{h_0(D_1)}{p_0(D_1)}\right)^2(1-S)\right\} +  \frac{2|\Delta D|^2 \xi_2^2}{E[|\Delta D|\, |S=1]^2},
\end{align*}
Then, as we have assumed that $cov(\eta,\xi_1|S=1,D_1,D_2)=0$,
\begin{align*}
E[S\eta \xi_1/\Delta D]=&E[S]E(\eta \xi_1/\Delta D|S=1)\\
=& E[E(\eta| S=1,D_1,D_2) E( \xi_1| S=1,D_1,D_2)  /\Delta D|S=1]\\
=&0,
\end{align*}
where the last equality follows from the fact that by Assumption \ref{hyp:parallel_trends}, $$E(\eta| S=1,D_1,D_2)=E[\Delta Y(D_1)|S=1,D_1,D_2]-E[ \Delta Y(D_1)|D_1]=0.$$ Therefore,
\begin{align}
V_1E[S]^2 & \ge \underline{\sigma}^2 E[S]E[1/\Delta D^2|S=1], \label{eq:prop_ineg1}\\
V_2E[S]^2 & \le \frac{2\overline{\sigma}^2}{E[|\Delta D|\,|S=1]^2}\left(E[S] + E[h_0(D_1)^2/p_0(D_1)] + E[|\Delta D|^2 \xi_2^2]\right).  \label{eq:prop_ineg2}
\end{align}
Now,  remark that
\begin{align}
E[1/\Delta D^2|S=1]=& \int_{\kappa}^\infty \frac{f_{|\Delta D||S=1}(x)}{x^2}dx\ge \int_{\kappa}^{\kappa+c} \frac{f_{|\Delta D||S=1}(x)}{x^2}dx\notag \\
\ge & \underline{f} \int_{\kappa}^{\kappa+c} \frac{1}{x^2}dx=\frac{\underline{f}c}{\kappa(\kappa+c)}. \label{eq:prop_ineg3}
\end{align}
With the same reasoning,
\begin{equation}
E[|\Delta D||S=1]^2\ge \left(\frac{\underline{f}}{2}c(2\kappa+c)\right)^2 \ge \frac{\underline{f}^2 c^4}{4}.
	 \label{eq:prop_ineg4}
\end{equation}
Also,
\begin{align*}
h_0(D_1)^2 = & E[S_+-S_-|D_1]^2 = E[S(S_+-S_-)|D_1]^2 = E[S]^2 E[S_+-S_-|D_1,S=1]^2  \\
\le & E[S],	
\end{align*}
where the last inequality follows since $|S_+-S_-|\le 1$. Since $p_0(D_1)\ge \zeta$, we obtain
\begin{equation}
E[h_0(D_1)^2/p_0(D_1)]\le E[S]/\zeta.	
	 \label{eq:prop_ineg5}
\end{equation}
Finally, as $E[|\Delta D| \xi_2|S=1]=0$ by definition of $\delta_2$,
\begin{equation}
E[|\Delta D|^2 \xi_2^2]=E[S|\Delta D|^2 \xi_2^2] = E[S] V[|\Delta D|(\xi-\delta_2)|S=1]\le E[S]M.
	 \label{eq:prop_ineg6}
\end{equation}
The result follows by combining \eqref{eq:prop_ineg1}-\eqref{eq:prop_ineg6} $_\Box$



\storecounter{equation}{nbeq}
\storecounter{lem}{nblem}
\storecounter{hyp}{nbhyp}
\storecounter{thm}{nbthm}

\newpage
\begin{center}
\huge{Web Appendix of ``Difference-in-Differences for Continuous Treatments and Instruments with Stayers''
}	
\end{center}
 
\begin{abstract}
	In this web appendix, we first 
	discuss in details two extensions sketched in the main text, to estimators with more than two time periods and estimators with covariates. Then, we present some additional results on the  paper's empirical application. Finally, the last two sections give the proofs of Theorems \ref{thm:WASasnormality_nonmonotonictreatment} and \ref{thm:main_IV} in the main paper and the proofs of the results given in this web appendix.
\end{abstract}

\newpage
\tableofcontents
\newpage

\section{Estimators with more than two time periods}\label{appendixsec:Tgeq3}

\subsection{Aggregate effects} 
\label{sub:aggregate_effects}

In this section, we assume the representative unit is observed at $T>2$ time periods. Let $(D_1,...,D_T)$ denote the unit's treatments and $\mathcal{D}_t=\Supp(D_t)$ for all $t\in \{1,...,T\}$. 
For any $t\in \{1,...,T\}$, and for any $d\in \mathcal{D}_t$ let $Y_t(d)$ denote the unit's potential outcome at period $t$ with treatment $d$. Finally, let $Y_t$ denote their observed outcome at $t$. For any $t\in \{2,...,T\}$, let $S_t=1\{D_t\ne D_{t-1}\}$ be an indicator equal to 1 if the unit's treatment switches from period $t-1$ to $t$.
Let also $S_{+,t}=1\{D_t> D_{t-1}\}$ and $S_{-,t}=1\{D_t< D_{t-1}\}$.
We assume that the assumptions made in the paper, rather than just holding for $t=1$ and $t=2$, actually hold for all pairs of consecutive time periods $(t-1,t)$. For instance, we replace Assumption \ref{hyp:parallel_trends} by:

\begin{hyp}
	(Parallel trends) For all $t\geq 2$, for all $d\in \mathcal{D}_{t-1}$,  $E(\Delta Y_t(d)|D_{t-1}=d,D_t)=E(\Delta Y_t(d)|D_{t-1}=d)$.
\label{hyp:parallel_trends_Tgeq3}
\end{hyp}

Assumption \ref{hyp:parallel_trends_Tgeq3} requires that $E(\Delta Y_t(d)|D_{t-1}=d,D_t=d')$ be constant across $d'$: groups of units with the same period-$t-1$ treatment but different period-$t$ treatments all have the same expected outcome evolution in the counterfactual where their period-$t-1$ treatment would not have changed. Importantly, note that Assumption \ref{hyp:parallel_trends_Tgeq3} only requires that some groups be on parallel trends over consecutive time periods, not over the entire duration of the panel. Specifically, because Assumption \ref{hyp:parallel_trends_Tgeq3} is conditional on $D_{t-1}$, it cannot be ``chained'' across pairs of time periods: for instance, under Assumption \ref{hyp:parallel_trends_Tgeq3}, two units $i$ and $i'$ such that $D_{i,1}=2,D_{i,2}=D_{i,3}=3$ and $D_{i',1}=D_{i',2}=D_{i',3}=2$ experience parallel trends from period one to two but not from period two to three, because they have the same treatment at period one but not at period two. To preserve space, we do not restate our other assumptions with more than two periods.

\medskip
Let
\begin{align*}
\delta_{1,t}& =E\left(\frac{Y_t(D_t) - Y_t(D_{t-1})}{D_t-D_{t-1}} \middle| S_t=1\right),\\
\delta_{2,t}&= \frac{E\left(\sgn(D_t-D_{t-1})(Y_t(D_t)-Y_t(D_{t-1}))\right)}{E(|D_t-D_{t-1}|)}.
\end{align*}
denote the AS and the WAS at period $t$. Then, let 
\begin{align*}
\delta^{T\geq 3}_{1}& =\sum_{t=2}^T \frac{P(S_t=1)}{\sum_{k=2}^T P(S_k=1)}\delta_{1,t},\\
\delta^{T\geq 3}_{2}&= \sum_{t=2}^T \frac{E(|\Delta D_t|)}{\sum_{k=2}^T E(|\Delta D_k|)}\delta_{2,t}
\end{align*}
be generalizations of the AS and WAS effects to applications with more than two periods. In line with the spirit of the two effects, we propose different weights to aggregate the AS and WAS across time periods. For the AS, the weights are just proportional to the proportion of switchers between $t-1$ and $t$. For the WAS, the weights are proportional to the average absolute value of the treatment switch from $t-1$ to $t$.

\medskip
Define the functions
\begin{align*}
&\mu_{0,t}(d):=E[\Delta Y_t|S_t=0,D_{t-1}=d],\\
&\alpha_{0,t}(d):=E\left(\left.\frac{S_t}{\Delta D_t}\right\vert D_{t-1}=d\right)\frac{1}{E(1-S_t|D_{t-1}=d)},\\
&\gamma_{0,t}(d):=\frac{E(S_{+,t}|D_{t-1}=d)-E(S_{-,t}|D_{t-1}=d)}{E(1-S_t|D_{t-1}=d)}.
\end{align*}
\begin{thm}
\begin{enumerate}
\item If Assumption \ref{hyp:parallel_trends_Tgeq3} and generalizations of Assumptions \ref{hyp:noanticipation}, \ref{hyp:nodynamic}, \ref{hyp:regularity}, \ref{hyp:support_condition0}, and \ref{hyp:noquasi-stayers} to more than two periods hold,
\begin{align*}
\delta^{T\geq 3}_{1}=&\sum_{t=2}^T \frac{P(S_t=1)}{\sum_{k=2}^T P(S_k=1)}E\left(\frac{1}{E[S_t]}\left(\frac{S_t}{\Delta D_t}-(1-S_t)\alpha_{0,t}(D_{t-1})\right)\left[\Delta Y_t-\mu_{0,t}(D_{t-1})\right]\right).
\end{align*}
\item If Assumption \ref{hyp:parallel_trends_Tgeq3} and generalizations of Assumptions \ref{hyp:noanticipation}, \ref{hyp:nodynamic}, \ref{hyp:regularity} and \ref{hyp:support_condition0} to more than two periods hold,
\begin{align*}
\delta^{T\geq 3}_{2}=&\sum_{t=2}^T \frac{E(|\Delta D_t|)}{\sum_{k=2}^T E(|\Delta D_k|)}\frac{E\left[\left(S_{+,t}-S_{-,t}-(1-S_t)\gamma_{0,t}(D_{t-1})\right)(\Delta Y_t-\mu_{0,t}(D_{t-1}))\right]}{E[\abs{\Delta D_t}]}.
\end{align*}
\end{enumerate}
\label{thm:main0_Tgeq3}
\end{thm}
Theorem \ref{thm:main0_Tgeq3} is a straightforward generalization of Theorems \ref{thm:main0} and \ref{thm:main} to settings with more than two time periods; accordingly, its proof is omitted.

\medskip
To estimate $\delta^{T\geq 3}_{1}$, we use
$$\widehat{\delta}^{T\geq 3}_{1}:=\sum_{t=2}^T \frac{\widehat{P}(S_t=1)}{\sum_{k=2}^T \widehat{P}(S_k=1)}\widehat{\delta}_{1,t,\mathsf{DR}},$$
where $\widehat{\delta}_{1,t,\mathsf{DR}}$ is defined exactly as $\hat{\delta}_{1,\mathsf{DR}}$ in Section \ref{subsec:ASestimation}, restricting the sample to periods $t-1$ and $t$. Similarly, to estimate $\delta^{T\geq 3}_{2}$, we use
$$\widehat{\delta}^{T\geq 3}_{2}:=\sum_{t=2}^T \frac{\widehat{E}(|\Delta D_t|)}{\sum_{k=2}^T \widehat{E}(|\Delta D_k|)}\widehat{\delta}_{2,t,\mathsf{DR}},$$
where $\widehat{\delta}_{2,t,\mathsf{DR}}$ is defined exactly as $\hat{\delta}_{2,\mathsf{DR}}$ in Section \ref{subsec:WASestimation}, restricting the sample to periods $t-1$ and $t$.
Let
\small
\begin{align*}
\psi_{1,t}=&\frac{1}{E(S_t)}\left\{\left(\frac{S_t}{\Delta D_t}-E\left(\frac{S_t}{\Delta D_t}\middle| D_{t-1}\right)\frac{(1-S_t)}{E[1-S_t|D_{t-1}]}\right)[\Delta Y_t-E(\Delta Y_t | D_{t-1}, S_t=0)]-\delta_{1,t}S_t\right\},\\
\psi_{2,t}=&\frac{1}{E(|\Delta D_t|)}\left\{\left(S_{+,t}-S_{-,t}-E(S_{+,t}-S_{-,t}|D_{t-1})\frac{(1-S_t)}{E(1-S_t|D_{t-1})}\right)(\Delta Y_t -E(\Delta Y_t | D_{t-1}, S_t=0))-\delta_{2,t}|\Delta D_t|\right\}.
\end{align*}
\normalsize
After some algebra, one can show that the influence function of $\widehat{\delta}^{T\geq 3}_{1}$ is
\begin{align}\label{eq:ASIFT>=3}
\psi^{T\geq 3}_{1}&:=\frac{\sum_{t=2}^T (P(S_t=1)\psi_{1,t}+(\delta_{1,t}-\delta^{T\geq 3}_{1})(S_t-P(S_t=1)))}{\sum_{t=2}^T P(S_t=1)},
\end{align}
while the influence function of $\widehat{\delta}^{T\geq 3}_{2}$ is
\begin{align}\label{eq:WASIFT>=3}
\psi^{T\geq 3}_{2}&:=\frac{\sum_{t=2}^T E(\abs{\Delta D_t})\psi_{2,t}+(\delta_{2,t}-\delta^{T\geq 3}_{2})(\abs{\Delta D_t}-E(\abs{\Delta D_t}))}{\sum_{t=2}^T E(|\Delta D_t|)}.
\end{align}
Importantly, those influence functions allow units' treatments and outcomes to be arbitrarily serially correlated.



\subsection{``Pre-trends'' or ``placebo'' tests} 
\label{sub:testing_for_pre_trends}

With several time periods, one can test the following condition.
\begin{hyp}\label{hyp:parallel_trends_Tgeq3_testable} (Testable parallel trends)
For all $t\geq 3,t\le T$, for all $d\in \mathcal{D}_{t-1}$,  $E(\Delta Y_{t-1}(d)|D_{t-2}=D_{t-1}=d,D_t)=E(\Delta Y_{t-1}(d)|D_{t-2}=D_{t-1}=d)$.
\end{hyp}
Assumption \ref{hyp:parallel_trends_Tgeq3_testable} is the analogue of Assumption \ref{hyp:parallel_trends_Tgeq3}, replacing $\Delta Y_{t}$ by $\Delta Y_{t-1}$, and restricting the sample to $t-2$ to $t-1$ stayers.

\medskip
To test Assumption \ref{hyp:parallel_trends_Tgeq3_testable}, one can compute a placebo version of the estimators $\widehat{\delta}^{T\geq 3}_{1}$ and $\widehat{\delta}^{T\geq 3}_{2}$, replacing $\Delta Y_t$ by $\Delta Y_{t-1}$, and restricting the sample, for each pair of consecutive time periods $(t-1,t)$, to units whose treatment did not change between $t-2$ and $t-1$. Thus, the placebo estimators compare the average $\Delta Y_{t-1}$ of the $t-1$-to-$t$ switchers and stayers, restricting attention to $t-2$-to-$t-1$ stayers. If one finds that from $t-2$-to-$t-1$, $t-1$-to-$t$ switchers and stayers are on parallel trends, this lends credibility to Assumption \ref{hyp:parallel_trends_Tgeq3}.

\medskip
Assumption \ref{hyp:parallel_trends_Tgeq3} can only be placebo tested among $t-2$-to-$t-1$ stayers: without restricting the placebo sample to $t-2$-to-$t-1$ stayers, one could have a non-zero placebo estimand even if switchers and stayers are on parallel trends, if $t-1$-to-$t$ switchers and stayers have different probabilities of being $t-2$-to-$t-1$ switchers and/or different treatment effects.

\medskip
As Assumption \ref{hyp:parallel_trends_Tgeq3} can only be placebo tested among $t-2$-to-$t-1$ stayers, one may recompute, as a robustness check, the actual estimators in that subsample, and we let $\widehat{\delta}^{T\geq 3}_{1,ps}$ and $\widehat{\delta}^{T\geq 3}_{2,ps}$ denote the corresponding estimators, where ``$ps$'' stands for placebo sample. $\widehat{\delta}^{T\geq 3}_{1,ps}$ and $\widehat{\delta}^{T\geq 3}_{2,ps}$ rely on the following identifying assumption:
\begin{equation}\label{eq:PT_placebosample}
\forall t\geq 3,t\le T,d\in \mathcal{D}_{t-1}:~E(\Delta Y_{t}(d)|D_{t-2}=D_{t-1}=d,D_t)=E(\Delta Y_{t}(d)|D_{t-2}=D_{t-1}=d),
\end{equation}
the exact analogue of Assumption \ref{hyp:parallel_trends_Tgeq3_testable} but one period ahead.

\medskip
Finally, if \eqref{eq:PT_placebosample} fails, placebo estimators may be used to calibrate the bias of $\widehat{\delta}^{T\geq 3}_{1,ps}$ and $\widehat{\delta}^{T\geq 3}_{2,ps}$.
For instance, assume that \eqref{eq:PT_placebosample} and Assumption \ref{hyp:parallel_trends_Tgeq3_testable} both fail, but $\forall t\geq 3,t\le T,d\in \mathcal{D}_{t-1}$,
\begin{align*}
E(\Delta Y_{t}(d)-\Delta Y_{t-1}(d)|D_{t-2}=D_{t-1}=d,D_t)=E(\Delta Y_{t}(d)-\Delta Y_{t-1}(d)|D_{t-2}=D_{t-1}=d),
\end{align*}
as would for instance happen with differential linear trends but deviations from linear trends unrelated to $D_t$. Then, the placebo AS (resp. WAS) estimator is consistent for the bias of $\widehat{\delta}^{T\geq 3}_{1,ps}$ (resp. $\widehat{\delta}^{T\geq 3}_{2,ps}$).

\subsection{Estimators robust to dynamic effects up to a pre-specified treatment lag.} 
\label{sub:dyn}

$\widehat{\delta}^{T\geq 3}_{1}$ and $\widehat{\delta}^{T\geq 3}_{2}$ may not be robust to dynamic treatment effects. They compare the outcome evolutions of $t-1$-to-$t$ switchers and stayers. But there may be, say, $t-1$-to-$t$ stayers whose treatment changed from $t-2$ to $t-1$. If lagged treatments affect units' current outcome, that change could still affect the $t-1$-to-$t$ outcome evolution of those stayers, thus leading them to violate the parallel-trends assumption underlying $\widehat{\delta}^{T\geq 3}_{1}$ and $\widehat{\delta}^{T\geq 3}_{2}$.

\medskip
Then, computing $\widehat{\delta}^{T\geq 3}_{1,ps}$ and $\widehat{\delta}^{T\geq 3}_{2,ps}$ may again be a useful robustness check, because these estimators are robust to dynamic effects up to one treatment lag. Specifically, assume that units' current treatment and their first treatment lag affect their current outcome, and let $Y_t(d_{t-1},d_t)$ denote their period-$t$ potential outcome if $(D_{t-1},D_t)=(d_{t-1},d_t).$ Then, $\widehat{\delta}^{T\geq 3}_{1,ps}$ and $\widehat{\delta}^{T\geq 3}_{2,ps}$ are consistent for average effects of the current treatment on the outcome under the following assumption:
\small
$$\forall t\geq 3,t\le T,d\in \mathcal{D}_{t-1}:~E(Y_{t}(d,d)-Y_{t-1}(d,d)|D_{t-2}=D_{t-1}=d,D_t)=E(Y_{t}(d,d)-Y_{t-1}(d,d)|D_{t-2}=D_{t-1}=d).$$
\normalsize
Similarly, if one wants to allow for effects of the first and second treatment lags on the outcome, one just needs to restrict the estimation sample to $t-3$-to-$t-1$ stayers. However, the more robustness to dynamic effects one would like to have, the smaller the estimation sample becomes.

\medskip
Finally, note that when the treatment is continuously distributed at period one, allowing for dynamic effects up to any lag would open up the so-called initial-conditions problem. As units receive heterogeneous doses at period one, they may have experienced treatment changes before period one. With dynamic effects up to any lag, such changes may still affect their outcome over the entirety of the study period, but they cannot be accounted for because they are not observed. Ruling out dynamic effects, at least up to a pre-specified number of lags, allows us to abstract from this thorny issue. But of course, this could yield misleading results if dynamic effects up to any lag are present.

\section{Identification and estimation with covariates}\label{appendixsec:covariates}


In this section we discuss the extension of our results to allow for covariates. Let $X_t$ denote a $c$-dimensional vector of covariates at period $t$ with support $\mathcal{X}_t$. We start by considering the case with two time periods, $T=2$. The target parameters are the same as in the paper. We consider the following assumptions.

\begin{hyp}\label{hyp:parallel_trendsX} (Parallel trends with covariates)
For all $d_1\in \mathcal{D}_1$,  $E(\Delta Y(d_1)|X_1,D_1=d_1,D_2)=E(\Delta Y(d_1)|X_1,D_1=d_1)$.
\end{hyp}

\begin{hyp}\label{hyp:boundedX} (Bounded treatment and regularity of potential outcomes with covariates)
\begin{enumerate}
\item $\mathcal{D}_1$ and $\mathcal{D}_2$ are bounded subsets of $\mathbb{R}$.
\item $\sup_{(d,x)\in\mathcal{D}_1\times\mathcal{X}_1}E[(\Delta Y(d_1))^2|D_1=d_1,X_1=x]<\infty$
\item $\forall$ $t\in \{1,2\}$ and $\forall$ $(d,d')\in \mathcal{D}_t^2$,
    there is a random variable $\overline{Y}\ge 0$ such that
    $|Y_t(d)-Y_t(d')|\leq \overline{Y}|d-d'|$, with $\sup_{(d_1,d_2,x_1,x_2)\in
    \mathcal{D}_1\times\mathcal{D}_2\times\mathcal{X}_1\times\mathcal{X}_2}E[\overline{Y}^2|D_1=d_1, D_2=d_2X_1=x_1,X_2=x_2]<\infty$.
\end{enumerate}
\end{hyp}

\begin{hyp}\label{hyp:regularity_AS_X} (Regularity conditions for AS with covariates)
\begin{enumerate}
\item Overlap: $P(S=1)>0$ and $P(S=0|D_1,X_1)>\zeta$ for some $\zeta>0$.
\item No quasi-stayers: for any $x$ such that $P(S=1|X_1=x)>0$, $P(\abs{\Delta D}>\kappa|S=1,X_1=x)=1$ for some $\kappa>0$.
\end{enumerate}
\end{hyp}

\begin{hyp}\label{hyp:overlap_WAS_X} (Overlap condition for WAS with covariates)
$P(S=1)>0$ and 
$P(S=0|D_1,X_1)>\zeta$ for some $\zeta>0$.
\end{hyp}

We first introduce an auxiliary result.
\begin{lem}\label{lem:lemmaX1}
If Assumptions \ref{hyp:noanticipation}, \ref{hyp:nodynamic}, and \ref{hyp:parallel_trendsX} hold, then for all $(d_1, d_2) \in \mathcal{D}_1\times\mathcal{D}_2$ such that $d_1\neq d_2$ and $P(S|D_1 = d_1, X_1 =x)<1 $ for any $x\in \mathcal{X}_1$,
\begin{equation*}
E\left(\frac{Y_2(d_2) - Y_2(d_1)}{d_2-d_1} \middle| D_1 = d_1, D_2 = d_2 \right) = E\left(\frac{\Delta Y - E(\Delta Y|D_1=d_1, S=0, X_1)}{d_2-d_1} \middle| D_1 = d_1, D_2 = d_2 \right)
\end{equation*}
\end{lem}

We have the following identification result.

\begin{thm}\label{thm:thmX1}
Suppose Assumptions \ref{hyp:noanticipation}, \ref{hyp:nodynamic}, \ref{hyp:parallel_trendsX}, and \ref{hyp:boundedX} hold. Then, under Assumption \ref{hyp:regularity_AS_X},
\begin{align*}
\delta_1&=E\left(\frac{1}{E(S)}\left(\frac{S}{\Delta D}-(1-S)\alpha_0(X_1,D_1)\right)(\Delta Y-\mu_0(X_1,D_1))\right)
\end{align*}
and under Assumption \ref{hyp:overlap_WAS_X},
\begin{align*}
\delta_2&=E\left(\frac{1}{E(\abs{\Delta D})}\left(S_+-S_--(1-S)\gamma_0(X_1,D_1)\right)(\Delta Y -\mu_0(X_1,D_1))\right)
\end{align*}
where
\begin{align*}
\alpha_0(x,d)&:=E\left(\left.\frac{S}{\Delta D}\right\vert X_1=x,D_1=d\right)\frac{1}{E(1-S|X_1=x,D_1=d)},\\
\gamma_0(x,d)&:=E(S_+-S_-|X_1,D_1)\frac{1}{E(1-S|X_1,D_1)},\\
\mu_0(x,d)&:=E(\Delta Y|X_1=x,D_1=d,S=0).
\end{align*}
\end{thm}

For estimation, we follow the same procedure as for the estimators without covariates. Let
\[g_0(x,d):=E(S/\Delta D|X_1=x,D_1=d),\quad p_0(x,d):=E(1-S|X_1=x,D_1=d).\]
Using the cross-fitting procedure described in the paper, we define
\[\hat\delta_{1,X,\mathsf{DR}}^{(1)}=\frac{1}{n_S^{(1)}}\sum_{i\in \mathcal{I}_1}\left(\frac{S_i}{\Delta D_i}-\frac{\hat g^{(2)}(X_{i,1},D_{i,1})}{\hat p^{(2)}(X_{i,1},D_{i,1})}(1-S_i)\right)\left(\Delta Y_i-\hat\mu^{(2)}(X_{i,1},D_{i,1})\right),\]
estimated using the sample $\mathcal{I}_1$ and where $\hat{g}^{(2)}$, $\hat{p}^{(2)}$ and $\hat\mu^{(2)}$ are estimated in $\mathcal{I}_2$. Similarly, $\hat\delta_{1,X,\mathsf{DR}}^{(2)}$ is estimated in $\mathcal{I}_2$ after estimating the nuisance functions in $\mathcal{I}_1$, and $\hat\delta_{1,X,\mathsf{DR}}$ is computed as a weighted average of these two estimators. Under the additional assumptions that $p_0(x,d)$ is uniformly bounded from below, $E[\Delta Y^2|X_1=x,D_1=d_1,D_2=d_2]<\infty$ and $\mathcal{I}_1\perp (D_{i,2},Y_{i,1},Y_{i,2})_{i=1,\ldots,n}|(X_{i,1},D_{i,1})_{i=1,\ldots,n}$, asymptotic normality of $\hat\delta_{1,X,\mathsf{DR}}$ follows under the consistency and rate conditions in Theorem \ref{thm:ASasnormality} applied to $\hat g(x,d)$, $\hat p(x,d)$ and $\hat \mu(x,d)$ with influence function
\begin{align*}
\psi_{1,X}&:=\frac{1}{E(S)}\left\{\left(\frac{S}{\Delta D}-\frac{g_0(X_1,D_1)}{p_0(X_1,D_1)}(1-S)\right)(\Delta Y-\mu_0(X_1,D_1))-\delta_1S\right\}.
\end{align*}
By the same logic we can obtain asymptotic normality of the WAS estimator with covariates $\hat\delta_{2,X,\mathsf{DR}}$ as in Theorem  \ref{thm:WASasnormality_nonmonotonictreatment} with influence function:
\begin{align*}
\psi_{2,X}&:=\frac{1}{E(\abs{\Delta D})}\left\{\left(S_+-S_--\frac{h_0(X_1,D_1)}{p_0(X_1,D_1)}(1-S)\right)(\Delta Y -\mu_0(X_1,D_{1}))-\delta_{2}\abs{\Delta D}\right\}
\end{align*}
where
\[h_0(x,d):=E(S_+-S_-|X_1=x,D_1=d).\]

With many covariates, classical nonparametric estimators of the nuisance functions, such as series estimators, may no longer satisfy the convergence rate requirements ensuring asymptotic normality of $\hat\delta_{1,X,\mathsf{DR}}$ and $\hat\delta_{2,X,\mathsf{DR}}$. Then, it may be preferable to use machine-learning estimators to estimate those nuisance functions.

\medskip
Finally, we note that extending our estimators with controls to more than three time periods is immediate: one just defines $(t-1,t)$-specific estimators following the formulas above and letting $t-1=1$ and $t=2$. Then, the aggregation of the $(t-1,t)$-specific estimators and their influence functions is exactly the same as in Section \ref{appendixsec:Tgeq3}.

\section{Supplementary material on the empirical application}

\subsection{Comparing switching cells to other cells in the empirical application}

Table \ref{table:balancing} below compares some observable
characteristics of cells in and not in $\mathcal{S}$. Switching cells seem slightly
over-represented in the later years of the panel: $t$ is on average 2.5 years
larger for switching than for non-switching cells, and the difference is significant. On
the other hand, switching cells are not more populated than other cells, and their
gasoline consumption and gasoline price in 1966 are not significantly
different. Thus, there is no strong indication that the
cells in $\mathcal{S}$ are a very selected subgroup.

\begin{table}[H]
   \begin{center}
   \caption{Comparing switching cells 
   	 to other cells}\label{table:balancing}
      \begin{tabular}{lcccc}
      \toprule
      Dependent Variables:                       & $t$             & Adult Population     & log(quantity)$_{1966}$  & log(price)$_{1966}$\\
      \midrule
      Constant                                   & 1,986.7 & 3,691,608.0  & -0.5161       & 3.471\\
                                                 & (0.2739)        & (577,164.0)          & (0.0210)              & (0.0054)\\
      $\mathbf{1}\{(i,t)\in \mathcal{S}\}$     & 2.481   & 39,588.0             & -0.0099               & 0.0014\\
                                                 & (0.7519)        & (320,342.1)          & (0.0096)              & (0.0029)\\
      \midrule
                                    Observations & 2,016           & 2,016                & 2,016                 & 2,016\\
\bottomrule
   \end{tabular}
   \end{center}
\footnotesize{Notes: The table show the results of regressions of some dependent variables on a constant and an indicator for switching cells. The standard errors shown in parentheses are clustered at the state level.}
\end{table}

\subsection{Reduced-form and first-stage effects in the placebo sample}

\begin{table}[H]
\begin{center}
			\begin{threeparttable}
				\caption{Reduced-form and first-stage effects (WAS) in the placebo sample}
				\label{table:RFandFS_placebosample}
				\begin{tabular}{lcc}
					\toprule   %
					{
		\begin{tabular}{lc}
Reduced-form effect of taxes on quantities                 &     -0.0027 (0.0012) \\
First-stage effect of taxes on prices  &      0.0051  (0.0012) \\
Observations        &       1,059 \\
	\end{tabular}
}
\\
					\bottomrule
				\end{tabular}
				\end{threeparttable}
\end{center}
\footnotesize
Notes: The table shows the WAS estimates of the reduced-form and first-stage effects of taxes on quantities and prices, estimated in the placebo subsample, whose taxes do not change from $t-2$ to $t-1$. Standard errors, clustered at the state level, are shown next to the estimates, between parentheses.	
\end{table}

\subsection{Reduced-form and first-stage robust to lagged effects}

\begin{table}[H]
\begin{center}
			\begin{threeparttable}
				\caption{Effects of gasoline tax on quantities and prices, robust to effects up to any observed lag}
				\label{table:RFandFS_unrestricted_dyn}
				\begin{tabular}{lcc}
					\toprule   %
					{
		\begin{tabular}{lc}
Reduced-form effect of taxes on quantities                 &   -0.0021   (0.0130) \\
First-stage effect of taxes on prices                 &      0.0107   (0.0084) \\
		Observations        &       215 \\
	\end{tabular}
}
\\
					\bottomrule
				\end{tabular}
\end{threeparttable}
\end{center}
\footnotesize
Notes: The table shows estimates of the reduced-form (resp. first-stage) effect of current taxes on gasoline quantities (resp. prices) that are robust to unrestricted dynamic effects of taxes, following \cite{de2020difference}, and computed using the \st{did\_multiplegt\_dyn} Stata command. Estimates are computed controlling for a polynomial of order 1 in $(Z_{t-1},D_{t-1})$. Bootstrap standard errors, clustered at the state level, are shown next to the estimates, between parentheses.
\end{table}

\begin{table}[H]
\centering
\caption{Effects of gasoline tax on quantities and prices, robust to effects up to a pre-specified number of lags}
\label{tab:distlaghet_gasoline}
\begin{tabular}{l|c|cc|ccc}
\toprule
Pre-specified number of lags & \multicolumn{1}{c|}{$K=0$} & \multicolumn{2}{c|}{$K=1$} & \multicolumn{3}{c}{$K=2$} \\
Parameter & $\overline{\beta}_0$ & $\overline{\beta}_0$ & $\overline{\beta}_1$ & $\overline{\beta}_0$ & $\overline{\beta}_1$ & $\overline{\beta}_2$ \\
\midrule
\multicolumn{7}{l}{\textit{Panel A: Reduced-form effects of taxes on quantities}} \\
Estimate & -0.0020 & -0.0019 & -0.0012 & -0.0019 & -0.0012 & -0.0015 \\
Std. err. & 0.0006 & 0.0007 & 0.0008 & 0.0007 & 0.0007 & 0.0006 \\
N.obs & 48 & 48 & 48 & 48 & 48 & 48 \\
\midrule
\multicolumn{7}{l}{\textit{Panel B: First-stage effects of taxes on prices}} \\
Estimate & 0.0052 & 0.0051 & 0.0036 & 0.0050 & 0.0035 & -0.0014 \\
Std. err. & 0.0006 & 0.0006 & 0.0007 & 0.0007 & 0.0007 & 0.0005 \\
N.obs & 48 & 48 & 48 & 48 & 48 & 48 \\
\bottomrule
\end{tabular}

\medskip
\begin{minipage}{\linewidth}
\footnotesize
Notes: The table shows estimates of the robust distributed-lag model of \cite{de2025treatmenteffect}, computed using the \st{DistLagHet} R command. $\overline{\beta}_k$ is the average effect of a one-unit increase in the gasoline tax $k$ years earlier on the log of gasoline consumption (Panel A) or on the tax-inclusive gasoline price (Panel B). All models include year fixed effects. ``N.obs'' denotes the number of states on which we average effects. Standard errors obtained by bootstrapping states (500 replications).
\end{minipage}
\end{table}

\section{Additional proofs of the results in the main paper}

\subsection{Theorem \ref{thm:WASasnormality_nonmonotonictreatment}}

We keep the proof of the asymptotic normality short, as it follows exactly the same steps as that of Theorem \ref{thm:ASasnormality}. For the same reason, we omit the proof of the consistency of $\hat V_2$. First, we have that:
\begin{align*}
	\sqrt{I_1}(\hat\delta_{2,\mathsf{DR}}^{(1)}-\delta_{2})=\frac{1}{E(\abs{\Delta D})+o_P(1)}\cdot\frac{1}{\sqrt{I_1}}\sum_{i\in \mathcal{I}_1}V_{i}
\end{align*}
where
\[V_{i}:=\left(S_{i+}-S_{i-}-\frac{\hat h^{(2)}_2(D_{i,1})}{\hat p^{(2)}(D_{i,1})}(1-S_i)\right)(\Delta Y_i-\hat\mu^{(2)}(D_{i,1}))-\delta_{2}\abs{\Delta D_i}.\]
Now,
\begin{align*}
	V_{i}&=\left(S_{i+}-S_{i-}-\frac{h_0(D_{i,1})}{p_0(D_{i,1})}(1-S_i)\right)(\Delta Y_i-\mu_0(D_{i,1}))-\delta_{2}\abs{\Delta D_i}\\
	&+\left(\frac{h_0(D_{i,1})}{p_0(D_{i,1})}-\frac{\hat h^{(2)}(D_{i,1})}{\hat p^{(2)}(D_{i,1})}\right)(1-S_i)(\Delta Y_i-\mu_0(D_{i,1}))\\
	&+\left(\frac{h_0(D_{i,1})}{p_0(D_{i,1})}-\frac{\hat h^{(2)}(D_{i,1})}{\hat p^{(2)}(D_{i,1})}\right)(1-S_i)(\mu_0(D_{i,1})-\hat\mu^{(2)}(D_{i,1}))\\
	&+\left(S_{i+}-S_{i-}-\frac{h_0(D_{i,1})}{p_0(D_{i,1})}(1-S_i)\right)(\mu_0(D_{i,1})-\hat\mu^{(2)}(D_{i,1}))\\
	&=:E(\abs{\Delta D})\psi_{2,i}+T_{1i}+T_{2i}+T_{3i},
\end{align*}
and thus, letting $T_j:=\sum_{i\in \mathcal{I}_1} T_{j,i}/\sqrt{I_1}$,
\[\sqrt{I_1}(\hat\delta_{2,\mathsf{DR}}^{(1)}-\delta_{2})=\frac{1}{E(\abs{\Delta D})+o_P(1)}\left(\frac{1}{\sqrt{I_1}}\sum_{i\in \mathcal{I}_1}E(\abs{\Delta D})\psi_{2,i}+T_1+T_2+T_3\right).\]

Then, by following exactly the same reasoning as in the proof of Theorem \ref{thm:ASasnormality}, simply replacing $S_i/\Delta D_i$ by $S_{i+}-S_{i-}$ and $g_0$ by $h_0$, we obtain
\[\sqrt{n}(\hat\delta_{2,\mathsf{DR}}-\delta_{2})=\frac{1}{\sqrt{n}}\sum_{i=1}^n \psi_{2,i}+o_P(1)\]
and the result follows by the CLT using the fact that $E\left[\psi_{2i}^2\right]<\infty$ under Assumptions \ref{hyp:parallel_trends}, \ref{hyp:regularity} and \ref{hyp:support_condition0}.

\medskip
Finally, we prove efficiency of the influence function.
As above, the tangent space $\mathcal{T}$ is the whole set of mean-zero, square-integrable random variables, implying that $\psi_2\in \mathcal{T}$. Next, we show that $\delta_2$ is pathwise differentiable. Let us consider a parametric submodel indexed by $\theta$, with the true DGP being at $\theta=\theta_0$ (we omit the indexation by $\theta_0$ hereafter). We denote its score function by $q(Y_1,Y_2, D_1,D_2)$. For all $\theta$, we have, by Theorem \ref{thm:main} used with $\gamma(D_1)=0$,
\[0=E_\theta\left[(S_+-S_-)(\Delta Y-\mu_\theta(D_1))-\abs{\Delta D}\delta_2\right].\]
As a result,
\begin{align*}
	0&=\frac{\partial }{\partial \theta}E_\theta\left[(S_+-S_-)(\Delta Y-\mu_\theta(D_1))-\abs{\Delta D}\delta_2(\theta)\right]\big\vert_{\theta_0}\\
	&=E\left[\left((S_+-S_-)(\Delta Y-\mu(D_1))-\abs{\Delta D}\delta_2\right)q(Y_1,Y_2,D_1,D_2)\right] \\ & -E\left[(S_+-S_-)\frac{\partial }{\partial \theta}\mu_\theta (D_1)\big\vert_{\theta_0}\right] -E[\abs{\Delta D}]\frac{\partial }{\partial \theta}\delta_2(\theta)\big\vert_{\theta_0}.
\end{align*}
Hence, using $E\left[(S_+-S_-)\deriv{\mu_\theta}{\theta}(D_1)\right]=E\left[h_0(D_1)\deriv{\mu_\theta}{\theta}(D_1)\right]$, we get
\begin{align}
	\frac{\partial }{\partial \theta}\delta_2(\theta)\big\vert_{\theta_0}&=\frac{1}{E[\abs{\Delta D}]}E\left[\left((S_+-S_-)(\Delta Y-\mu(D_1))-\abs{\Delta D}\delta_2\right)q(Y_1,Y_2,D_1,D_2)\right] \notag \\
	&-\frac{1}{E[\abs{\Delta D}]}E\left[h_0(D_1)\frac{\partial }{\partial \theta}\mu_\theta (D_1)\right]. \label{eq:for_deriv_mu2}
\end{align}
As shown in the proof of Theorem \ref{thm:ASasnormality}, for any function $a(D_1)$,
\begin{align*}
	E\left[a(D_1)(1-S)(\Delta Y-\mu(D_1))q(Y_1,Y_2, D_1,D_2)\right]&=E\left[a(D_1)(1-S)\frac{\partial}{\partial \theta}\mu(D_1)\big\vert_{\theta_0}\right]\\
	&=E\left[a(D_1)p_0(D_1))\frac{\partial}{\partial \theta}\mu(D_1)\big\vert_{\theta_0}\right].
\end{align*}
Thus, setting $a(D_1):=h_0(D_1)/p_0(D_1)$, we obtain
\begin{align*}
	E\left[\frac{h_0(D_1)}{p_0(D_1)}(1-S)(\Delta Y-\mu(D_1))q(Y_1,Y_2, D_1,D_2)\right]&=E\left[h_0(D_1)\frac{\partial }{\partial \theta}\mu_\theta (D_1)\right].
\end{align*}
Plugging it back into \eqref{eq:for_deriv_mu2}, we obtain
\begin{align*}
	\frac{\partial }{\partial \theta}\delta_2(\theta)\big\vert_{\theta_0}&=\frac{1}{E[\abs{\Delta D}]}E\left[\left((S_+-S_-)(\Delta Y-\mu(D_1))-\abs{\Delta D}\delta_2\right)q(Y_1,Y_2,D_1,D_2)\right]\\
	&-\frac{1}{E[\abs{\Delta D}]}E\left[\frac{h_0(D_1)}{p_0(D_1)}(1-S)(\Delta Y-\mu(D_1))q(Y_1,Y_2, D_1,D_2)\right]\\
	&=E[\psi_2q(Y_1,Y_2,D_1,D_2)].
\end{align*}
Hence, $\delta_2$ is pathwise differentiable, implying also \cite[see, e.g.][p.363]{vanderVaart2000} $\psi_2$ is the efficient influence function $_\Box$

\subsection{Theorem \ref{thm:main_IV}}

Let us start by simplifying the expression of $\delta_{\mathsf{IV}}$.
\begin{align}
	\delta_{\mathsf{IV}}=&E\left(\frac{|D_2(Z_2)-D_2(Z_1)|}{E(|D_2(Z_2)-D_2(Z_1)||SC=1)}\times \frac{Y_2(D_2(Z_2))-Y_2(D_2(Z_1))}{D_2(Z_2)-D_2(Z_1)}\middle|SC=1\right)\nonumber\\
	=&\frac{E[\sgn(D_2(Z_2)-D_2(Z_1))\left(Y_2(D_2(Z_2))-Y_2(D_2(Z_1))\right)|SC=1]}{E[\sgn(D_2(Z_2)-D_2(Z_1))(D_2(Z_2)-D_2(Z_1))|SC=1]}\nonumber\\	
	=&\frac{E[\sgn(\Delta Z)\left(Y_2(D_2(Z_2))-Y_2(D_2(Z_1))\right)|SC=1]}{E[\sgn(\Delta Z)(D_2(Z_2)-D_2(Z_1))|SC=1]}\nonumber\\		
	=&\frac{E[\sgn(\Delta Z)\left(Y_2(D_2(Z_2))-Y_2(D_2(Z_1))\right)]}{E[\sgn(\Delta Z)(D_2(Z_2)-D_2(Z_1))]}.\label{eq:param_IV}
\end{align}
The third equality follows from the fact that by the monotonicity assumption $\sgn(D_2(Z_2)-D_2(Z_1))=\sgn(\Delta Z)$ if $D_2(Z_2)\ne D_2(Z_1)$. The fourth equality follows from the fact that $Y_2(D_2(Z_2))-Y_2(D_2(Z_1))=0$ and $D_2(Z_2)-D_2(Z_1)=0$ if $SC=0$. We now simplify the expression of the estimand identifying $\delta_{\mathsf{IV}}$. Its denominator satisfies
\begin{align*}
	&E\left[\left(S_+^I-S_-^I-(1-S^I)\gamma^I(Z_1,D_1)\right)(\Delta D-\mu_0^D(Z_1,D_1))\right]\\
	=&E\left[\left(S_+^I-S_-^I\right)(\Delta D-\mu_0^D(Z_1,D_1))\right]-E\left[(1-S^I)\gamma^I(Z_1,D_1)(\Delta D  -\mu_0^D(Z_1,D_1))\right]\\
	=&E\left[\left(S_+^I-S_-^I\right)(\Delta D-\mu_0^D(Z_1,D_1))\right]-E\left[E(S_+^I-S_-^I|Z_1,D_1)(E[\Delta D|S^I=0,Z_1,D_1]\right. \\
	& \left. -\mu_0^D(Z_1,D_1))\right]\\
	=&E\left[\sgn(\Delta Z)(\Delta D-\mu_0^D(Z_1,D_1))\right]
\end{align*}
and similarly, its numerator satisfies
\begin{align*}
	E&\left[\left(S_+^I-S_-^I-(1-S^I)\gamma^I(Z_1,D_1)\right)(\Delta Y-\mu_0^Y(Z_1,D_1))\right]=E\left[\sgn(\Delta Z)(\Delta Y-\mu_0^Y(Z_1,D_1))\right].
\end{align*}
Therefore, the result follows if
\begin{align*}
	E[\sgn(\Delta Z)\left(Y_2(D_2(Z_2))-Y_2(D_2(Z_1))\right)] & = E\left[\sgn(\Delta Z)(\Delta Y-\mu_0^Y(Z_1,D_1))\right], \\
	E[\sgn(\Delta Z)(D_2(Z_2)-D_2(Z_1))] & = E\left[\sgn(\Delta Z)(\Delta D-\mu_0^D(Z_1,D_1))\right].
\end{align*}
First, for the denominator,
\begin{align*}
	& E\left[\sgn(\Delta Z)\left(\Delta D-\mu_0^D(Z_1,D_1)\right)\right] \\
	=& E\left[\sgn(\Delta Z)\left(D_2(Z_2)-D_2(Z_1)\right)\right] +E\left[\sgn(\Delta Z)\left(D_2(Z_1)-D_1(Z_1)-\mu_0^D(Z_1,D_1)\right)\right]\\
	=&E\left[\sgn(\Delta Z)\left(D_2(Z_2)-D_2(Z_1)\right)\right]
\end{align*}
because
\begin{align*}
	& E\left[\sgn(\Delta Z)\left(D_2(Z_1)-D_1(Z_1)-\mu_0^D(Z_1,D_1)\right)\right]\\
	=&E\left\{E\left[\sgn(\Delta Z)\left(D_2(Z_1)-D_1(Z_1)-\mu_0^D(Z_1,D_1)\right)|Z_1,Z_2,D_1\right]\right\}\\
	=&E\left\{\sgn(\Delta Z)\left(E\left(\Delta D(Z_1)|Z_1,Z_2,D_1\right)-E(\Delta D(Z_1)|Z_1,S^I=0,D_1)\right)\right\}\\
	=&0,
\end{align*}
by Point 2 of Assumption \ref{hyp:parallel_trends_IV}. Second, for the numerator,
\begin{align*}
	& E\left[\sgn(\Delta Z)\left(\Delta Y-\mu_0^Y(Z_1,D_1)\right)\right] \\
	=& E\left[\sgn(\Delta Z)\left(Y_2(D_2(Z_2))-Y_2(D_2(Z_1))\right)\right] +E\left[\sgn(\Delta Z)\left(Y_2(D_2(Z_1))-Y_1(D_1(Z_1))\right.\right.\\
	&\left.\left. -\mu_0^Y(Z_1,D_1)\right)\right]\\
	=& E\left[\sgn(\Delta Z)\left(Y_2(D_2(Z_2))-Y_2(D_2(Z_1))\right)\right],
\end{align*}
where the last equality follows from Point 1 of Assumption \ref{hyp:parallel_trends_IV}. This completes the proof $_\Box$

\section{Proofs of the results of the web appendix}

\subsection{Lemma \ref{lem:lemmaX1}}

We have
\begin{align*}
	& E((Y_2(d_2) - Y_2(d_1))\mid D_1 = d_1, D_2 = d_2) \\
	= & E(([Y_2(d_2) - Y_1(d_1)] -[Y_2(d_1) -  Y_1(d_1)])\mid D_1 = d_1, D_2 = d_2) \\
	= & E(\Delta Y\mid D_1 = d_1, D_2 = d_2) - E(\Delta Y(d_1)\mid D_1 = d_1, D_2 = d_2) \\
	= & E(\Delta Y\mid D_1 = d_1, D_2 = d_2) - E\left\{E(\Delta Y(d_1)\mid D_1 = d_1, D_2 = d_2,X)\mid D_1 = d_1, D_2 = d_2\right\} \\
	= & E(\Delta Y\mid D_1 = d_1, D_2 = d_2) - E\left\{E(\Delta Y(d_1)\mid D_1 = d_1, D_2 = d_1,X)\mid D_1 = d_1, D_2 = d_2\right\} \\
	= & E(\Delta Y\mid D_1 = d_1, D_2 = d_2) - E\left\{E(\Delta Y\mid D_1 = d_1, S = 0,X)\mid D_1 = d_1, D_2 = d_2\right\} \\
	= & E\left\{(\Delta Y- E(\Delta Y\mid D_1 = d_1, S = 0,X))\mid D_1 = d_1, D_2 = d_2\right\}.
\end{align*}
which completes the proof. $_\Box$

\subsection{Theorem \ref{thm:thmX1}}

Start by noting that:
\[E\left((1-S)\alpha_0(X_1,D_1)(\Delta Y-\mu_0(X_1,D_1))\right)=0\]
and thus the result can be rewritten as:
\[\delta_1=E\left(\left.\frac{\Delta Y-\mu_0(X_1,D_1)}{\Delta D}\right\vert S=1\right).\]
Then, by the law of iterated expectations and Lemma \ref{lem:lemmaX1},
\begin{align*}
\delta_1:&=E\left(\frac{Y_2(D_2) - Y_2(D_1)}{D_2-D_1} \middle| S=1\right) \\
& =  E\left(E\left(\frac{Y_2(D_2) - Y_2(D_1)}{D_2-D_1}\middle| D_1,D_2,S=1\right) \middle| S=1\right) \\
& =  E\left(E\left\{\frac{\Delta Y- E(\Delta Y\mid D_1, S = 0,X_1)}{\Delta D}\mid D_1, D_2,S=1\right\} \middle| S=1\right) \\
& =  E\left(\frac{\Delta Y- E(\Delta Y\mid D_1, S = 0,X_1)}{\Delta D} \middle| S=1\right)\\
& = E\left(\frac{\Delta Y- \mu_0(X_1,D_1)}{\Delta D} \middle| S=1\right)
\end{align*}
Similarly, for $\delta_2$, using that:
\[E\left((1-S)\gamma_0(X_1,D_1)(\Delta Y -\mu_0(X_1,D_1))\right)=0,\]
we can rewrite the result as:
\[\delta_2=\frac{E\left(\sgn(\Delta D)(\Delta Y -\mu_0(X_1,D_1))\right)}{E(\abs{\Delta D})}\]
and thus by the law of iterated expectations and Lemma \ref{lem:lemmaX1},
\begin{align*}
\delta_{2}:&=\frac{E\left(\text{sgn}(\Delta D) (Y_2(D_2)-Y_2(D_1))\right)}{E(|\Delta D|)} \\
& =\frac{ P(S=1)E\left(E\left\{\text{sgn}(\Delta D) (Y_2(D_2)-Y_2(D_1))\mid D_1,D_2,S=1\right\}\mid S=1\right)}{ E(|\Delta D|)} \\
& =\frac{ P(S=1)E\left(E\left\{\text{sgn}(\Delta D) (\Delta Y-E(\Delta Y\mid D_1, S=0, X))\mid D_1,D_2,S=1\right\}\mid S=1\right)}{ E(|\Delta D|)} \\
& =\frac{E\left(\text{sgn}(\Delta D) (\Delta Y-E(\Delta Y\mid D_1, S=0, X))\right)}{ E(|\Delta D|)}\\
&=\frac{E\left(\text{sgn}(\Delta D) (\Delta Y-\mu_0(X_1,D_1))\right)}{ E(|\Delta D|)}
\end{align*}
where the first equality follows by definition of $S= \mathbf{1}\{\sgn(\Delta D)\neq 0\}$ and using the law of iterated expectations, and the second equality follows from Lemma \ref{lem:lemmaX1}. $_\Box$

\end{document}